\documentclass[review]{elsarticle}

\usepackage{lineno,hyperref}
\usepackage{amsmath}
\usepackage{amssymb}
\usepackage{cancel}
\usepackage{bm}
\usepackage{caption}
\usepackage{subcaption,color}
\usepackage{xcolor}
\usepackage{soul}
\usepackage{verbatim}  
\usepackage{makecell}
\usepackage{adjustbox}
\usepackage[noabbrev]{cleveref}
\biboptions{sort&compress}
\modulolinenumbers[5]

\newcommand{\sigmat}{\sigma_\mathrm{t}}
\newcommand{\sigmas}{\sigma_\mathrm{s}}

\newcommand{\SN}{S$_N$}
\newcommand{\PN}{P$_N$}
\renewcommand{\vec}[1]{\bm{#1}} 
\newcommand{\vd}{\bm{\cdot}} 
\newcommand{\grad}{\vec{\nabla}} 

\begin{document}

\begin{frontmatter}

\title{A sweep-based low-rank method for the discrete ordinate transport equation} 

\author[mymainaddress]{Zhuogang Peng}
\ead{zpeng5@nd.edu}

\author[mymainaddress]{Ryan G.\ McClarren\corref{mycorrespondingauthor}}
\cortext[mycorrespondingauthor]{Corresponding author}
\ead{rmcclarr@nd.edu}
 
\address[mymainaddress]{Department of Aerospace and Mechanical Engineering, University of Notre Dame, Notre Dame, IN, 46545, USA}

\begin{abstract}
The dynamical low-rank (DLR) approximation is an efficient technique to approximate the solution to matrix differential equations. Recently, the DLR method was applied to radiation transport calculations to reduce memory requirements and computational costs. This work extends the low-rank scheme for the time-dependent radiation transport equation in 2-D and 3-D Cartesian geometries with discrete ordinates discretization in angle (SN method). The reduced system that evolves on a low-rank manifold is constructed via an “unconventional” basis update \& Galerkin integrator to avoid a substep that is backward in time, which could be unstable for dissipative problems. The resulting system preserves the information on angular direction by applying separate low-rank decompositions in each octant where angular intensity has the same sign as the direction cosines. Then, transport sweeps and source iteration can efficiently solve this low-rank-SN system. The numerical results in 2-D and 3-D Cartesian geometries demonstrate that the low-rank solution requires less memory and computational time than solving the full rank equations using transport sweeps without losing accuracy.
\end{abstract}


\end{frontmatter}


\section{Introduction}
The radiation transport equation (RTE) models the movement of particles through materials and their interactions with background media. The application of RTE among various science and engineering communities, such as optical imaging \cite{KIM2003}, neutron transport \cite{Azmy2010}, rarefied gas dynamics \cite{cercignani2000rarefied}, to name a few, require solving RTE accurately and efficiently. The accurate yet computationally efficient RTE solution has been an active research question for decades but remains open because of its rich dimensional spaces. The solution to the RTE is radiation intensity, which is determined by up to seven independent variables, including three in space, two in angle, one in time, and one in energy. Thus, the computation could consume a large amount of computer memory, challenging proposed exascale computing systems \cite{ShalfJohn2020Tfoc}. This work aims to design a fast and memory-efficient numerical scheme for solving the RTE. 

The angular treatment is essential to solve the RTE. One commonly used numerical method to discretize the angular variable in RTE is the spherical harmonic (\PN) method \cite{pomraning2005equations, heizler2010asymptotic, mcclarren2007quasilinear} that expresses the angular dependence of the radiation density using orthogonal bases on the unit sphere. This method has many desired properties, such as rotational invariance, but it suffers from oscillations that destroy the robustness of the method \cite{mcclarren2008solutions}. 

Another popular angular treatment is the discrete ordinates (\SN) method  \cite{LewisMiller} that solves the radiation intensity along with particular directions and uses quadrature to estimate moments of the intensity. Given that the direction of movement of radiation intensity is pre-selected, the resulting discrete ordinates system can be solved by a very efficient method called a transport sweep. This makes the \SN~method popular in high-performance computing applications because it is computationally efficient. For this reason, we seek to develop novel methods based on \SN that is compatible with these transport sweeps. 

This work follows the development of the dynamical low-rank (DLR) method applied in transport calculations to reduce the computer memory requirements and the computational cost. The DLR method aims to approximate large time-dependent matrices determined by matrix differential equations \cite{Koch2007}. The desired approximation has three components similar to factors in singular value decomposition (SVD). Each of them is solved by integrating the matrix differential equation projected onto the tangent space of the low-rank manifold. We refer to \cite{Nonnenmacher2008,Lubich2014,Kieri2016} for more background. In previous work, the authors applied the DLR method to transport calculations with a spherical harmonics expansion and an explicit time scheme \cite{peng2020lowrank}. Later, a high-order/low-order algorithm was developed in \cite{peng2021holo} to overcome the conservation loss in the low-rank evolution. For more analytical details, there is an error analysis for the backward Euler and Crank-Nicholson methods \cite{ding2021error}. In \cite{Einkemmer2021ap} the asymptotic-preserving property is achieved by a macro-micro decomposition of the transport equation. 

In this work, we develop a practical computation scheme with the discrete-ordinates model. By applying the time integrator proposed recently in \cite{ceruti2020unconventional}, we avoid a substep that is backward in time, which could be unstable for dissipative problems as described in \cite{ding2021error}. We then solve the low-rank equation with the iterative approach and transport sweeps. 

The remainder of this paper begins with a brief review of the \SN~formulation of the radiation transport equation. Then the low-rank representation of the \SN~equations is derived. Section 4 will be the numerical implementation details, including the spatial discretization and the computational method for the resulting matrix equations. The efficiency and accuracy of our low-rank algorithms are demonstrated in section 5. Section 6 presents a discussion.

\section{Discrete Ordinates Radiation Transport Equation}
We begin with the time-dependent radiative transfer equation with one energy group given by 
\begin{linenomath*}
	\begin{equation} \label{eq: RadiativeTransfer} 
		\begin{split}
			\frac{1}{c} \frac{\partial \psi (\vec{r}, \hat{\Omega}, t)}{\partial t} + \hat{\Omega} \vd \grad \psi (\vec{r}, \hat{\Omega}, t)  +  \sigmat(\vec{r}) \psi (\vec{r}, \hat{\Omega}, t) = 
			\frac{1}{4\pi}\sigmas(\vec{r}) \phi(\vec{r}, t) + Q(\vec{r},t).
		\end{split}
	\end{equation}
\end{linenomath*}
In this equation, $\psi(\vec{r}, \hat{\Omega}, t)$ is the radiation intensity with units of  particles per area per steradian per time. We use the standard notation with $\vec{r} = (x, y, z) \in \mathbb{R}^3$ being the position, the unit angular vector $\hat{\Omega}(\mu, \varphi) \in \mathbb{S}^2$ specified by the cosine of the polar angle $\mu \in [-1, 1]$ and the azimuthal angle $\varphi \in [0, 2\pi]$, and $t$ as the time. Additionally, $\sigma_s(\vec{r}) $ and $\sigma_t(\vec{r}) $ are isotropic scattering and total macroscopic cross-sections with units of inverse length; $Q(\vec{r},t)$ is a prescribed source. We integrate $\psi(\vec{r}, \hat{\Omega}, t)$ over all angles to obtain the scalar intensity:
\begin{linenomath*}
	\begin{equation}
		\phi(\vec{r},t) = \int_{4 \pi}\psi(\vec{r}, \hat{\Omega}, t) \, d \hat{\Omega}.
	\end{equation}
\end{linenomath*}
The scalar intensity is important because it can be used to compute reaction rate densities (e.g., absorption rate density) that determine the coupling with other physical operators in a given system.

We apply the \SN~discretization to the angular variables with a finite quadrature set $\{(\hat{\Omega}_n, w_n)\: |\: 1 \leq n \leq N\}$, and solve Eq.~\eqref{eq: RadiativeTransfer} along these angular directions as
\begin{linenomath*}
	\begin{equation} \label{eq: RadiativeTransfer_SN} 
		\frac{1}{c} \frac{\partial \psi{_n} (\vec{r}, t)}{\partial t} + \hat{\Omega}_n \vd \grad \psi{_n} (\vec{r}, t)  +  \sigmat(\vec{r}) \psi_{n} (\vec{r}, t) 
		= \frac{1}{4\pi} \sigmas(\vec{r}) \phi(\vec{r}, t) + Q(\vec{r},t).
	\end{equation}
\end{linenomath*}
where the discrete direction $\hat{\Omega}_n$ is specified by the direction cosines $\mu_n$, $\eta_n$ and $\xi_n$. The angular integral is approximated by quadrature rules, e.g., suppose there are $N_{\Omega}$ directions, the scalar intensity can be written as 
\begin{equation}\label{eq: scalarfluxSN}
	\phi(\vec{r}, t) = \sum_{n}^{N_{\Omega}}\omega_n \psi_n(\vec{r}, t),
\end{equation}
Many kinds of quadrature sets can be implemented to the \SN~method, such as level-symmetric \cite{Jenal1977}, Legendre-Chebyshev, and Legendre-equal weight \cite{Longoni2004}, to name a few. We refer to \cite{Hunter2013} for a comprehensive comparison. We use the Legendre-Chebyshev quadrature sets throughout this work because it enables a high quadrature order, which is not available with level-symmetric quadrature sets \cite{Endo2007}. Then we have the formula for the number of discrete ordinates in each octant as  $N = \frac{SN^2}{4}$ and $N_{\Omega} = Nn_o$, where $SN$ is the order of discrete ordinates, $n_o$ is the number of octants, e.g., $n_o = 4$ for 2-D space and $n_o = 8$ for 3-D space, and $N$ is the number of angles per octant. 

One most commonly used method to solve Eq.~\eqref{eq: RadiativeTransfer_SN} is source iteration and transport sweep equipped with the implicit Euler method for time discretizations. To simplify the notation we define several operators that only used in this section \cite{warsaKrylov}
\begin{linenomath*}
	\begin{equation} \label{eq: collosion_operator} 
		L = \hat{\Omega} \vd \grad  +  \sigmat(\vec{r}) 
	\end{equation}
\end{linenomath*}
\begin{linenomath*}
	\begin{equation} \label{eq: moment_to_discrete} 
		M\phi = \frac{1}{4\pi} \phi, \qquad 
		D \psi =  \sum_{n=1}^{N_\Omega} w_n \psi_n\, d \hat{\Omega}, \qquad 
		S \phi = \sigmas(\vec{r}) \phi.
	\end{equation}
\end{linenomath*}
where $L$ denotes the streaming and collision operator that operates on the angular intensity at the $N_\Omega$ angles and outputs a vector of the same size, $M$ is known as the moment-to-discrete operator that projects the scalar intensity to the angular intensity, $S$ is the scattering operator, and $D$ is the discrete-to-moment operator that represents the angular quadrature. Using these operators we obtain the abstract form of Eq.~\eqref{eq: RadiativeTransfer_SN} as 
\begin{linenomath*}
	\begin{equation} \label{eq: RadiativeTransfer_SN_BE} 
		\frac{1}{c \Delta t} \psi^{\ell+1} + L \psi^{\ell+1} 
		= MSD\psi^{\ell+1} + Q + \frac{1}{c \Delta t} \psi^{\ell},
	\end{equation}
\end{linenomath*}
where $\psi^{\ell}$ is the solution for the radiation intensity at time step $\ell$ at each angle $\Omega_n$. We can further simplify this equation into a quasi-steady form by defining  
\begin{linenomath*}
	\begin{equation} \label{eq: collosion_operator_further} 
		L^{*} = \hat{\Omega}_n \vd \grad  +  \left(\sigmat(\vec{r}) + \frac{1}{c \Delta t}\right), \qquad q = Q + \frac{1}{c \Delta t} \psi^{\ell}.
	\end{equation}
\end{linenomath*}
to get
\begin{linenomath*}
	\begin{equation} \label{eq: RadiativeTransfer_SN_BE_abs} 
		L^{*} \psi^{\ell+1} 
		= S\psi^{\ell+1} + q.
	\end{equation}
\end{linenomath*}

Discrete ordinates with implicit Euler time integration can be computed using an efficient algorithm called a ``transport sweep," which is highly desirable in computation. The idea is that we can solve for the unknowns by marching along the direction of the flow of particles from the given boundary conditions and iterating over a lagged scattering source. After spatial discretizations, $\left(L^*\right)^{-1}$ is a triangular matrix for each angle so that 
\begin{equation}
	\phi^{\ell+1,c+1} = D\left(L^*\right)^{-1} MS\phi^{\ell+1,c} + Dq
\end{equation}
can be computed by 
a simple lower or upper triangular solve in each angle. This iterative method is known as the source iteration (SI) method \cite{LewisMiller}. These iterations are repeated until the scalar intensity $\phi$ converges, and we denote $c$ as the iteration index. A further benefit of this method is that $\psi^{\ell+1,c}$ does not need to be stored during the iteration process. After convergence, it can be computed using one more application of $\left(L^*\right)^{-1}$.

\section{Low-rank Representations of the Solution}
\begin{figure}[h!] 
	\begin{minipage}{\textwidth}
		\centering
		\includegraphics[width=\textwidth]{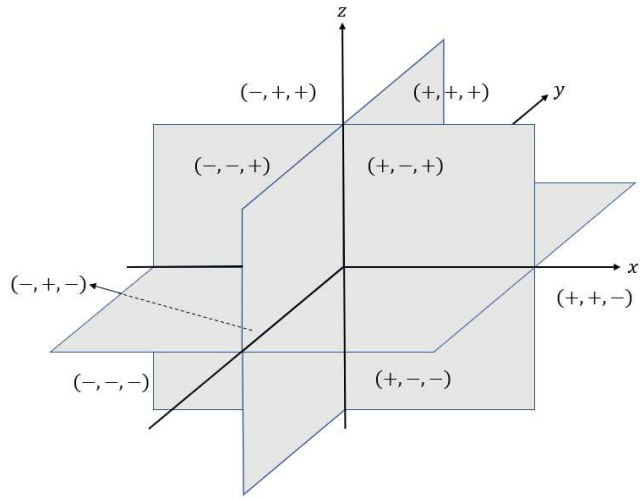}
	\end{minipage}%
	\caption{The sign of direction cosines in the form ($x$,$y$,$z$) in the 3D Cartesian geometry for each octant. }
	\label{fig: octants}
\end{figure}
The preliminary requirement for transport sweeps is the known direction of particle advection. As mentioned above, the original \SN~equations \eqref{eq: RadiativeTransfer_SN} can be updated with a transport sweep using a triangular solve along one discrete ordinate at a time. This property is not guaranteed to be preserved by low-rank equations if the low-rank procedure combines solutions along angles from different octants. As shown later, the low-rank formulation reduces the number of equations by coupling them with linear combinations, so different directions are mixed and solved together. The key idea of the low-rank method proposed in this work is to apply a separate low-rank decomposition to the unknowns with different sign combinations of the direction cosines, as shown in Figure \ref{fig: octants}. In other words, we group the equations in the same octant, so they have the same propagation directions even after being coupled.

We begin the derivation by writing the low-rank approximation of rank $r$ to the solution of \eqref{eq: RadiativeTransfer_SN} in the form of a linear combination of $r$ basis functions similar as \cite{peng2020lowrank}
\begin{linenomath*}
	\begin{equation} 
		\label{eq: lra_psi}
		\vec{\psi}^{k}(\vec{r}, t) = \sum_{ij}^{r}X^{k}_i(\vec{r}, t) S_{ij}^{k}(t)V_j^{k}(t)
	\end{equation}
\end{linenomath*}
where $\vec{\psi}^{k} = [\psi^k_1, \psi^k_2, ..., \psi^k_{N}]^T$ collect all the intensities in octant $k$, and $X_i^k(\vec{r}, t)$ and $V_j^k(t)$ are orthonormal bases, e.g., $\langle X^k_i, X^k_j \rangle_{\vec{r}} = V^k_i V^k_j = \delta_{ij}$.  The inner product over space is defined as $\langle f,g\rangle _{\vec{r}} = \int_{\mathbb{R}^3} f  g \, d\vec{r}$. We build the low-rank equation in each octant by using $ \bar{X}^k = \{X^k_1, X^k_2, ..., X^k_{r}\} $ and $\bar{V}^k = \{V^k_1, V^k_2, ..., V^k_{r}\} $ as ansatz spaces. Then we define orthogonal projectors using the bases:
\begin{linenomath*}
	\begin{equation}\label{eq: SN_Projection1}
		P(\bar{X}^k) \, g = \sum_{i=1}^{r}X^k_i \langle X^k_i g \rangle_{\vec{r}},
	\end{equation}
\end{linenomath*}
\begin{linenomath*}
	\begin{equation}\label{eq: SN_Projection2}
		P(\bar{V}^k) \, g = \sum_{j=1}^{r}V^k_j  V^k_j g .
	\end{equation}
\end{linenomath*}
The projector onto the tangent space of the low-rank manifold {$\mathcal{M}_r$} is given by
\begin{linenomath*}
	\begin{equation}
		\label{projector0}
		{P(\bar{X}, \bar{V}) \, g = P_{\bar{V}}g  - P_{\bar{X}} P_{\bar{V}} g + P_{\bar{X}} g}.
	\end{equation}
\end{linenomath*}
Using this projector, we can define the low-rank governing equations as
\begin{linenomath*}
	\begin{equation}
		\label{eq: lra_transport}
		\partial_t \vec{\psi}^{k} = P(\bar{X}^{k}, \bar{V}^{k}) F(t, \vec{\psi}^{k})
	\end{equation}
\end{linenomath*}
where the $n_{th}$ component of $F(t,\vec{\psi}^k)$ is
\begin{linenomath*}
	\[
	F (t, \vec{\psi}^{k} )_n = -\hat{\Omega}_n \vd \grad {\psi}_n^{k}  -  \sigmat(\vec{r}) \vec{\psi}_n^{k} + 
	\frac{1}{4\pi}\sigmas(\vec{r}) \phi + Q.
	\]
\end{linenomath*}

We apply the integrating method proposed in \cite{ceruti2020unconventional} to solve the Eq.\eqref{eq: lra_transport} at $t_0+h$ from $t_0$ implicitly where $h$ is the step size. The initial condition is given by the low-rank formulation of $\vec{\psi}^{k}(\vec{r}, t_0)$
\begin{linenomath*}
	\begin{equation}\label{eq: SN_ini_cond}
		\vec{\psi}^{k}(\vec{r},t_0) = \sum_{i,j = 1}^r X^{k}_i(\vec{r}, t_0)S^{k}_{ij}(t_0)V^{k}_j(t_0) = \sum_{i,j = 1}^r X^{k, 0}_iS^{k, 0}_{ij}V^{k, 0}_j.
	\end{equation}
\end{linenomath*}
and the intended solution at $t_0+h$ is denoted as 
\begin{linenomath*}
	\begin{equation}
		\begin{aligned}\label{eq: SN_ini_sol}
			\vec{\psi}^{k}(\vec{r},t_0+h) &= \sum_{i,j = 1}^r X^{k}_i(\vec{r}, t_0+h)S^{k}_{ij}(t_0+h)V^{k}_j(t_0+h)\\  &= \sum_{i,j = 1}^r X^{k, 1}_i S^{k, 1}_{ij}V^{k, 1}_j.
		\end{aligned}
	\end{equation}
\end{linenomath*}

The first part of the solving procedure computes $X^{k, 1}_i$ and $V^{k, 1}_j$ by solving 
\begin{linenomath*}
	\begin{equation}
		\label{eq: lra_transport_K1}
		\partial_t \vec{\psi}^{k} = P_{\bar{V}} F(t, \vec{\psi}^{k})
	\end{equation}
\end{linenomath*}
and
\begin{linenomath*}
	\begin{equation}
		\label{eq: lra_transport_L1}
		\partial_t \vec{\psi}^{k} = P_{\bar{X}} F(t, \vec{\psi}^{k})
	\end{equation}
\end{linenomath*}
in parallel. In Eq.~\eqref{eq: lra_transport_K1} the basis $V^{k}_j$ does not change with time and is evaluated at the initial value $V^{k, 0}_j$. We simplify the notation by writing $K^{k}_j(\vec{r},t) = \sum^{r}_{i} X^{k}_{i}(\vec{r},t)S^{k}_{ij}(t)$, then the initial condition \eqref{eq: SN_ini_cond} becomes
\begin{linenomath*}
	\begin{equation}\label{eq: SN_ini_cond2}
		\vec{\psi}^{k, 0} = \sum_{j=1}^{{r}} K^{k, 0}_j V^{k, 0}_j.
	\end{equation}
\end{linenomath*}
We plug Eq. \eqref{eq: SN_ini_cond2} into Eq.~\eqref{eq: lra_transport_K1} and multiply the both sides by $V^{k, 0}_\ell$ to get
\begin{linenomath*}
	\begin{equation}
		\label{eq: psi1}
		\partial_t K^{k}_j = -\sum_{l=1}^{{r}} \grad K^{k}_\ell \, V^k_\ell \vec{\hat{\Omega}}^k V^k_j - \sigmat K^k_j + \frac{1}{4\pi}\sigmas \phi V^{k}_j + Q V^{k}_j.
	\end{equation}
\end{linenomath*}

Note that we use $\vec{\hat{\Omega}}^k$ to collect all the ordinates in octant $k$. This equation can be solved by the source iteration method and we will return to it with more details in the next subsection. The solution to Eq \eqref{eq: psi1} is factored into $K^{k}_j(\vec{r}, t_0 + h) = \sum^{r}_{i} X^{k, 1}_{i}\Tilde{T}^{k}_{ij}$ by a QR decomposition. 

Similarly, by writing $L^{k}_i(t) = \sum_j^{r} S^{k}_{ij}(t)V^{k}_{j}(t)$ we can expand the Eq.\eqref{eq: lra_transport_L1} as
\begin{linenomath*}
	\begin{multline}\label{eq: psi2}
		\frac{d}{dt} L^{k}_{i} = -\vec{\hat{\Omega}}^k \sum_{l}^{r} \langle \grad X^{k}_\ell \, X^{k}_i \rangle_{\vec{r}} L^{k}_\ell - \sum_{l}^{r} \langle \sigma_t X^{k}_\ell X^{k}_i \rangle_{\vec{r}} L^{k}_{l} 
		\\ + \frac{1}{4 \pi} \sigmas  \langle X^k_i \phi \rangle_{\vec{r}} +  \langle X^{k}_i Q \rangle_{\vec{r}}.
	\end{multline}
\end{linenomath*}
With the initial condition $L^{k, 0}_i = \sum_j^{r} S^{k, 0}_{ij}V^{k, 0}_{j}$, we obtain the basis $V^{k, 1}_j$ by factoring the solution $L^{k, 1}_i$ into $\sum_{j=1}^{r} \Tilde{R}^{k}_{ij}V^{k, 1}_j$ using a QR decomposition. We mention that $\Tilde{R}^{k}_{ij}$ and $\Tilde{T}^{k}_{ij}$ are not kept in these two steps. 

Last we update $S^{k}_{ij}(t)$ by solving the matrix differential equation
\begin{linenomath*}
	\begin{multline}\label{eq: psi3}
		\frac{d}{dt} S^{k}_{ij} = -\sum_{ml}^{r} \langle \partial_{\vec{r}} X^{k}_m \, X^{k}_i \rangle_{\vec{r}} S^{k}_{ml}  V^{k}_\ell \vec{\hat{\Omega}}^{k} V^{k}_j
		- \sum_{m}^{r} \langle \sigma_t X^{k}_m X^{k}_i \rangle_{\vec{r}} S^{k}_{mj} \\ + \frac{1}{4 \pi} \sigmas \langle X^k_i \phi \rangle_{\vec{r}} V^k_j + \langle X^k_i Q \rangle_{\vec{r}} V^k_j
	\end{multline}
\end{linenomath*}
with the initial condition $S^{k, 0}_{ij} = \sum_{m,l = 1}^{r} \langle X^{k, 1}_i \, X^{k, 0}_m \rangle_{\vec{r}} S^{k, 0}_{ml}V^{k, 0}_\ell V^{k, 1}_j.$

\section{Implementation Details}
This section presents the numerical scheme for solving Eqs.\eqref{eq: psi1}-\eqref{eq: psi3}. Specifically, we apply the finite volume method for the spatial discretization and the backward Euler for the implicit time integration. For simplicity we present the method in 1-D slab geometry. We begin with the governing equations discussed above  
\begin{linenomath*}
	\begin{equation}
		\label{eq: psi1_1D}
		\partial_t K^{k}_j = -\sum_{l=1}^{r} \partial_x K^{k}_\ell \, V^k_\ell \vec{\mu}^k V^k_j - \sigmat K^k_j + \frac{1}{2} \sigmas \phi V^{k}_j + \frac{1}{2}Q V^{k}_j,
	\end{equation}
\end{linenomath*}
\begin{linenomath*}
	\begin{multline}\label{eq: psi2_1D}
		\frac{d}{dt} L^{k}_{i} = -\vec{\mu}^k \sum_{l}^{r} \langle \partial_x X^{k}_\ell \, X^{k}_i \rangle_{x} L^{k}_\ell - \sum_{l}^{r} \langle \sigma_t X^{k}_\ell X^{k}_i \rangle_{x} L^{k}_{l}  \\ + \frac{1}{2} \sigmas  \langle X^k_i \phi \rangle_{x} +  \langle X^{k}_i Q \rangle_{x},
	\end{multline}
\end{linenomath*}
and
\begin{linenomath*}
	\begin{multline}
		\label{eq: psi3_1D}
		\frac{d}{dt} S^{k}_{ij} = -\sum_{ml}^{r} \langle \partial_x X^{k}_m \, X^{k}_i \rangle_{x} S^{k}_{ml}  V^{k}_\ell \vec{\mu}^{k} V^{k}_j
		- \sum_{m}^{r} \langle \sigma_t X^{k}_m X^{k}_i \rangle_{x} S^{k}_{mj}\\ + \frac{1}{2} \sigmas \langle X^k_i \phi \rangle_{x} V^k_j + \langle X^k_i Q \rangle_{x} V^k_j,
	\end{multline}
\end{linenomath*}
where the superscript $k = +$ or $-$ denotes directions moving in positive or negative $x$-direction, and $\vec{\mu}^k$ collects the cosine of corresponding polar angles.

\subsection{Numerical Scheme}
We define the orthonormal bases $X_i$ and $V_j$ as 
\begin{linenomath*}
	\begin{equation}\label{eq: discretization1}
		X^k_i(t,x) = \sum_{m=1}^{M}Z_m(x)u^k_{mi}(t),
	\end{equation}
\end{linenomath*}
\begin{linenomath*}
	\begin{equation}\label{eq: discretization2}
		V^k_j(t) = v^k_{*, j}(t),
	\end{equation}
\end{linenomath*}
where $Z_m(x) = \frac{1}{\sqrt{\Delta x}}$ with $x \in [x_{m-\frac{1}{2}},x_{m+\frac{1}{2}}]$ are based on a finite volume discretization in space with a constant mesh spacing $\Delta x$ and $M$ zones; $m$ is the cell number. Here $u^{k}_{mi}$ are components of the time dependent matrix $U^{k}(t) \in \mathbb{R}^{M \times r}$, and $v^k_{*, j}$ refers to the $j_{th}$ column of the $V^k(t) \in \mathbb{R}^{N \times r}$. Note that the ${N} \times 1$ vector $v^{k}_{*,j}$ is a group of linear factors of all positive or negative ordinates, so there are ${r}$ combinations for each set of angles.  

Next we describe procedures to solve Eq.~\eqref{eq: psi1_1D} using source iteration and transport sweeps. By applying the standard upwinding technique, we write the spatial derivative term using upwind finite differences as
\begin{linenomath*}
	\begin{equation}\label{eq: term_K}
		\begin{aligned}
			\partial_x K^{+}_{\ell} \, V^+_\ell \vec{\mu}^+ V^+_j \approx &  \frac{1}{\Delta x}(K^+_{i, \ell} - K^+_{i-1, \ell}) \, V^+_\ell \vec{\mu}^+ V^+_j, \\
			\partial_x K^{-}_{\ell} \, V^-_\ell \vec{\mu}^- V^-_j \approx &  \frac{1}{\Delta x}(K^-_{i+1, \ell} - K^-_{i, \ell}) \, V^-_\ell \vec{\mu}^- V^-_j .
		\end{aligned}
	\end{equation}
\end{linenomath*}
Note that $ V^k_\ell \vec{\mu}^k V^k_j $ forms a matrix $R^k = (r^k_{\ell j})\in \mathbb{R}^{r \times r}$ that is precomputed. Thus we can discretize Eq.~\eqref{eq: psi1_1D} using backward Euler method for the time integration
\begin{linenomath*}
	\begin{equation}
		\begin{aligned}
			\frac{1}{h} K^{+, 1}_{i, j} &= -\frac{1}{\Delta x}\sum_{l=1}^{r} (K^{+, 1}_{i, \ell} - K^{+, 1}_{i-1, \ell}) \, r^+_{\ell j} - \sigmat K^{+, 1}_{i, j} + s^{+}_{ij} \qquad \text{for $\mu > 0$}\\
			\frac{1}{h} K^{-, 1}_{i, j} &= -\frac{1}{\Delta x}\sum_{l=1}^{r} (K^{-, 1}_{i+1, \ell} - K^{-, 1}_{i, \ell}) \, \, r^-_{lj} - \sigmat K^{-, 1}_{i, j} + s^{-}_{ij} \qquad \text{for $\mu < 0$}
		\end{aligned}
		\label{eq: K_dis}
	\end{equation}
\end{linenomath*}
where $(s^k_{i j})^\pm = \frac{1}{2} \sigmas \phi V^{\pm} + \frac{1}{2}Q V^{\pm} + \frac{1}{h} K^{\pm, 0} \in \mathbb{R}^{M \times r}$ is the source term and the superscript 1 denotes the current time step and 0 denotes the previous time step. For maximal clarity we write out Eq.~\eqref{eq: K_dis2} in full, matrix form:
\begin{linenomath*}
	\begin{equation}
		\begin{aligned}
			\begin{bmatrix}
				P^+_1&0&0&...&0
				\\
				-\bar{R}^+&P^+_2&0&...&0
				\\
				0&-\bar{R}^+&P^+_3&...&0
				\\
				...&...&...&...&...
				\\
				0&0&...&-\bar{R}^+&P^+_M)
			\end{bmatrix}
			\begin{bmatrix}
				K^{+, 1}_{1\,*}\\K^{+, 1}_{2\,*}\\K^{+, 1}_{3\,*}\\...\\K^{+, 1}_{M ,\, *}
			\end{bmatrix}&= \begin{bmatrix}
				s^+_{1, *} + b^{+}_L
				\\
				s^+_{2, *}
				\\
				s^+_{3, *}
				\\
				...
				\\
				s^+_{M, *} + b^{+}_R
			\end{bmatrix}\qquad \text{for $\mu > 0$},
			\\
			\\
			\begin{bmatrix}
				P^-_1&\bar{R}^-&0&...&0
				\\
				0&P^-_2&\bar{R}^-&...&0
				\\
				...&...&...&...&...
				\\
				0&0&...&P^-_3&\bar{R}^-
				\\
				0&0&...&0&P^-_M
			\end{bmatrix}
			\begin{bmatrix}
				K^{-, 1}_{1\,*}\\K^{-, 1}_{2\,*}\\K^{-, 1}_{3\,*}\\...\\K^{-, 1}_{M ,\, *}
			\end{bmatrix} &= \begin{bmatrix}
				s^-_{1, *} + b^{-}_L
				\\
				s^-_{2, *}
				\\
				s^-_{3, *}
				\\
				...
				\\
				s^-_{M, *} + b^{-}_R
			\end{bmatrix} \qquad \text{for $\mu < 0$}.
		\end{aligned}
		\label{eq: K_matrix}
	\end{equation}
\end{linenomath*}
Here $L^+_i = \frac{1}{\Delta x}R^+ + \left (\sigma_t(i) + \frac{1}{h} \right )I_r$, $L^-_i = -\frac{1}{\Delta x}R^- + \left (\sigma_t(i) + \frac{1}{h} \right )I_r$, $I_r$ is the $r \times r$ identity matrix, $\bar{R}^k = \frac{1}{\Delta x} R^k$, $\sigma_t(i)$ denotes the value of total cross-section in cell $i$, $b^k_L$ and $b^k_R$ are the left and the right boundary values. We set $b^k_L = b^k_R = 0$ to correspond to vacuum boundary conditions. A compact form of the marching scheme for Eq.\eqref{eq: K_matrix} is  
\begin{linenomath*}
	\begin{equation}
		\begin{aligned}
			K^{+, 1}_{i,*} &= \left (P_i^+ \right)^{-1} \left ( \bar{R}^+ K^{+, 1}_{i-1, *} + s^+_{i, *}\right) \qquad \text{for $\mu > 0$}\\
			K^{-, 1}_{i,*} &= \left ( P_i^- \right)^{-1} \left ( -\bar{R}^- K^{-, 1}_{i+1, *} + s^-_{i, *}\right) \qquad \text{for $\mu < 0$}
		\end{aligned}       
	\end{equation}\label{eq: K_dis2}
\end{linenomath*}


Eqs.~\eqref{eq: psi2_1D} and \eqref{eq: psi3_1D} are matrix ordinary differential equations and can be solved by an implicit Runge–Kutta method. 


\subsection{Computational Cost}
This work focuses on reducing the computer memory cost for radiative transfer simulations. The memory consumption for solving Eq.~\eqref{eq: RadiativeTransfer_SN} using the classical iteration solution procedure is $2 \times M \times (N_{\Omega} + 1)$ during each time step to store the angular and scalar intensities.  
Here we do not include the simulation parameters such as the scattering/absorption cross-sections and the prescribed source based on the assumption that they can be stored efficiently during the computation.

In the low-rank algorithm the memory requirements are determined by the size of matrices $U$, $S$ and $V$. The initial conditions $U^{k,0}$, $V^{k,0}$, and $S^{k,0}$ take $(M\times r + N \times r + r^2) \times n_o$ floating point numbers. Solving Eq.~\eqref{eq: psi1_1D} requires $M\times r\times n_o + M$ more memory to store the scattering source and the updated matrix $U$ (factorized from the updated matrix $K$). Eq.\eqref{eq: psi2_1D} is solved next and uses $N \times r \times n_{o}$ memory for the updated $V$ for a total of $(M\times r + N \times r + r^2) \times n_o$ for the updated basis. Lastly, the $n_o$ equations in Eq.~\eqref{eq: psi3_1D} are $r \times r$ Sylvester matrix equations which require $n_o \times \mathcal{O}(r^3)$ of  memory when using the algorithm proposed in \cite{Bouhamidi2008ANO}. In this step the memory usage is $(M\times r + N \times r + r^2) \times n_o + n_o \times \mathcal{O}(r^3)$. In practice, we choose the rank $r \ll \min(M, N)$, so we can assume $\mathcal{O}(r^3) + r^2  < M\times r + N \times r$. To summarize we use $2\times (M\times r\times n_o + N \times r \times n_{o} + M)$ to approximate the largest memory usage during the low-rank calculations. We can see that the low-rank solution requires much smaller memory than the full solution, when $r \ll \min(M, N)$. 

In numerical experiments we measure the low-rank memory occupied by double precision floating point numbers in megabytes (MB) as
\begin{linenomath*}
	\begin{equation}\label{eq: lowrank_Memory}{
			\mathrm{memory} = 8 \time 2 \times (M\times r\times n_o + N \times r \times n_{o} + M) \times 10^{-6}},
	\end{equation}
\end{linenomath*}
and for the full-rank solution the memory required is
\begin{linenomath*}
	\begin{equation}\label{eq: full_Memory}
		{\mathrm{memory} = 8 \times 2 \times M \times (N_{\Omega} + 1) \times 10^{-6}}.
	\end{equation}
\end{linenomath*}

\subsection{Initial Condition}
Usually, the initial conditions $U^{k,0}$, $V^{k,0}$, and $S^{k,0}$ for the low-rank evolution are obtained by taking the SVD of the given initial intensity and then truncating the singular values smaller than the $r$th largest in matrix $S$ and removing the corresponding rows and columns in $U$ and $V$. Nevertheless, setting the initial condition is not always straightforward. If the intensities are zero or a constant initially (as in many common test problems), the initial condition will be of rank 0 or 1. One way to deal with this issue is to add $r$ randomly generated orthogonal basis to $U$ and $V$ and set $S$ to a $r \times r$  matrix of zeros \cite{Lubich2014}. We find that this approach does not work well for implicit schemes with large time steps. In this work, we choose to let the low-rank basis evolve for a very small time step, such as CFL$ = 0.01$, to obtain initial conditions with a basis that is in the range of the low-rank operator. After this treatment, we can set time steps as intended. 

\section{Numerical Results}
We present results for four 2-D and one 3-D benchmark problems. In all of the test results, we set the particle speed to 1 cm/s. The CFL condition is defined as CFL $= \min(\frac{\Delta t}{\Delta x}, \frac{\Delta t}{\Delta y}, \frac{\Delta t}{\Delta z})$. We implement the low-rank algorithm in MATLAB. For several problems, we measured the running memory in MATLAB with the built-in function “memory” and record the running time with the stopwatch timer functions. 

\subsection{Double Chevron problem} 
\begin{figure}[h!]
    \centering
	\includegraphics[width=0.5\textwidth]{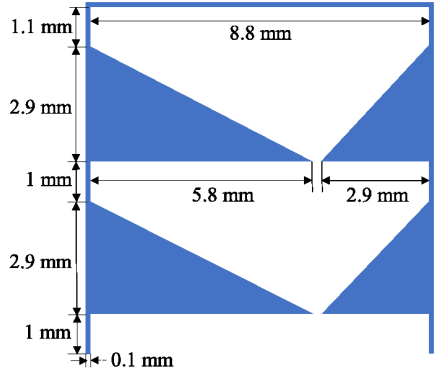}
	\caption{{The layout for the double chevron problem is shown. The blue are dense materials and the blank are purely-scattering region. There is an isotropic source at the bottom and other three sides have vacuum boundary conditions. }}
	\label{fig: dc_layout}
\end{figure}

To demonstrate the accuracy of our low-rank algorithm, we consider a multi-material 2-D problem with an asymmetric layout originally presented in \cite{peng2020lowrank}. As detailed in Figure \ref{fig: dc_layout}, this problem has purely scattering zones with $\sigmat = \sigmas = 0.01\, \mathrm{cm}^{-1}$, absorbing zones $\sigmat = 100\, \mathrm{cm}^{-1}, \sigmas = 0.1\, \mathrm{cm}^{-1}$, and an isotropic source at the bottom with $Q = 1$ of thickness 0.1 $\mathrm{mm}$. We solve this problem using the spatial grid of size $90 \times 90$ for the computational domain $[0, 9\,\mathrm{mm}] \times [0, 9 \,\mathrm{mm}]$, and the simulation time $t = 0.9$ s. The CFL number is set to be 3. 

\begin{figure}[h!]
    \centering
	\begin{subfigure}[b]{.495\linewidth}\centering
	\includegraphics[width=\textwidth]{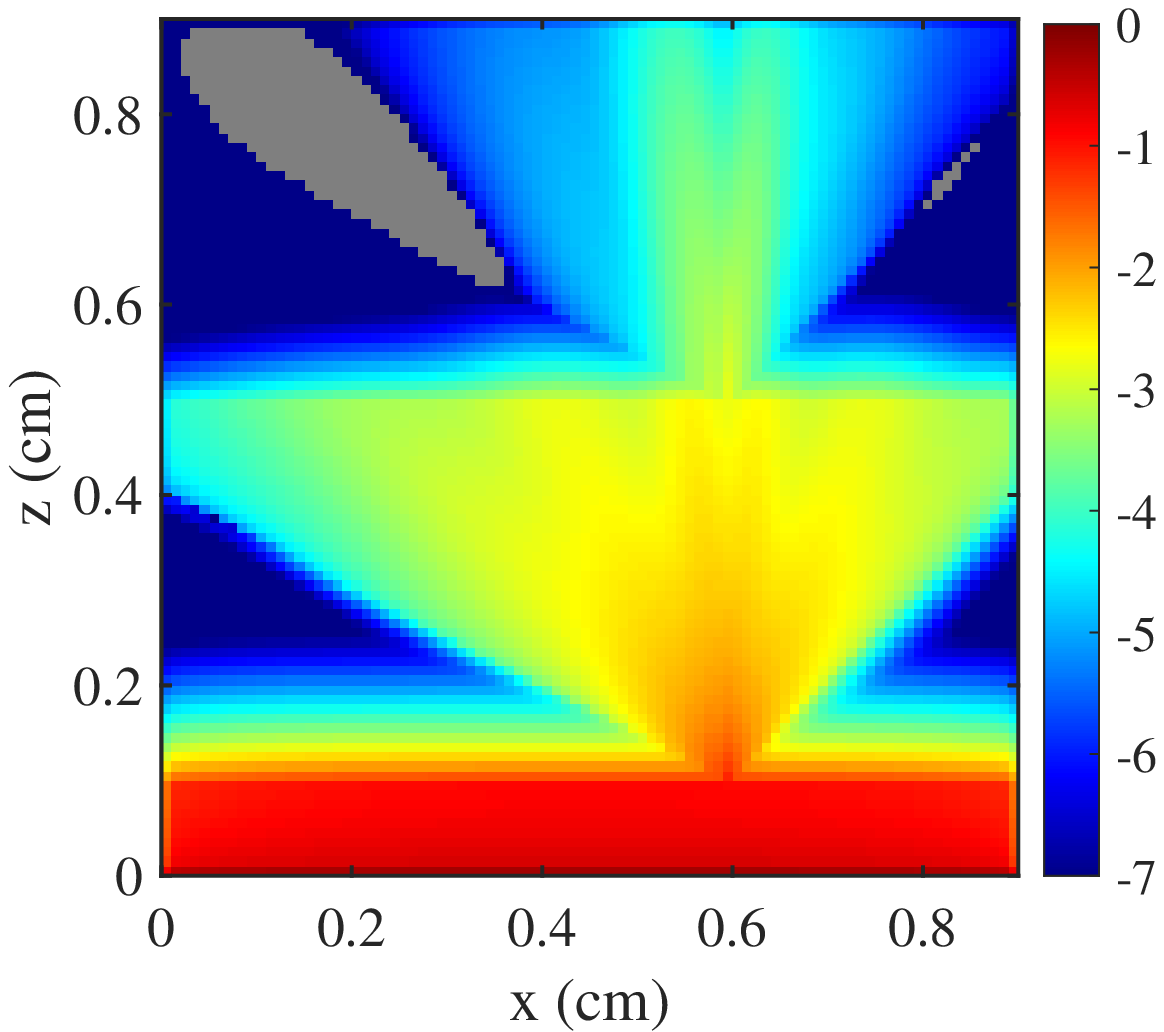}
	\subcaption{S$_{32}$, rank 16}
	\end{subfigure}
	\hfill
	\begin{subfigure}[b]{.495\linewidth}\centering
	\includegraphics[width=\textwidth]{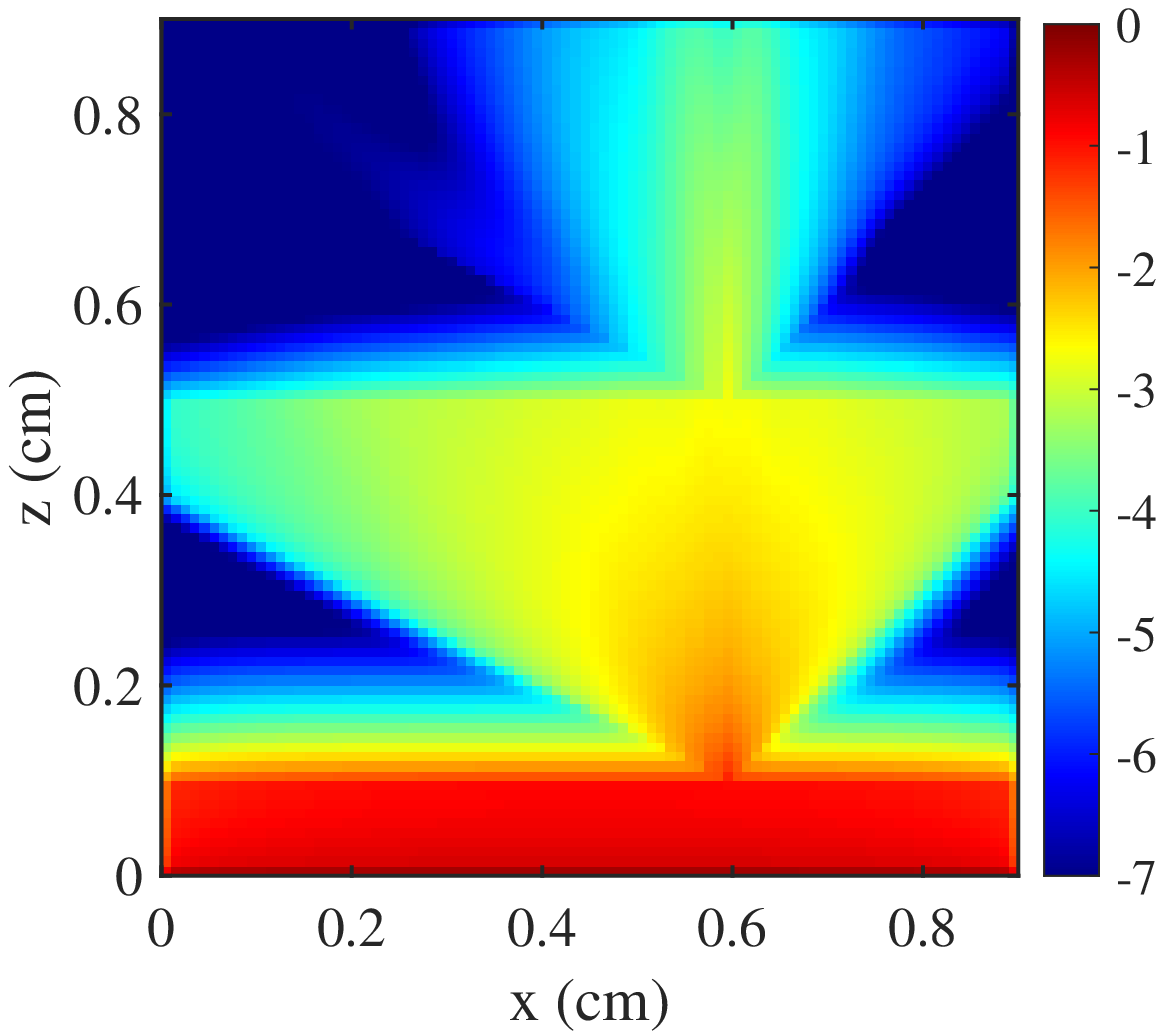}
	\subcaption{S$_{32}$, rank 25}
	\end{subfigure}
	\begin{subfigure}[b]{.495\linewidth}\centering
	\includegraphics[width=\textwidth]{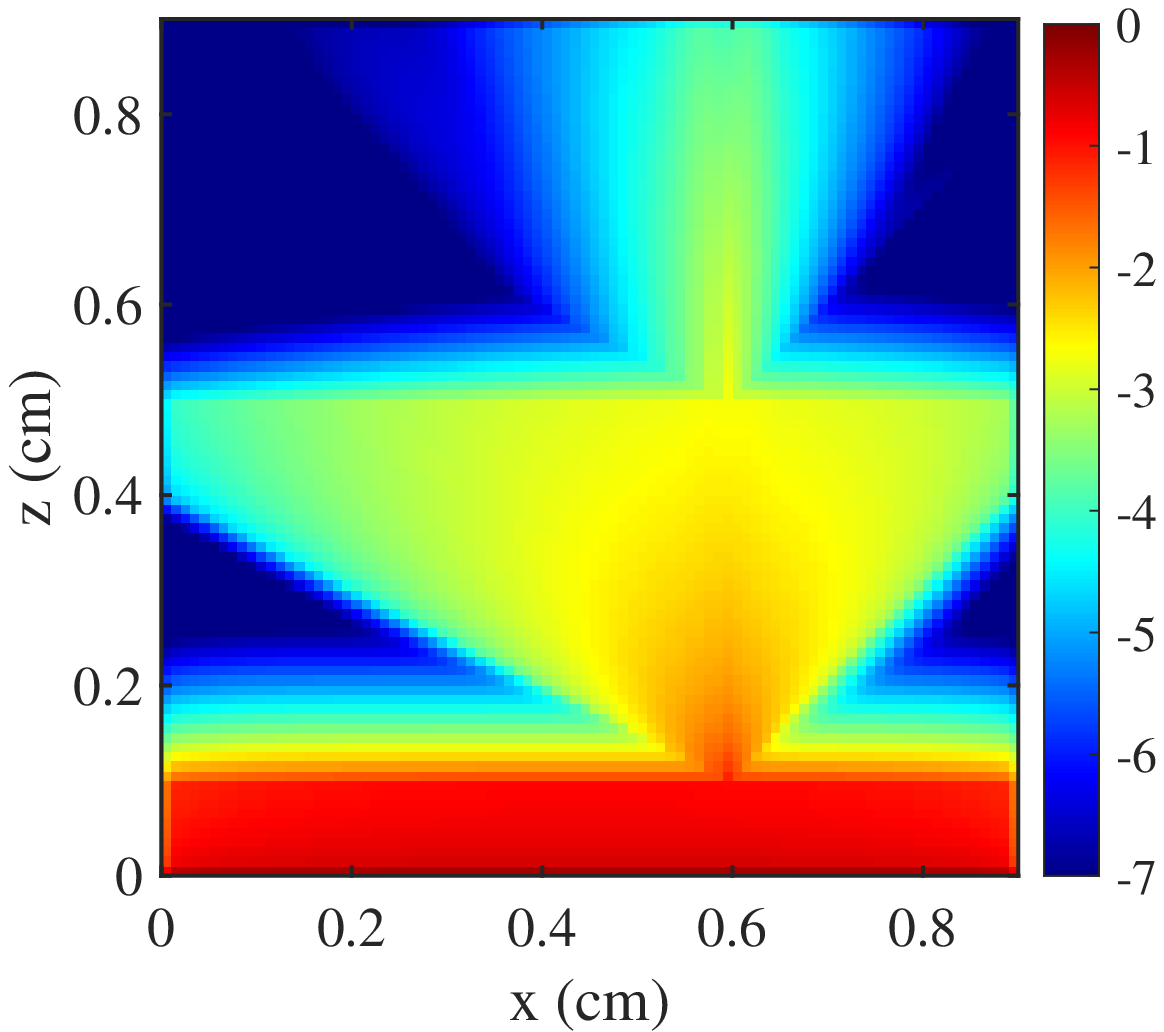}
	\subcaption{S$_{32}$, rank 36}
	\end{subfigure}
	\hfill
	\begin{subfigure}[b]{.495\linewidth}\centering
	\includegraphics[width=\textwidth]{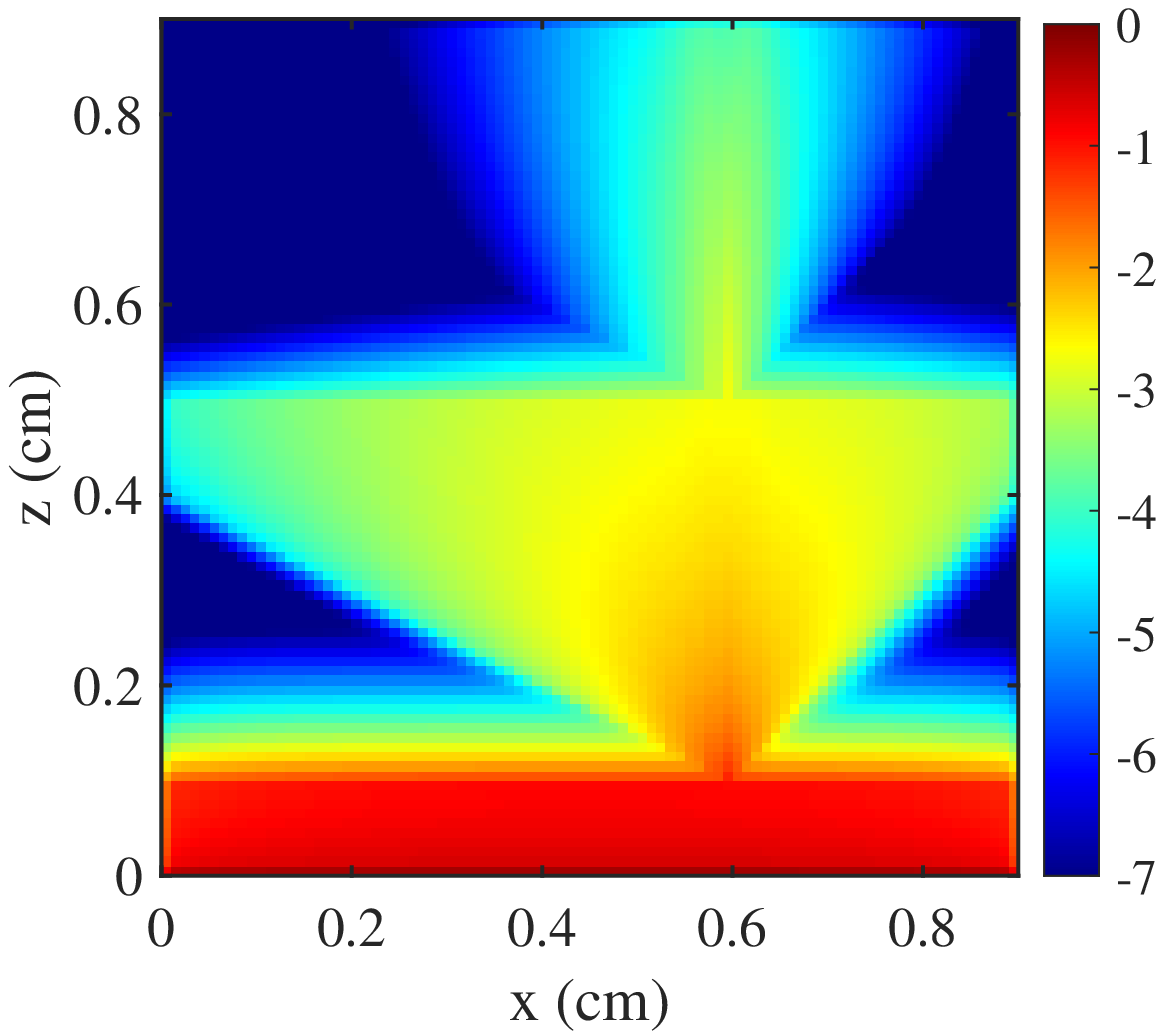}
	\subcaption{S$_{32}$, rank 49}
	\end{subfigure}
	\hfill
	\begin{subfigure}[b]{0.495\linewidth}
	\includegraphics[width=\textwidth]{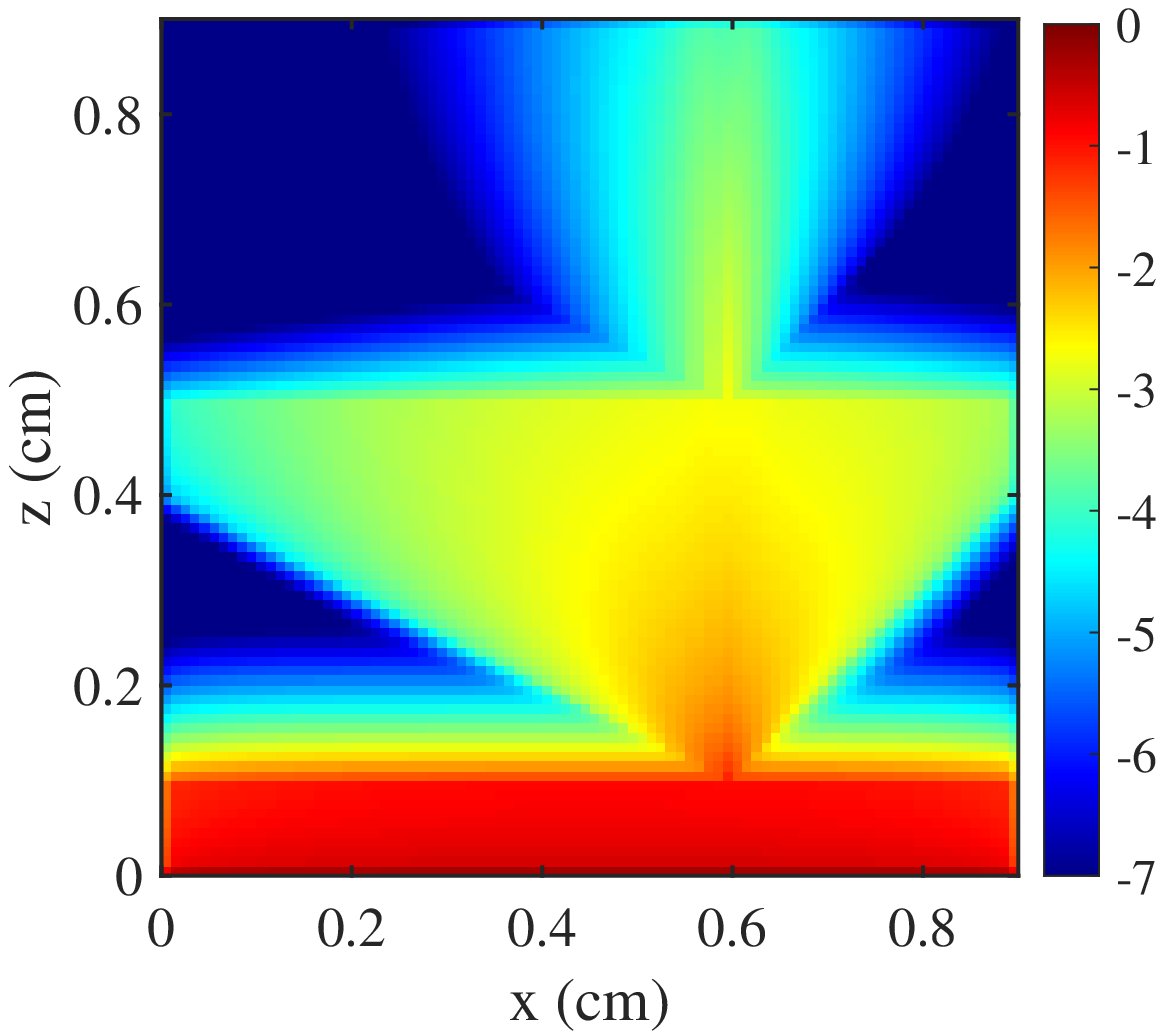}
	\subcaption{S$_{32}$, rank 256 (full rank)}
	\end{subfigure}
	\hfill	
	\begin{subfigure}[b]{0.495\linewidth}
	\includegraphics[width=\textwidth]{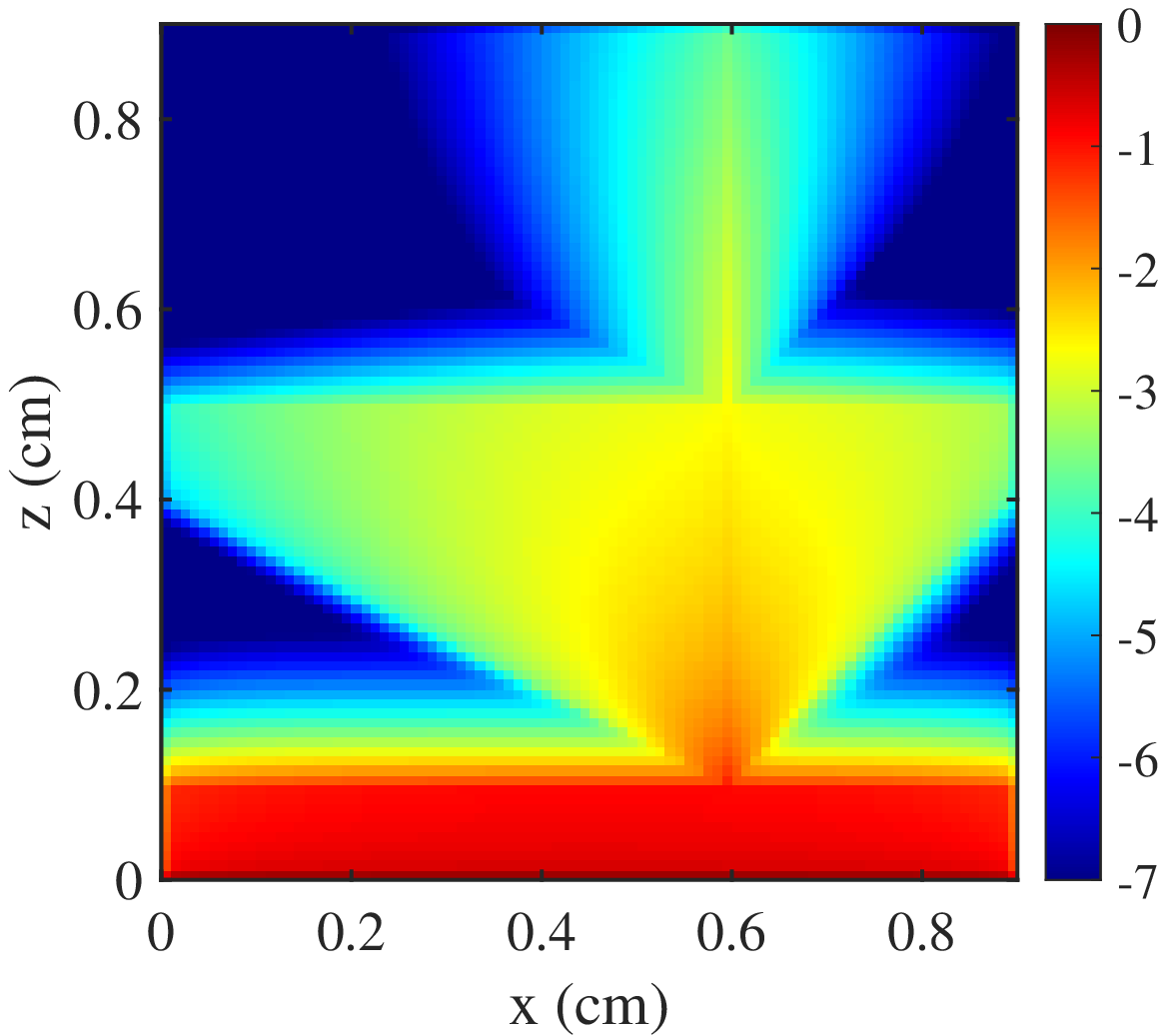}
	\subcaption{S$_{100}$, rank 2500 (full rank)}
	\end{subfigure}
	\caption{{Solutions to the double chevron problem at $t = 0.9 \,$s. The color scale is logarithmic and negative regions are shaded gray.}}
	\label{fig: dc_compare}
\end{figure}

Figure \ref{fig: dc_compare} compares the low-rank solutions for the scalar intensity with S$_{32}$ and varying rank to the full rank S$_{32}$ solution and the S$_{100}$ benchmark solution. As we can see S$_{32}$ with rank 36 is close to the full rank S$_{32}$ solution but it is not enough to resolve the particle distribution behind the second obstacle and the negative results are observed. The low-rank solution with rank 25 is nearly identical to the full rank S$_{32}$ solution, which means rank 25 is sufficient to capture the important features in this problem. 

\begin{figure} 
	\centering
    \begin{subfigure}[b]{0.49\linewidth}\centering
	\includegraphics[width=\textwidth]{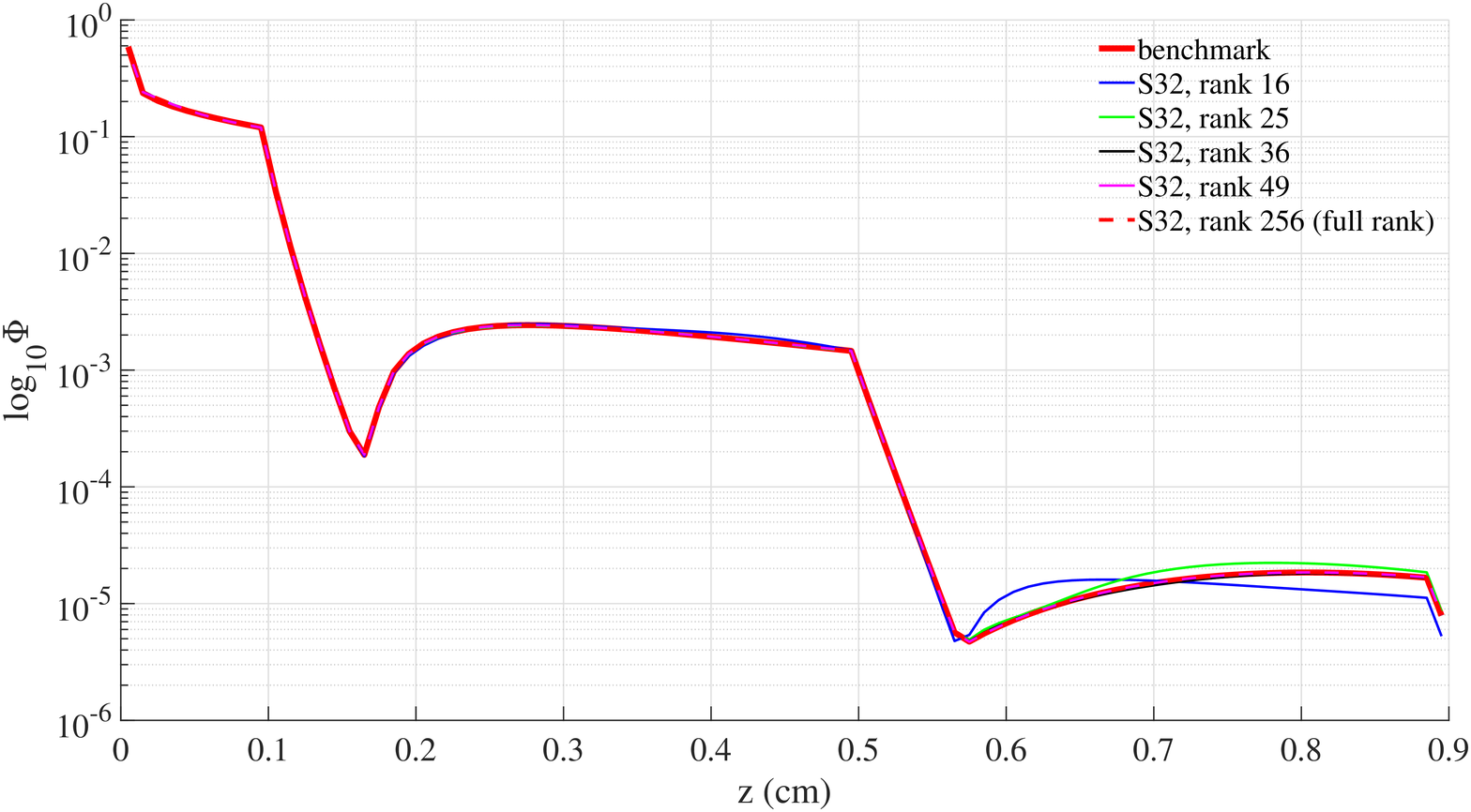}
	\subcaption{Cut at $x = 0.45$ cm}
	\end{subfigure} 
	\begin{subfigure}[b]{0.49\linewidth}\centering
	\includegraphics[width=\textwidth]{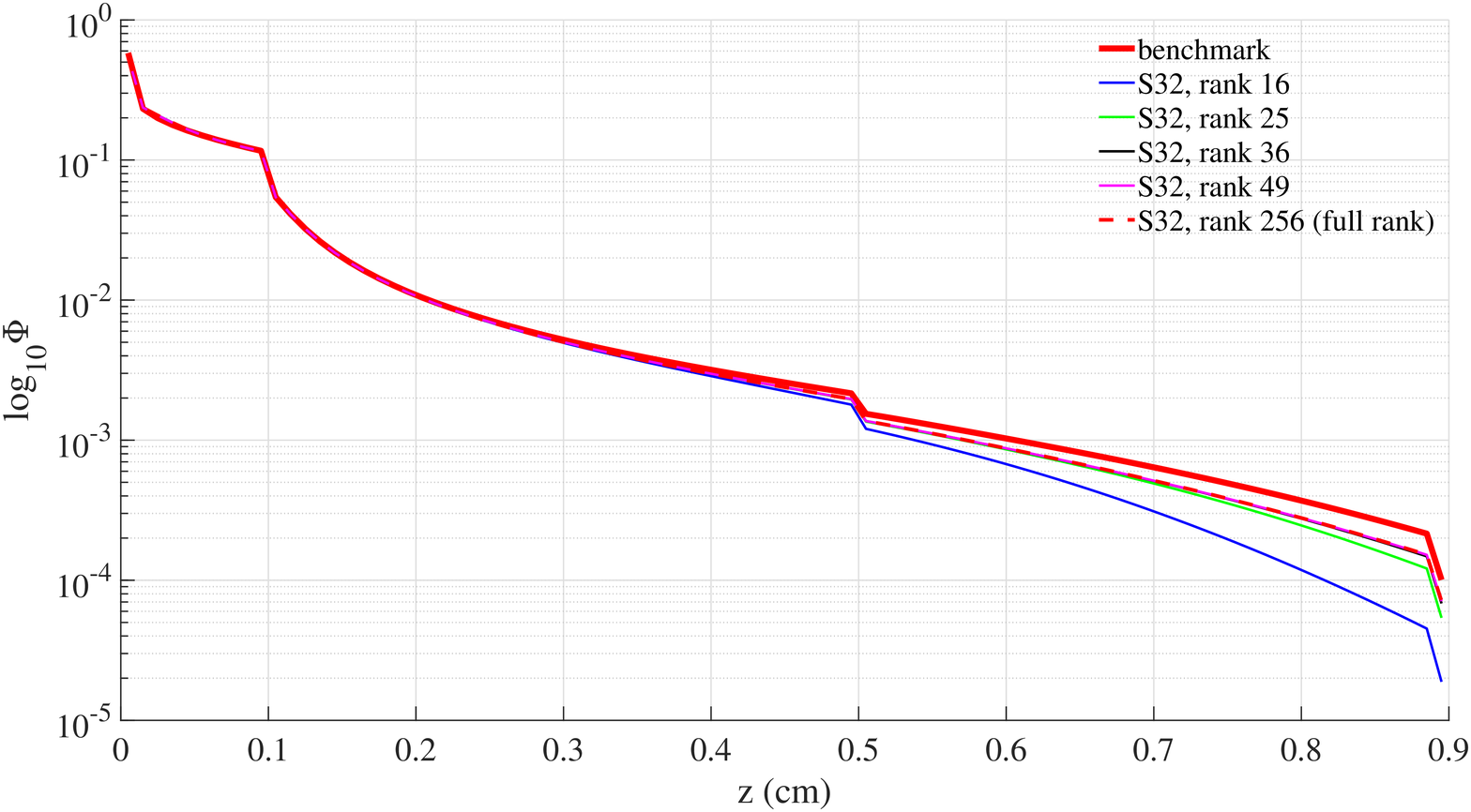}
	\subcaption{Cut at $x = 0.6$ cm}
	\end{subfigure}
	\caption{Logarithmic scalar insentityes to the double chevron problem at $t = 0.9 \,$s.}
	\label{fig: dc_cut}
\end{figure}

Figure \ref{fig: dc_cut} presents a comparison for scalar intensities along $x = 0.45$ cm and $x = 0.6$ cm. The left plot emphasizes their different behavior behind the second chevron, and the right plot shows the particle distribution along the gap that particles can travel through. We find that the low-rank solution with rank 36 has a similar error to the full rank S$_{32}$ solution when compared to the benchmark. We also notice that S$_{32}$ is not enough to capture the particle stream along the gap and it requires high angular resolution to obtain. 

\begin{figure} 
	\centering
    \begin{subfigure}[b]{\linewidth}\centering
	\includegraphics[width=\textwidth]{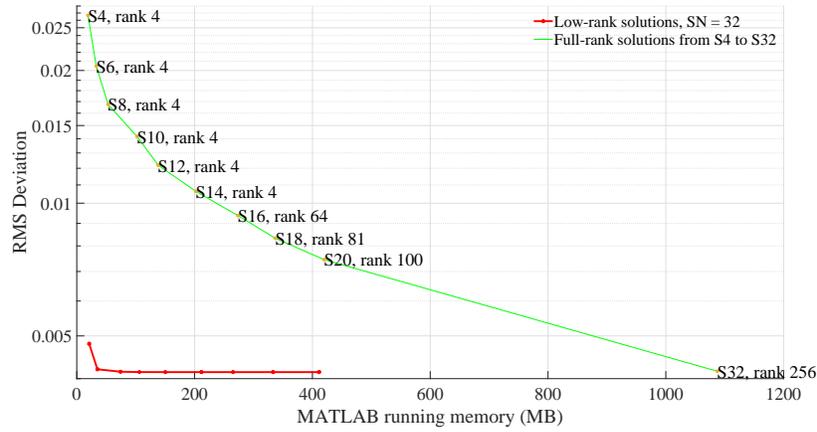} 
	\subcaption{Error versus running memory}
	\label{fig: dc_memory}
	\end{subfigure} 
	\begin{subfigure}[b]{\linewidth}\centering
	\includegraphics[width=\textwidth]{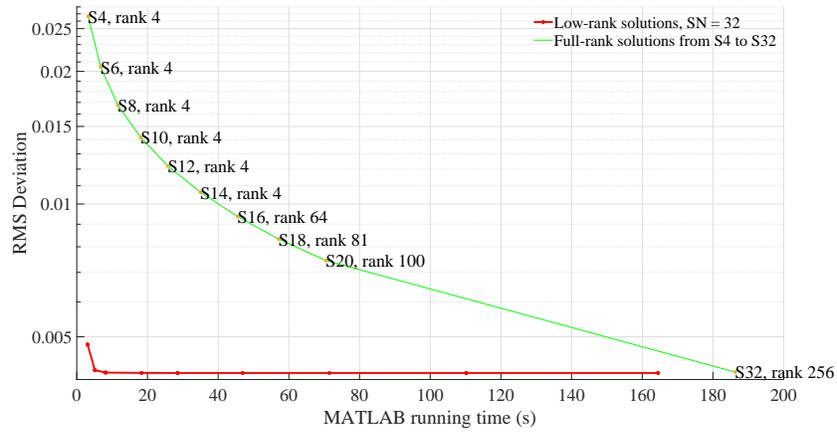}
	\subcaption{Error versus running time}
	\label{fig: dc_time}
	\end{subfigure}
	\caption{The comparison of errors for the low-rank and full-rank solutions of double chevron problem with different memory usage and computational time is shown. The red dots represents low-rank solution with rank 4, 9, 16, 25, 49, 64, 81, and 100. }
	\label{fig: dc_memory_time}
\end{figure}

The computational efficiency of our low-rank algorithm is presented in Figure \ref{fig: dc_memory_time}. Our solutions' accuracy is measured by the root mean square (RMS) error to the benchmark solution. In Figure \ref{fig: dc_memory} we plot the error of our solutions versus the running memory in the sweeping process. As we can see, the third point in the red dotted line, which indicates the solution with rank 16, has already achieved the same error level as the full rank S$_{32}$ solution, but with only $10\%$ of the memory. It also can be seen that the low-rank and the full rank solution use a similar amount of memory for the same rank. As explained by our memory formula \eqref{eq: lowrank_Memory}, the memory usage depends on the rank when $M$ is fixed, and $N$ is much smaller than $M$.  

Figure \ref{fig: dc_time} makes the similar comparison with the running time per one time step ($\Delta t = 0.03$ s for this test). It appears that our rank 16 solution could save $95\%$ of the computational time compared to the full rank solution. We also point out that our low-rank method will take more time than the full rank solution with the same rank. For example, the rightmost dot in the red line is the solution with rank 100, the same as the full rank S$_{20}$ solution, but it runs twice as long. But when rank is properly chosen and relatively small, we can still save a large amount of memory and time. 

\begin{figure} 
\centering
	\includegraphics[width=\textwidth]{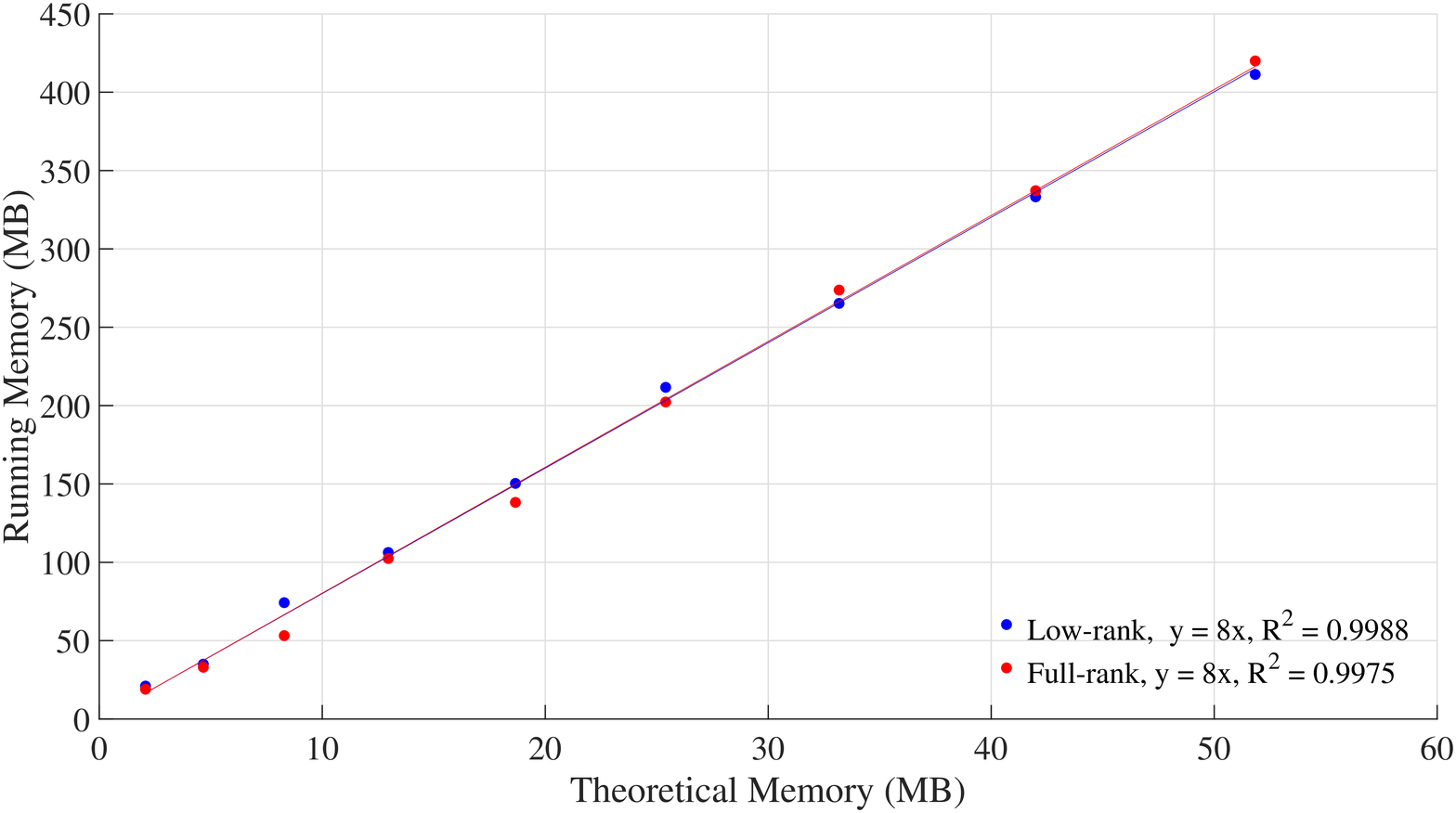}
	\caption{The relation between the memory calculate by formulas \eqref{eq: lowrank_Memory} and \eqref{eq: full_Memory} and the running memory in MATLAB for the double chevron problem. The angular discretization for the low-rank solutions is S$_{32}$ with ranks varying from 4 to 100, the full-rank solutions are with S$_{4}$ to S$_{20}$. The spatial and time discretizations are the same as previous simulations.}
	\label{fig: dc_memory_compare}
\end{figure}

The computer memory occupied in MATLAB during the calculation and the theoretical values are shown in Figure {\ref{fig: dc_memory_compare}}. The coefficients of determination ($R^2$) for the linear fit are approximately one, which indicates a strong linear relationship between the theoretical and actual memory. We conclude that Eqs.~\eqref{eq: lowrank_Memory} and \eqref{eq: full_Memory} are valid estimations for the actual memory usage, and the low-rank algorithm reduces the memory requirement proportionally to the reduction in rank.

\subsection{Lattice problem}
\begin{figure}[h!]
    \centering
	\begin{subfigure}[b]{.38\linewidth}\centering
	\includegraphics[width=\textwidth]{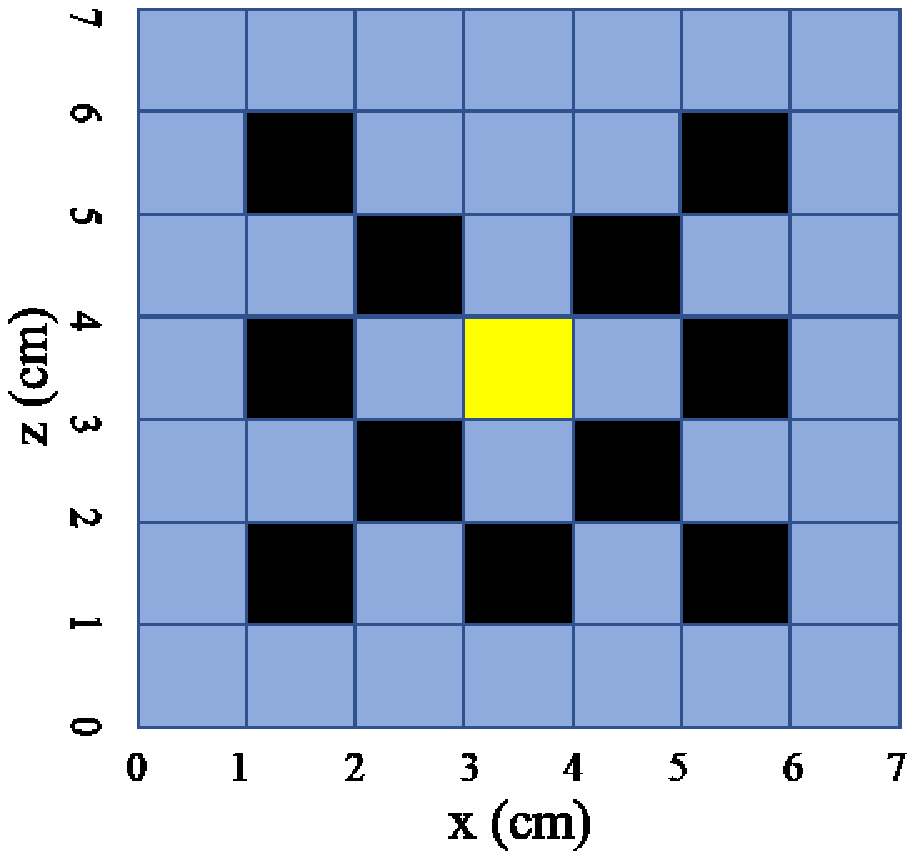}
	\subcaption{Geometry for Lattice test}
	\end{subfigure}
	\hfill
	\begin{subfigure}[b]{.49\linewidth}\centering
	\includegraphics[width=\textwidth]{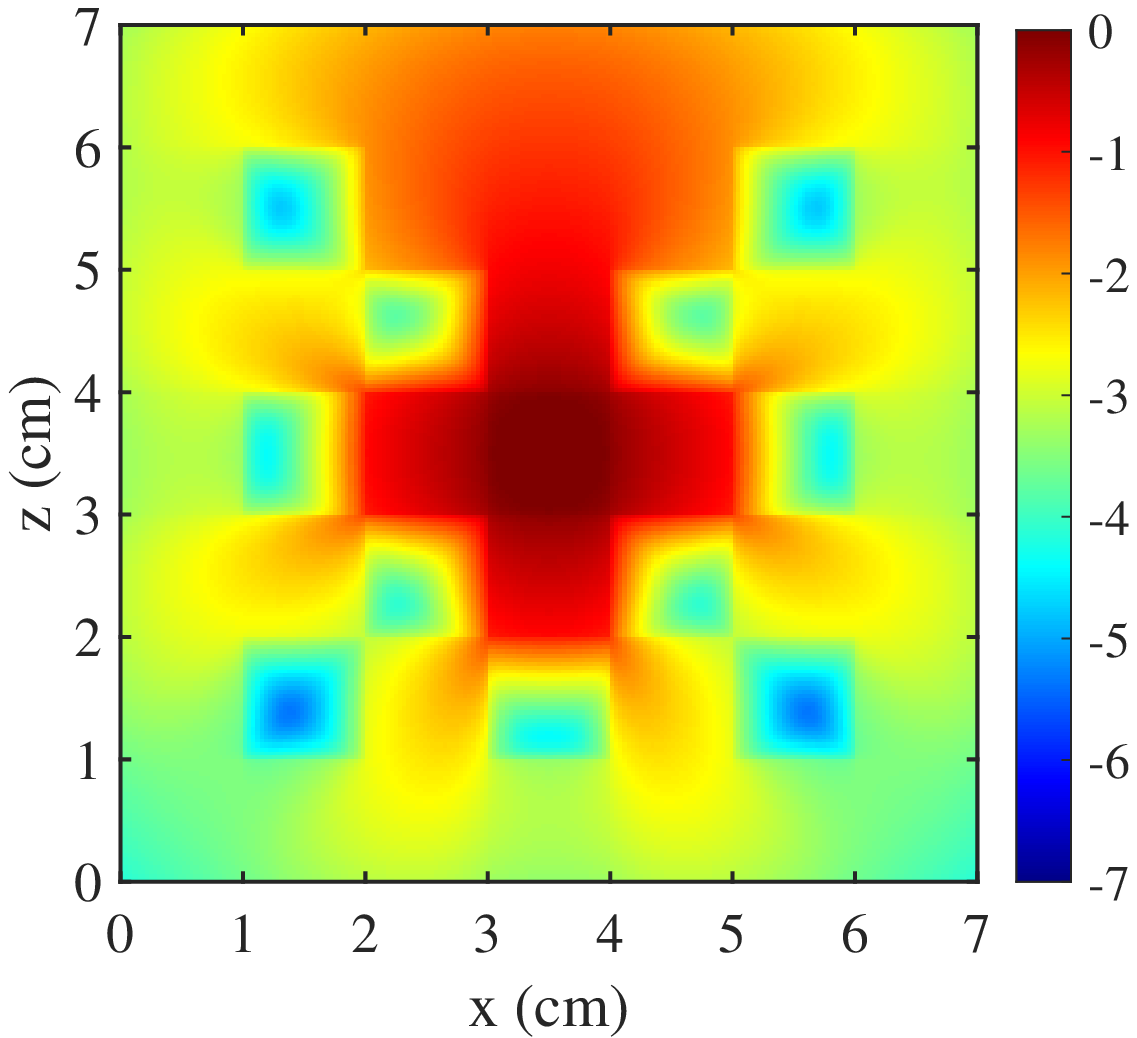}
	\subcaption{S$_{64}$, rank 1024 (full rank)}
	\end{subfigure}
  	\caption{{The left plot shows the layout of the lattice problem, where the blue zones are purely scattering region with $\sigma_s = \sigma_t = 1 \ \mathrm{cm^{-1}}$, the black are absorbing region with $\sigma_s = 0$, $\sigma_t = 10 \ \mathrm{cm^{-1}}$ and the yellow is the scattering region with an isotropic source $Q = 1$. The right plot is the benchmark solution with the logarithmic color scale.}}
	\label{fig: la_layout}
\end{figure}

Next, we consider the lattice problem: a $7 \times 7$ cm checkerboard consisting of pure absorbers, purely scattering regions, and a strong source turned on at $t = 0$ with a zero initial condition. We use a computational domain of $[0, 7] \times [0, 7]$ with a $210 \time 210$ spatial grid. We solve this problem with a single, large time step using CFL $= 10^4$. The layout and the reference solution calculated with S$_{64}$ for this problem are shown in Figure \ref{fig: la_layout}.

\begin{figure}[h!]
    \centering
	\begin{subfigure}[b]{.495\linewidth}\centering
	\includegraphics[width=\textwidth]{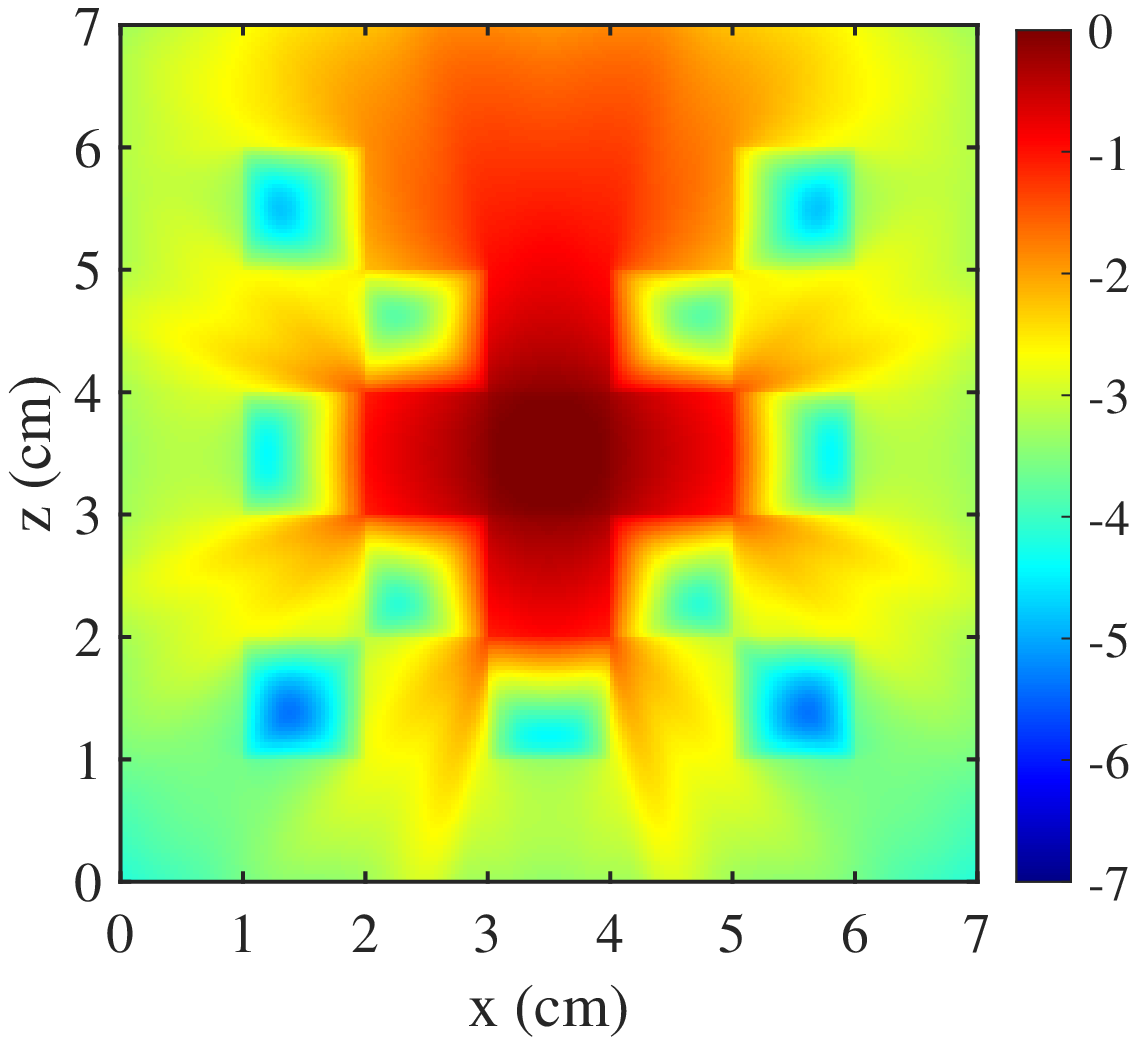}
	\subcaption{S$_{6}$, rank 9 (full rank)}
	\end{subfigure}
	\begin{subfigure}[b]{.495\linewidth}\centering
	\includegraphics[width=\textwidth]{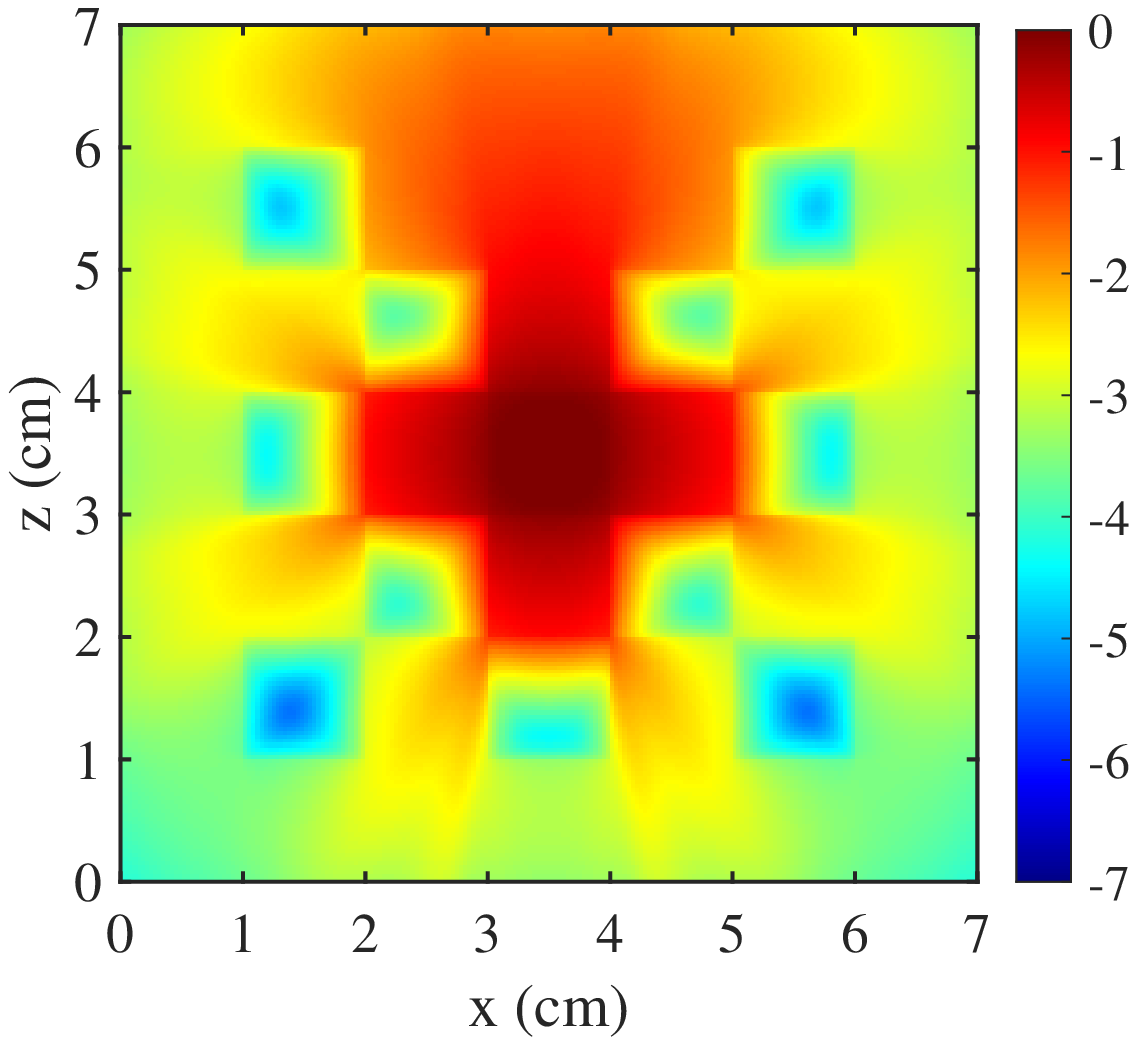}
	\subcaption{S$_{8}$, rank 16 (full rank)}
	\end{subfigure}
	\hfill
	\begin{subfigure}[b]{.495\linewidth}\centering
	\includegraphics[width=\textwidth]{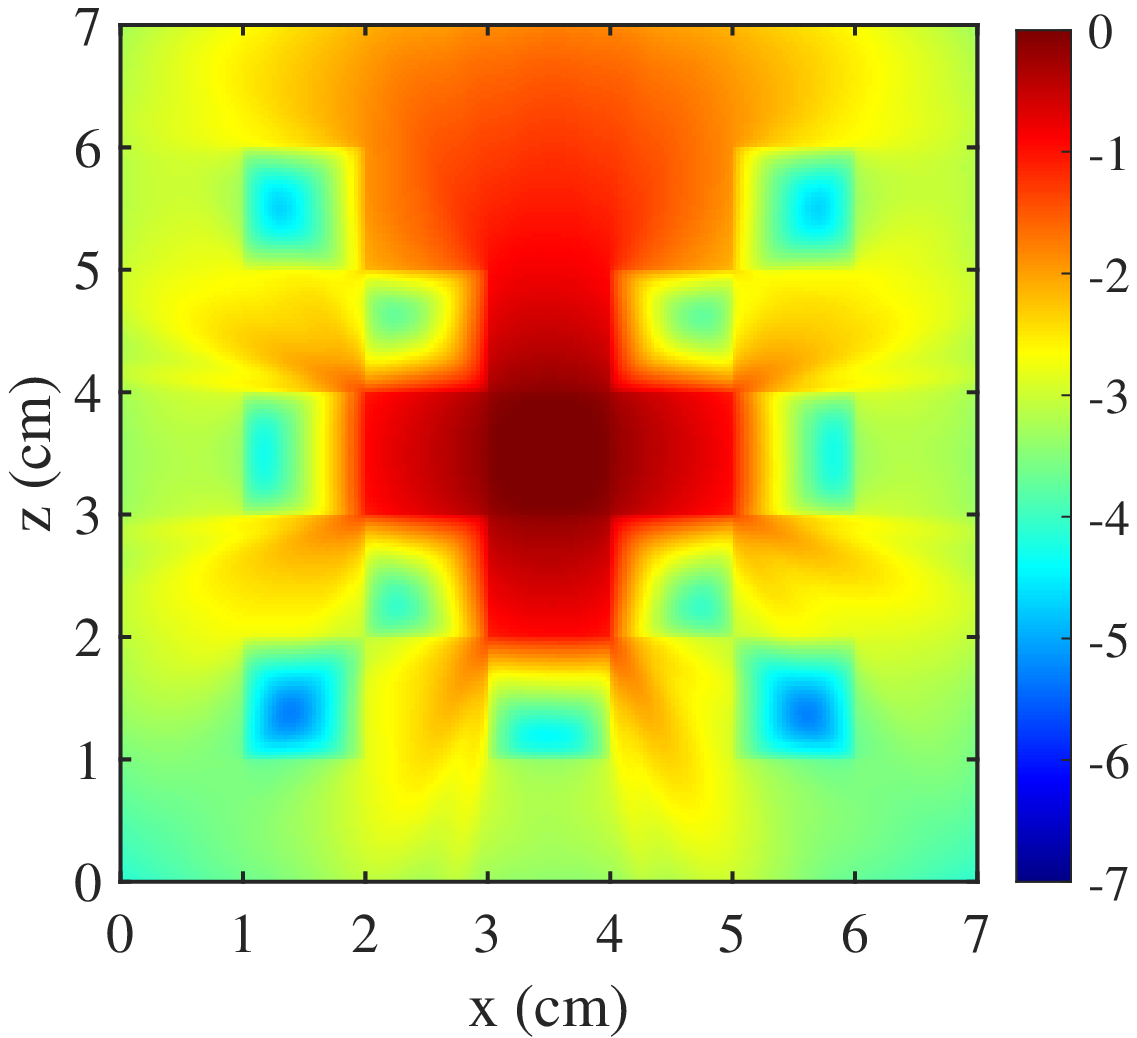}
	\subcaption{S$_{64}$, rank 9}
	\end{subfigure}
	\hfill
	\begin{subfigure}[b]{.495\linewidth}\centering
	\includegraphics[width=\textwidth]{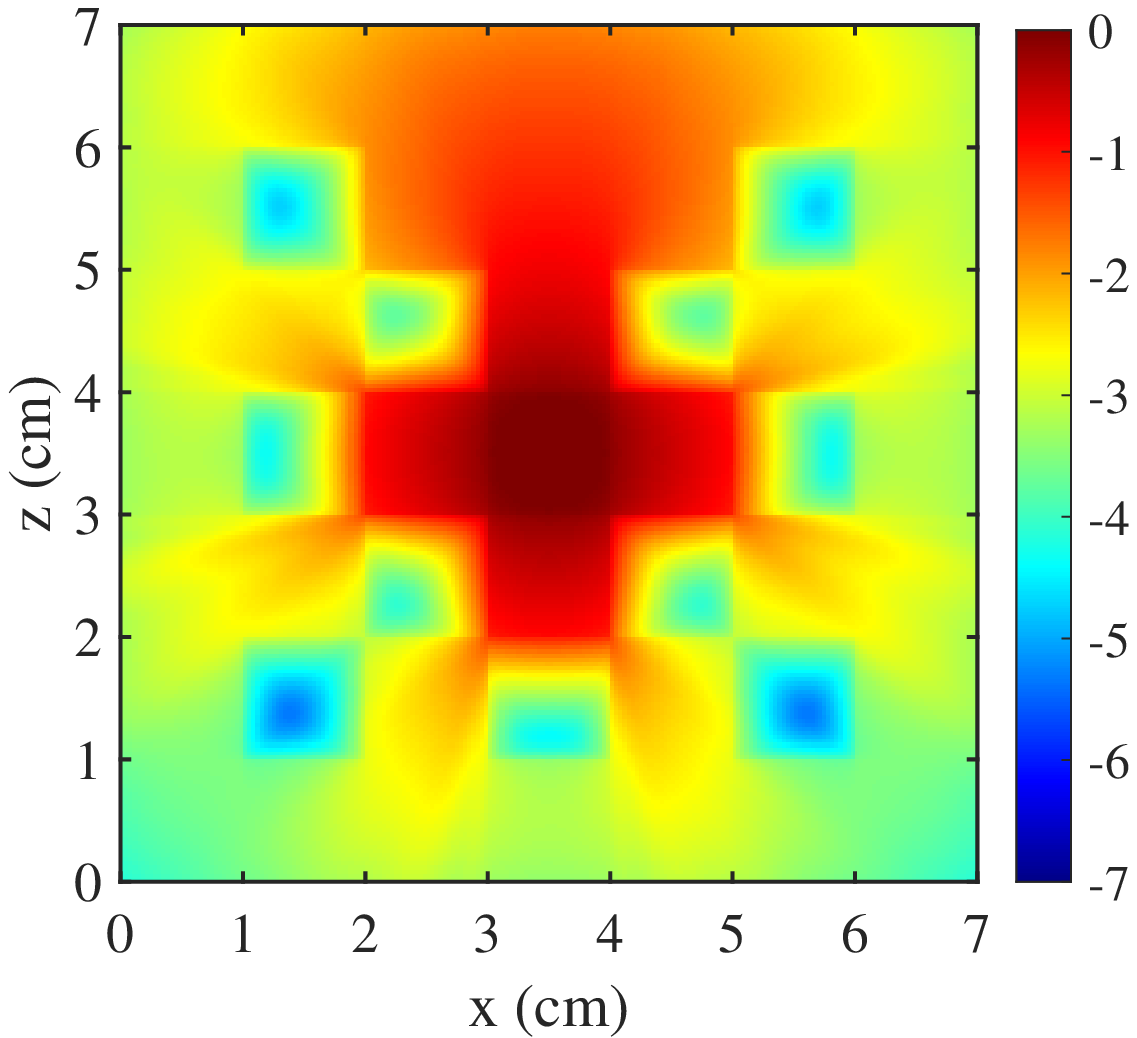}
	\subcaption{S$_{64}$, rank 16}
	\end{subfigure}
	\hfill
	\begin{subfigure}[b]{\linewidth}\centering
	\includegraphics[width=\textwidth]{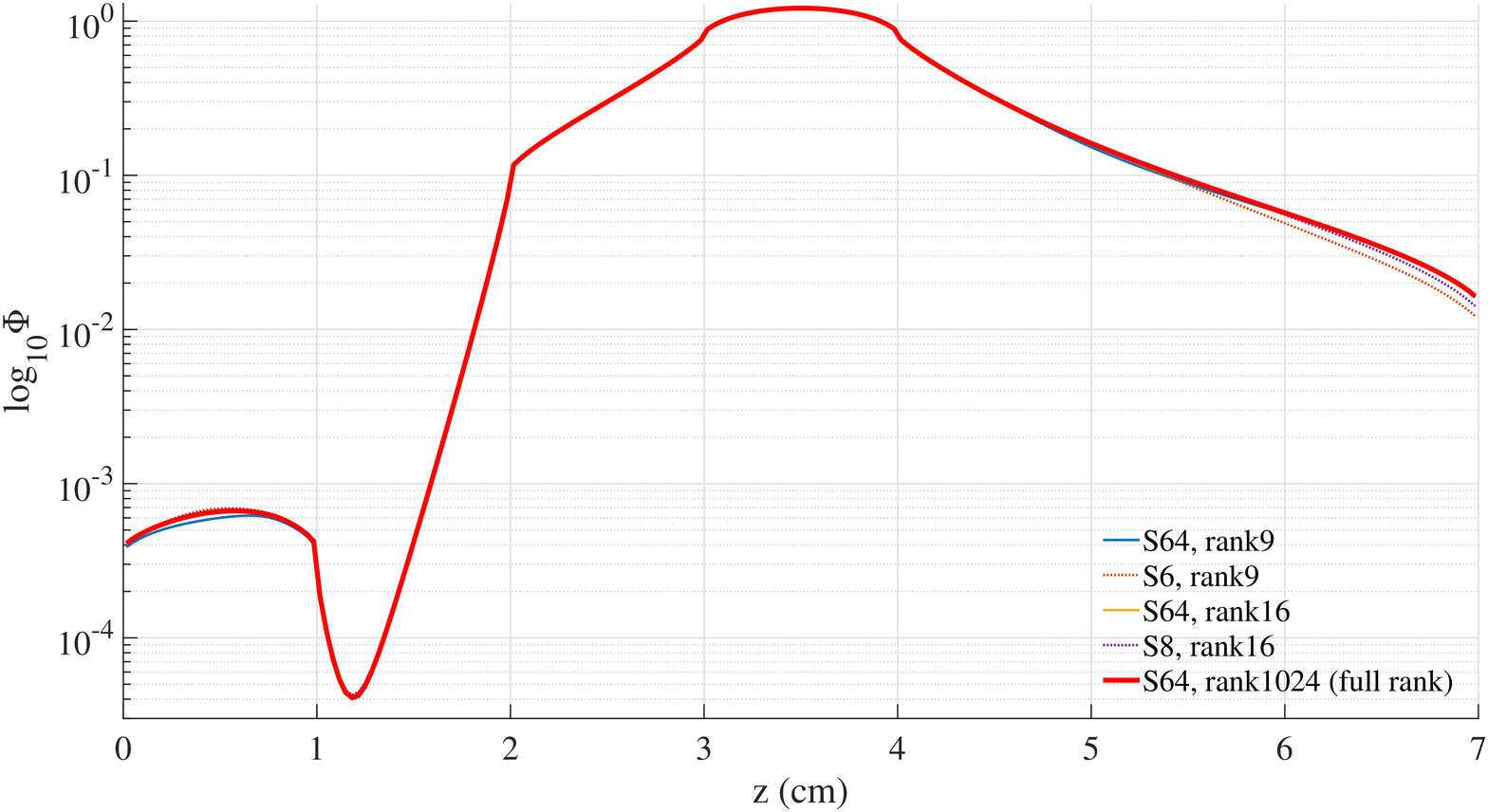}
	\subcaption{Cut at $x = 3.5$ cm}
	\end{subfigure}
	\caption{The logarithmic scalar insentity to the lattice problem calculated by the low-rank method with S$_{64}$ are compared to the full rank solutions with the same rank.}
	\label{fig: la_compare}
\end{figure}

This test aims to show the accuracy improvement by adding more angular directions while keeping the rank fixed. As shown in Figure \ref{fig: la_compare}, the full rank S$_{6}$ solution has ray effects in the solution near the bottom, right and left sides. These can be alleviated by using more discrete ordinates as observed in solutions with S$_{64}$ and rank 9. Similar phenomena are found in solutions with rank 16, where the full rank solution suffered from ray effects while the low-rank solution is closer to our reference. We also plot the solution along $x = 3.5$ cm, where we can see that low-rank solutions are closer to the benchmark solution than the full rank solution.  

\begin{figure} 
	\centering
	\includegraphics[width=\textwidth]{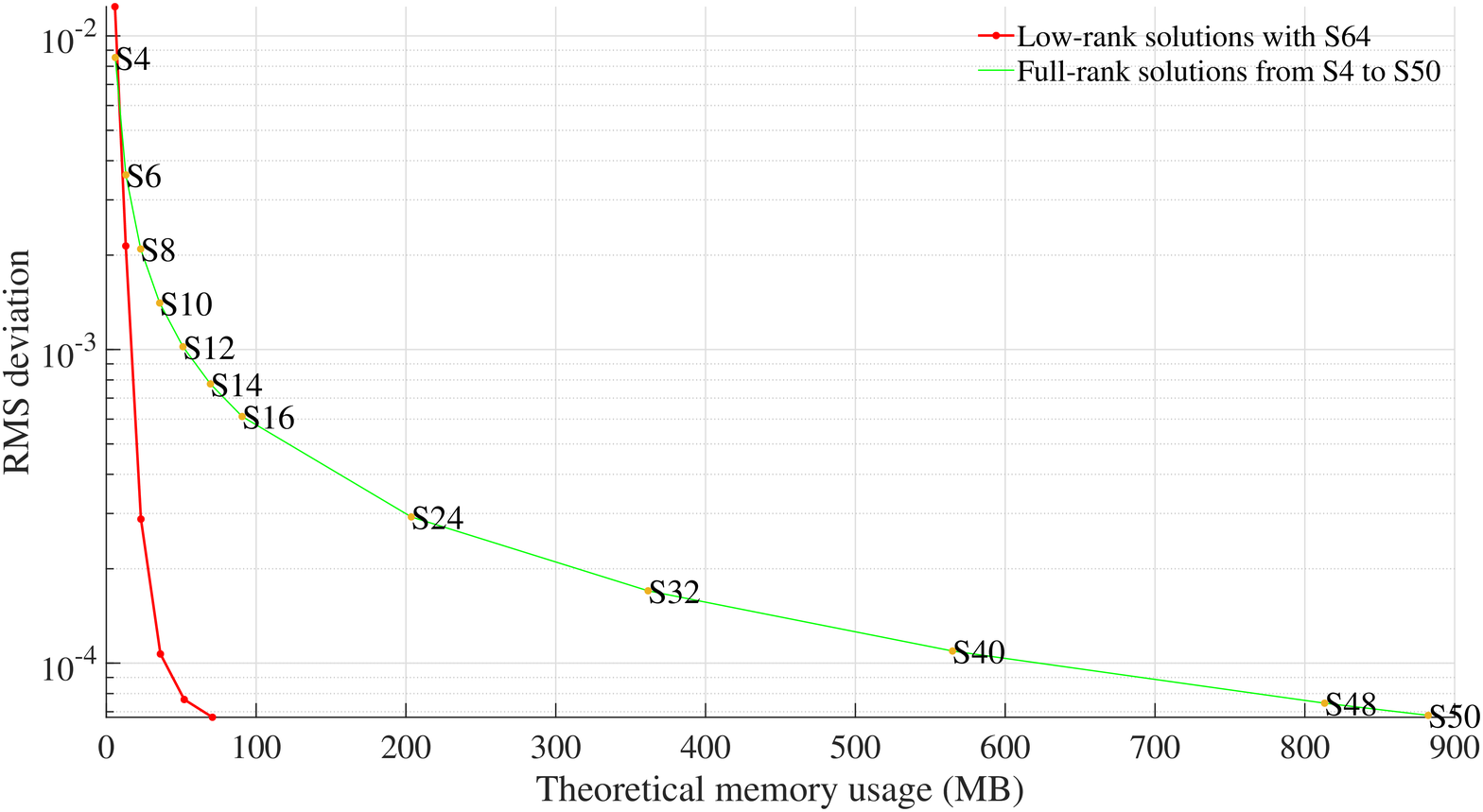}
	\caption{{The comparison of errors for the lattice problem with different memory usage. The red dot line represents the error of the full rank solution that varies the number of discrete ordinates. The solid line represents the error of the low-rank solutions with S$_{64}$ that varies the rank (rank = $[4, 9, 16, 25, 36, 49]$).}}
	\label{fig: la_memory}
\end{figure}

Figure \ref{fig: la_memory} gives a quantitative comparison between low-rank and full rank solutions in terms of their memory usage and error measured by RMS deviations to the reference. We notice that the low-rank solutions converge faster than the full rank solutions, where the low-rank solution with rank 49 achieves the same accuracy as the full rank S$_{50}$ solution, but only requires $10\%$ of the memory. 

\subsection{Line source problem}
\begin{figure}[h!] 
\begin{minipage}{0.49\textwidth}
  \centering
\includegraphics[width=\textwidth]{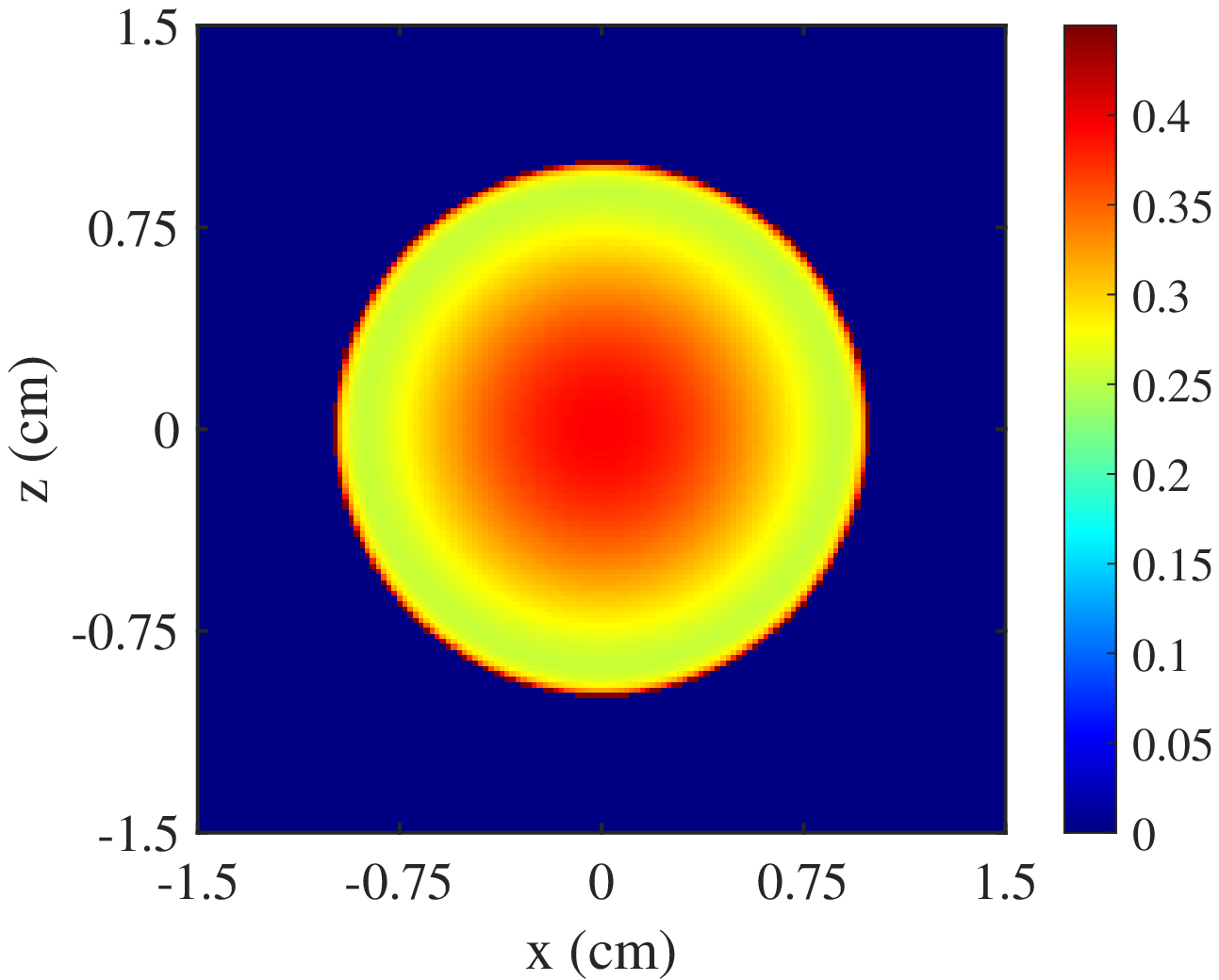}
\subcaption{Analytic Solution}\label{fig: LS_a}
\end{minipage}%
\begin{minipage}{0.49\textwidth}
  \centering
\includegraphics[width=\textwidth]{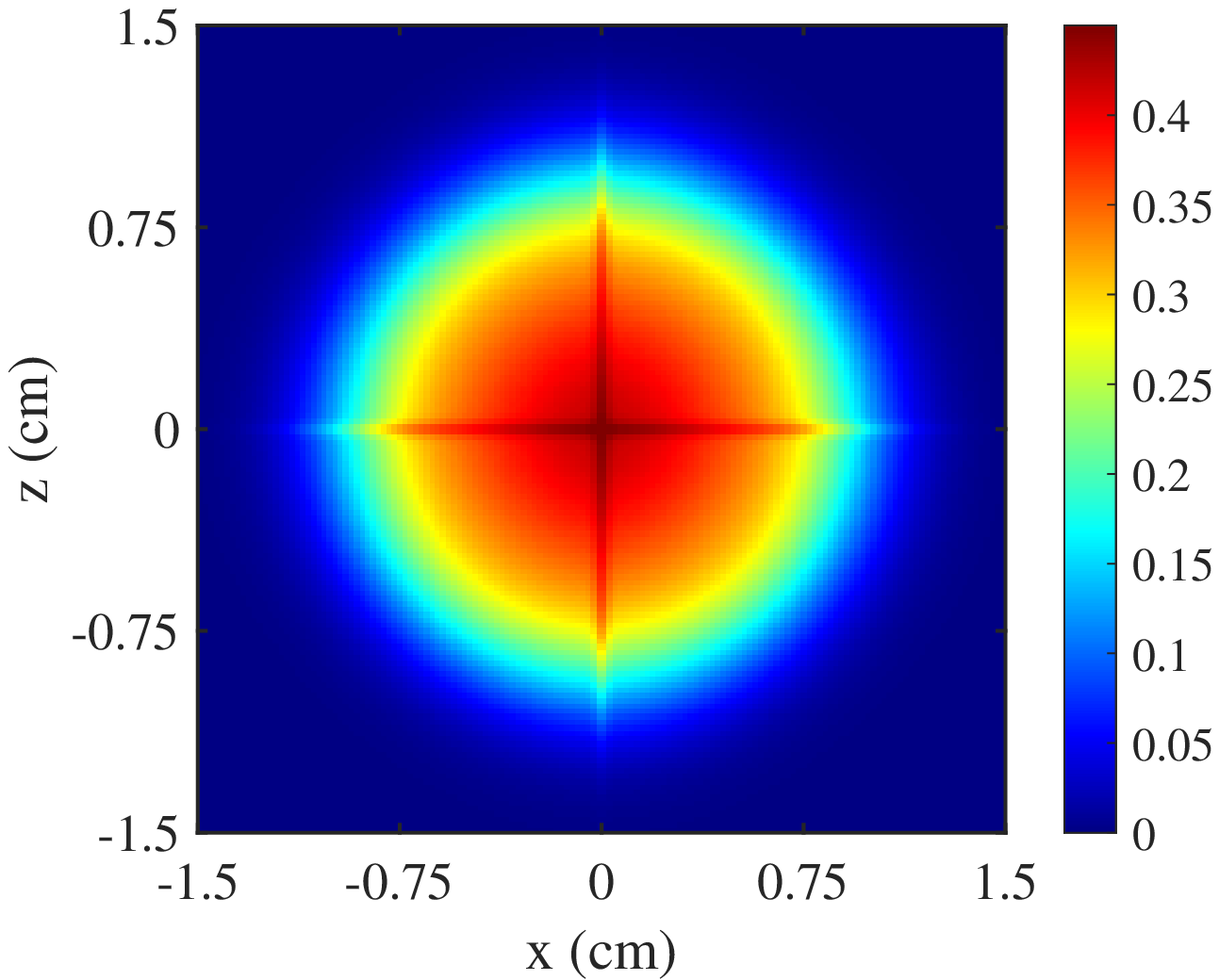}
\subcaption{S$_{100}$, rank 2500 (full rank)}\label{fig: LS_b}
\end{minipage}%
\caption{The scalar insentity $\phi$ of the line source problem at $t = 1$ s calculated by full-rank $S_{100}$ is compared to the benchmark solution.}
\label{fig: ls_bench}
\end{figure}

The line source problem describes a beam of radiation is spreading out in a purely scattering plane. In this problem we have the initial condition $\psi(x, z, \mu, \varphi, t) = \delta(x)\delta(z)$, and the purely scattering medium $\sigmas = \sigmat = 1$ with no source $Q = 0$. We use a computational domain of $[-1.5,1.5] \times [-1.5,1.5]$ for the simulation time $t=1 \,$s, while the spatial grid is set to be $150 \times 150$. Figure \ref{fig: ls_bench} shows the analytic solution and  our benchmark solution. 

\begin{figure}[h!]
    \centering
	\begin{subfigure}[b]{.327\linewidth}\centering
	\includegraphics[width=\textwidth]{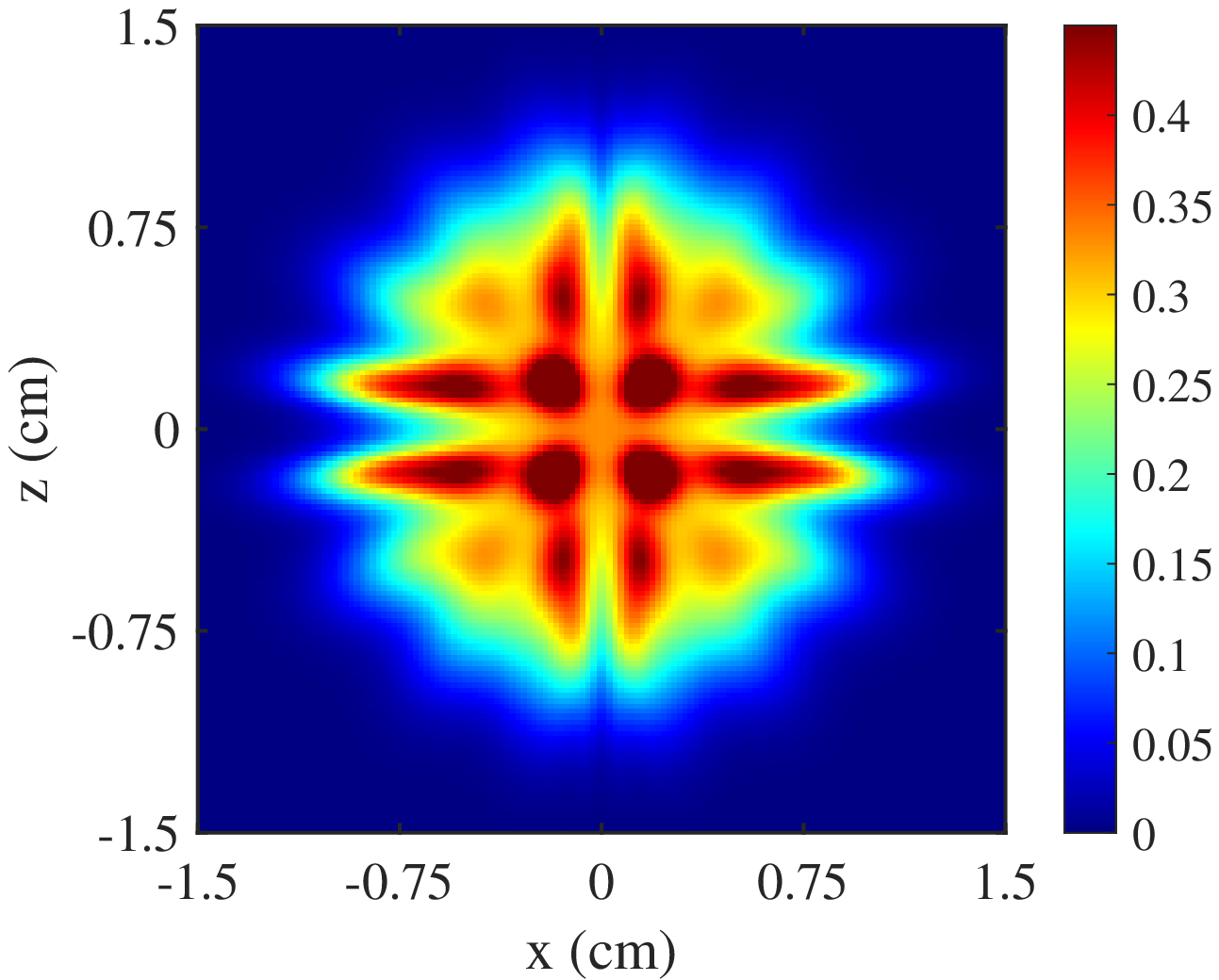}
	\subcaption{S$_{8}$, rank 16 (full rank)}
	\end{subfigure}
	\hfill
	\begin{subfigure}[b]{.327\linewidth}\centering
	\includegraphics[width=\textwidth]{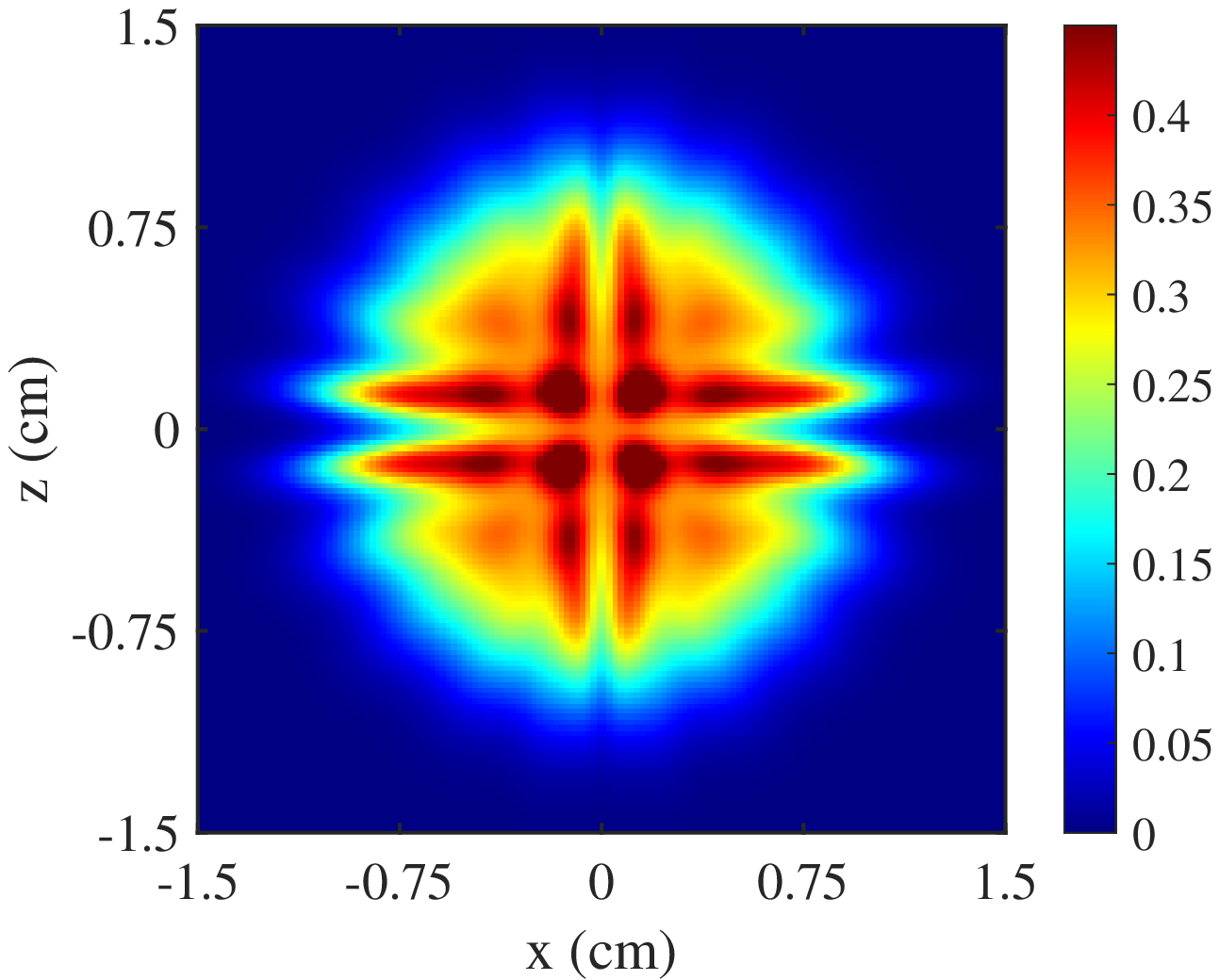}
	\subcaption{S$_{10}$, rank 25 (full rank)}
	\end{subfigure}
	\begin{subfigure}[b]{.327\linewidth}\centering
	\includegraphics[width=\textwidth]{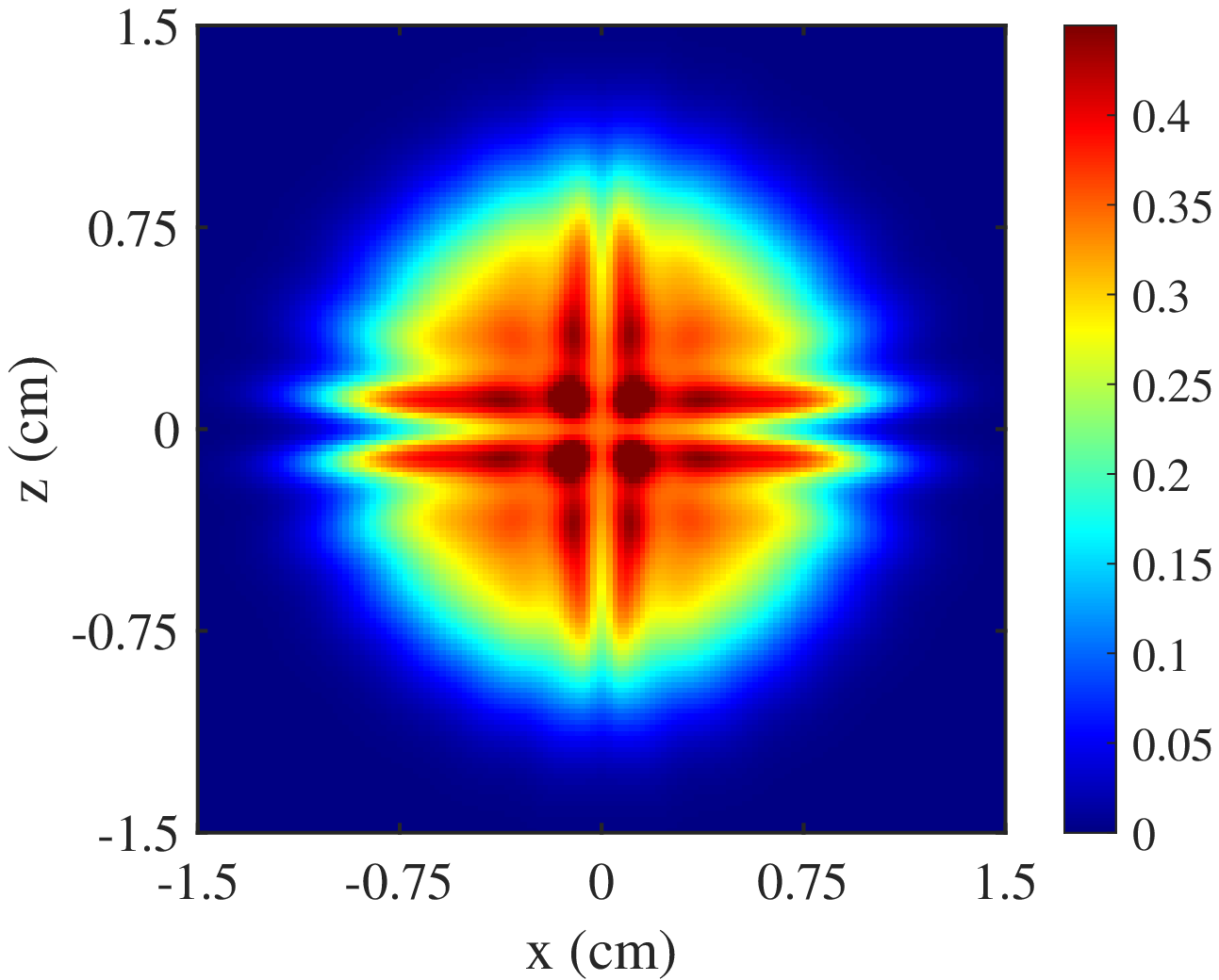}
	\subcaption{S$_{12}$, rank 36 (full rank)}
	\end{subfigure}
	\hfill
	\begin{subfigure}[b]{.327\linewidth}\centering
	\includegraphics[width=\textwidth]{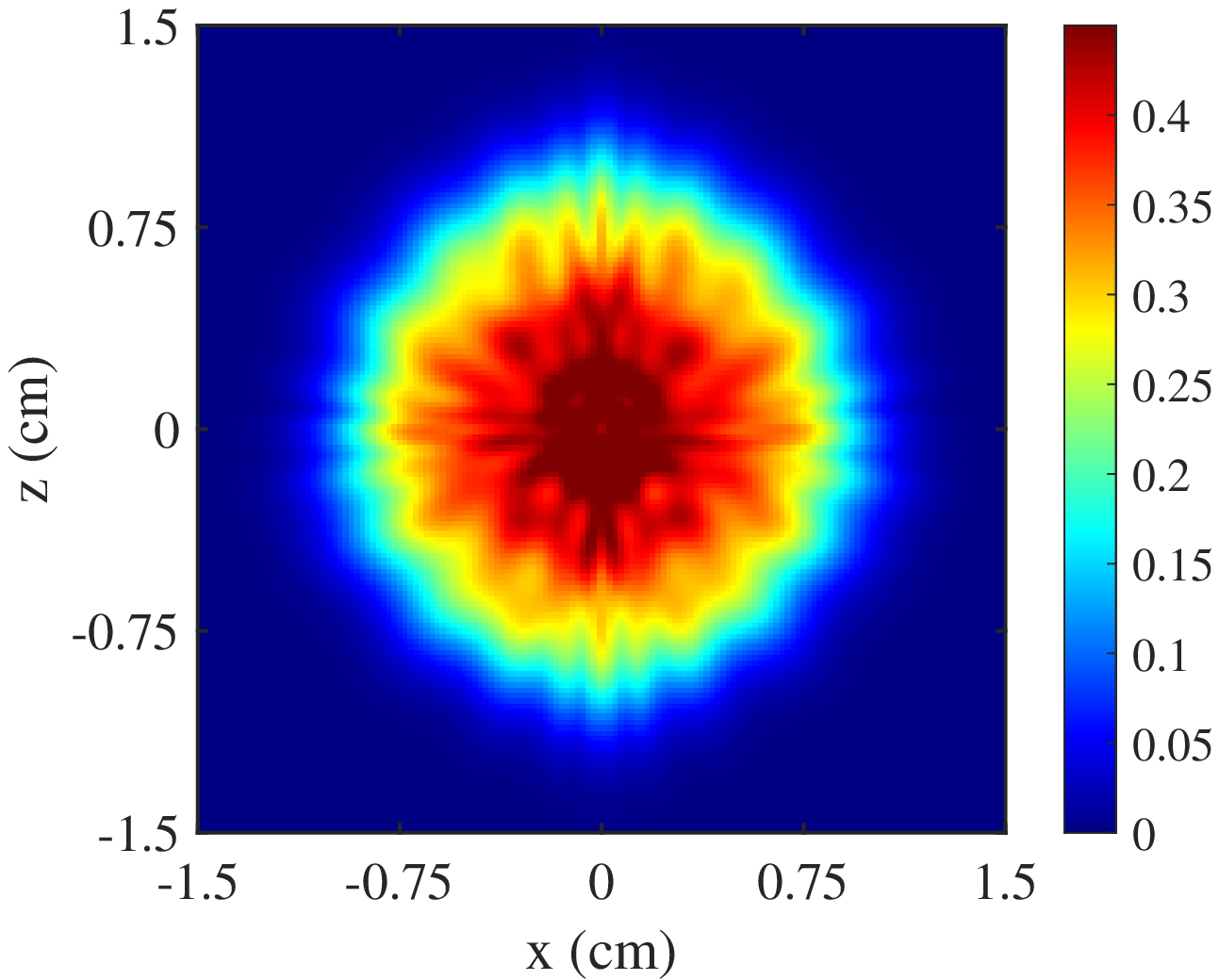}
	\subcaption{S$_{100}$, rank 16}
	\end{subfigure}
	\begin{subfigure}[b]{.327\linewidth}\centering
	\includegraphics[width=\textwidth]{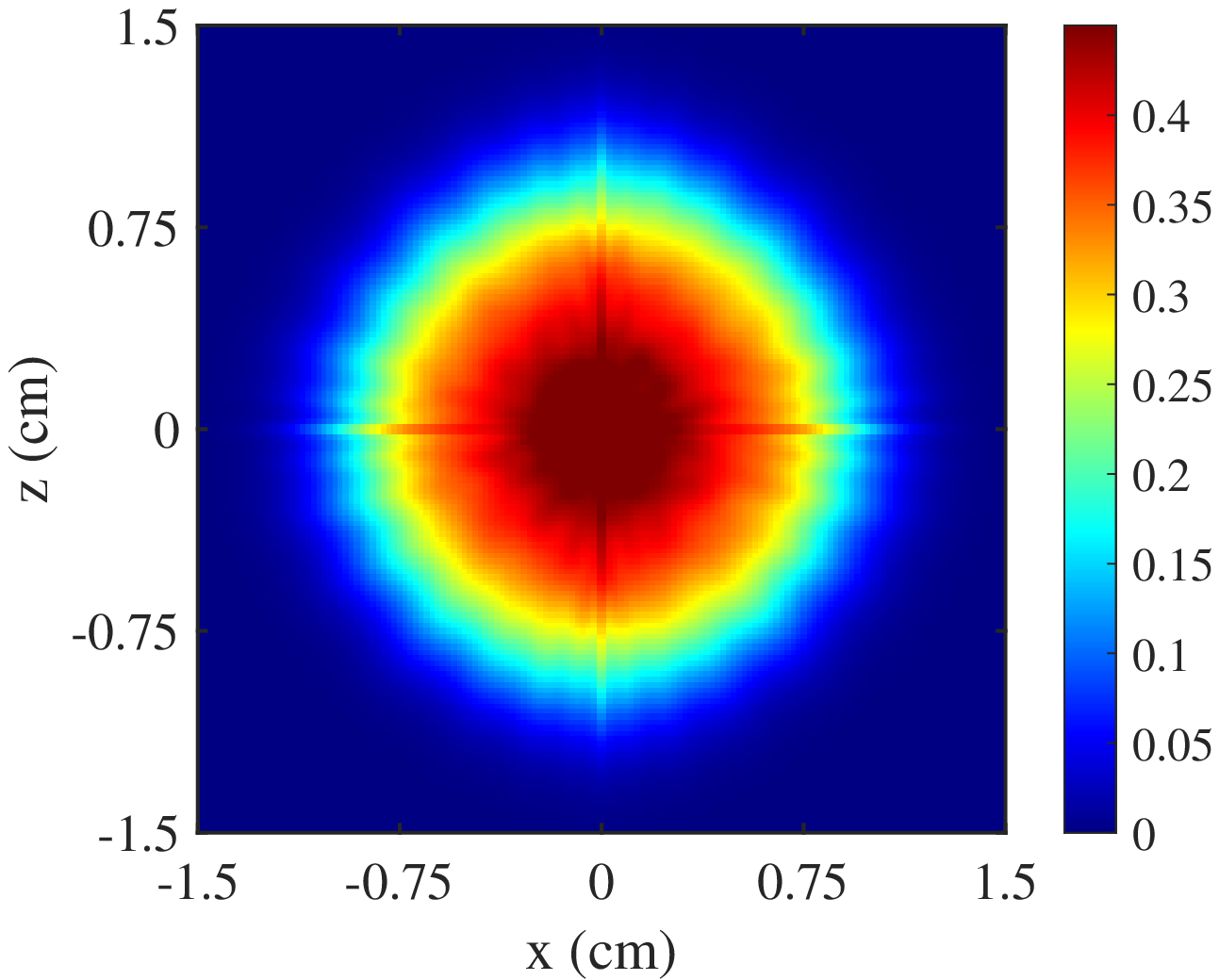}
	\subcaption{S$_{100}$, rank 25}
	\end{subfigure}
	\hfill
	\begin{subfigure}[b]{.327\linewidth}\centering
	\includegraphics[width=\textwidth]{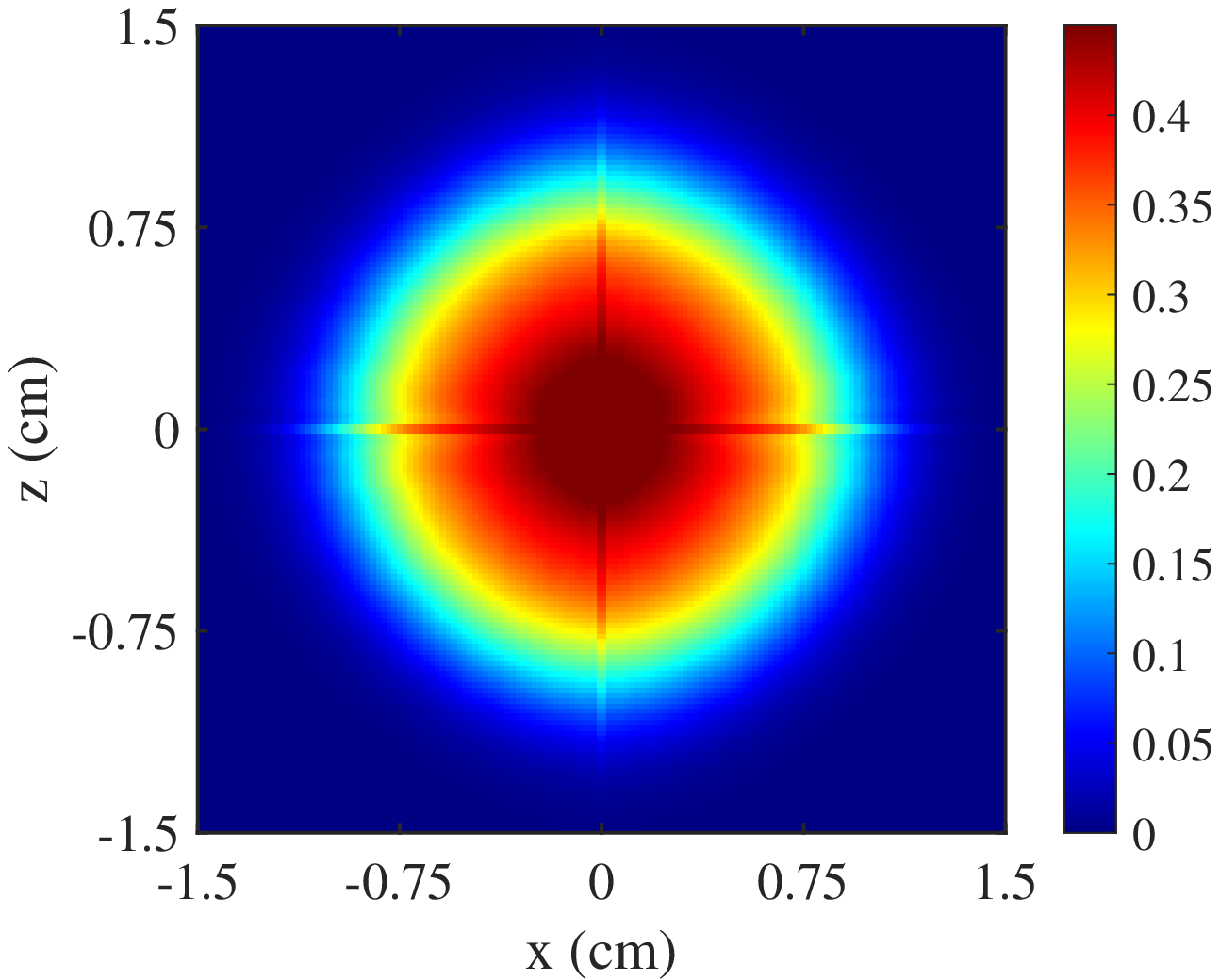}
	\subcaption{S$_{100}$, rank 36}
	\end{subfigure}
	\caption{{The scalar insentity to the line source problem calculated by the low-rank method are compared to the full rank solution with the same rank.}}
	\label{fig: ls_compare}
\end{figure}

We use the line source test to demonstrate the benefits of the low-rank method with high angular resolutions. Figure \ref{fig: ls_compare} compares the full rank solution to low-rank solutions with the same rank. We notice that all the full-rank solutions have remarkable ray effects, which means the number of angular directions are insufficiently dense to move particles to all the regions . The low-rank method could significantly improve the full rank solution by using a small amount of extra memory. As we can see, the ray effects are alleviated in the solution with rank 16, and the solution with rank 36 is comparable to the benchmark solution.

\subsection{Hohlraum problem}
\begin{figure}[h!]
    \centering
	\begin{subfigure}[b]{.38\linewidth}\centering
	\includegraphics[width=\textwidth]{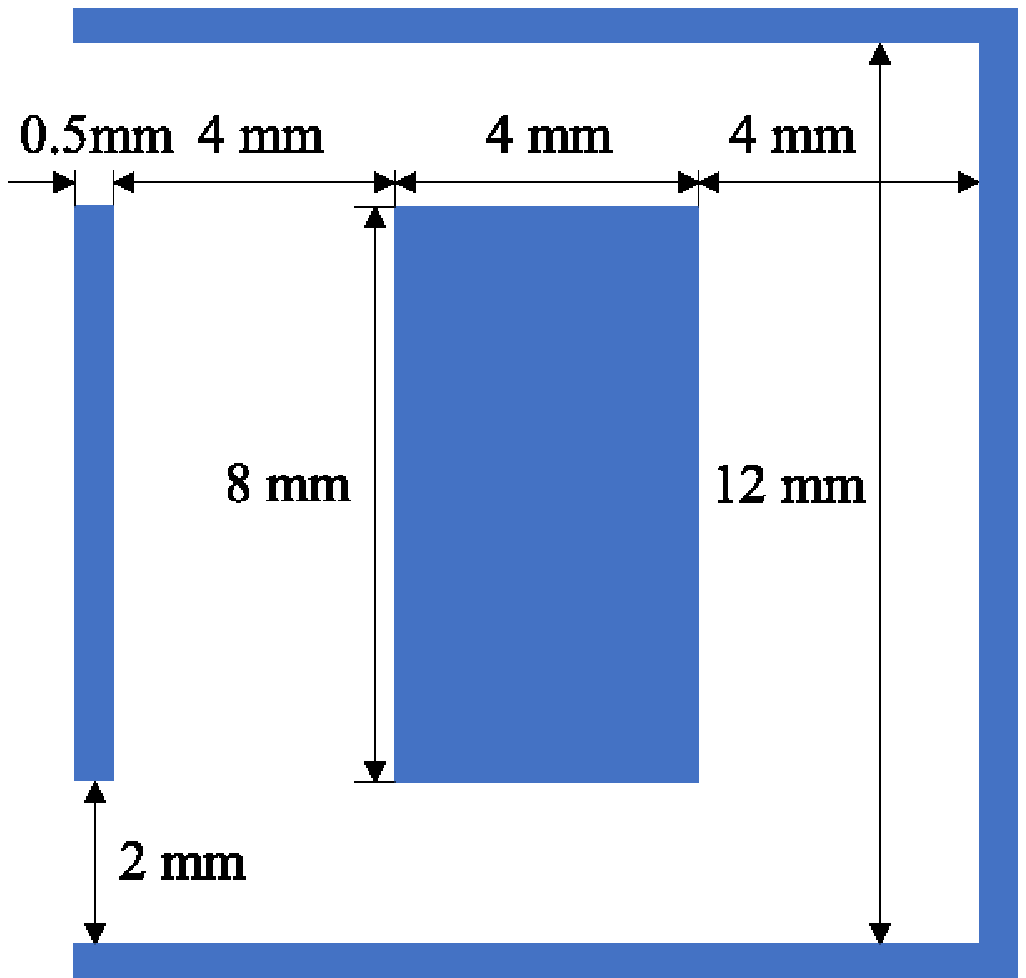}
	\subcaption{Geometry for Hohlraum test}
	\end{subfigure}
	\hfill
	\begin{subfigure}[b]{.5\linewidth}\centering
	\includegraphics[width=\textwidth]{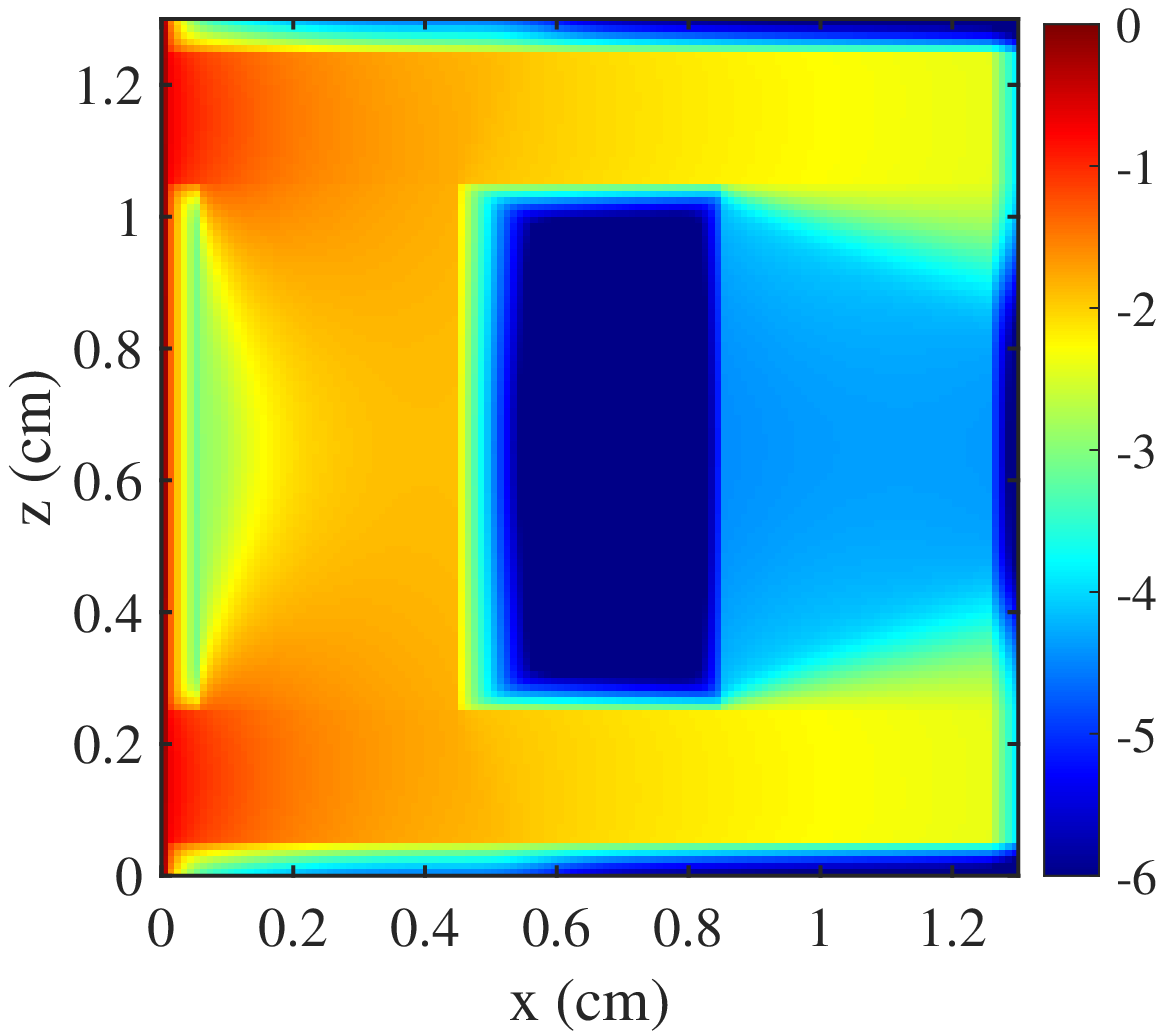}
	\subcaption{S$_{100}$, rank 2500 (full rank)}
	\end{subfigure}
	\caption{{The layout of the Hohlraum problem and the high-order benchmark solution.}}
	\label{fig: hr_layout}
\end{figure}

The last 2-D problem is a modified Hohlraum problem as shown in Figure \ref{fig: hr_layout}. The hohlraum is surrounded by vacuums, except for the incoming source on the left. In this problem, we are interested in the particle distribution behind the block that particles cannot reach directly, which will require more angular direction samples to resolve. For this problem we use a computational domain of $[0,1.3] \times [0, 1.3]$ for the simulation time $t=0.9 \,$s, and set the spatial grid to $150 \times 150$.

\begin{figure}[h!]
    \centering
	\begin{subfigure}[b]{.327\linewidth}\centering
	\includegraphics[width=\textwidth]{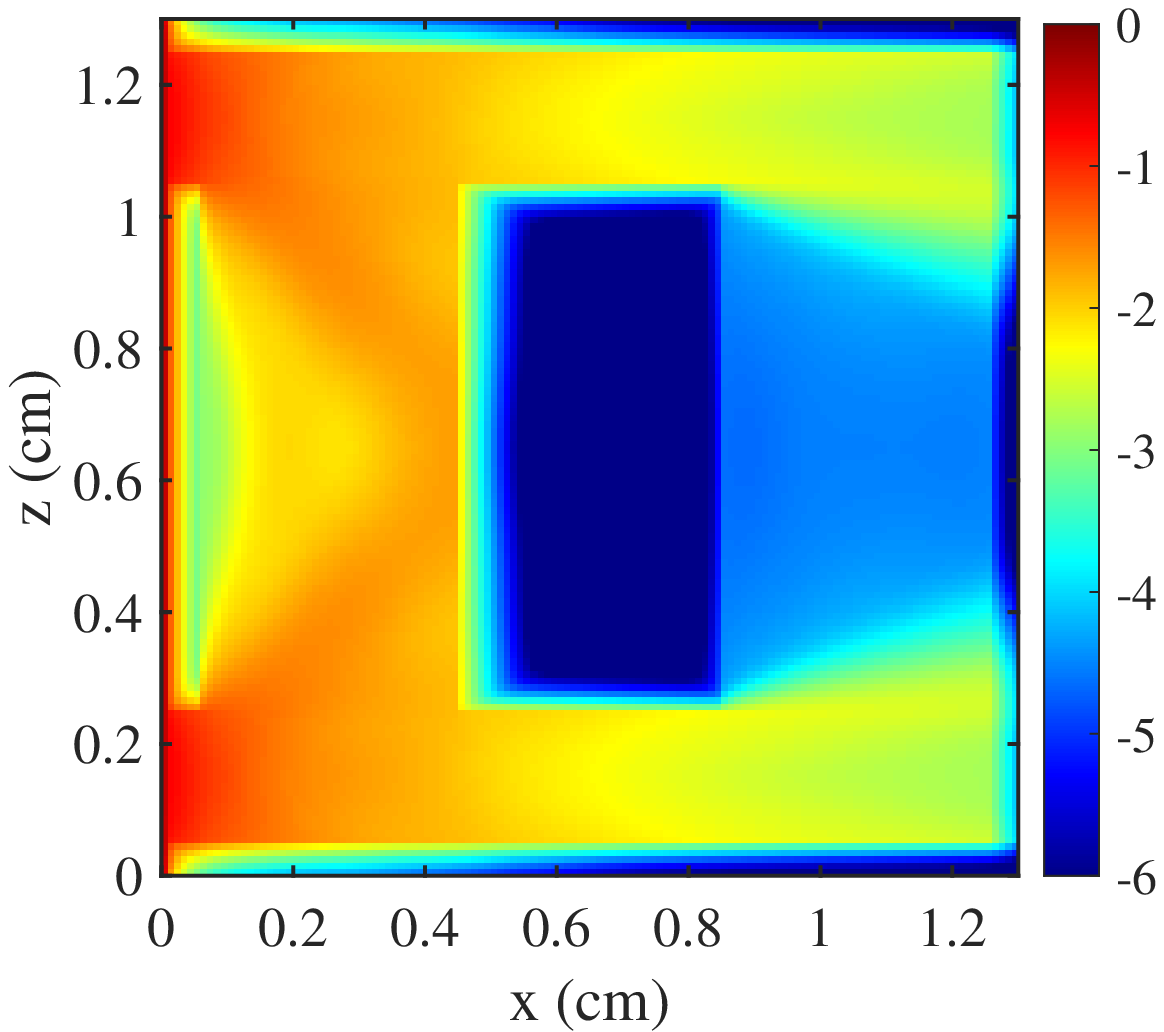}
	\subcaption{S$_{8}$, rank 16 (full rank)}
	\end{subfigure}
	\hfill
	\begin{subfigure}[b]{.327\linewidth}\centering
	\includegraphics[width=\textwidth]{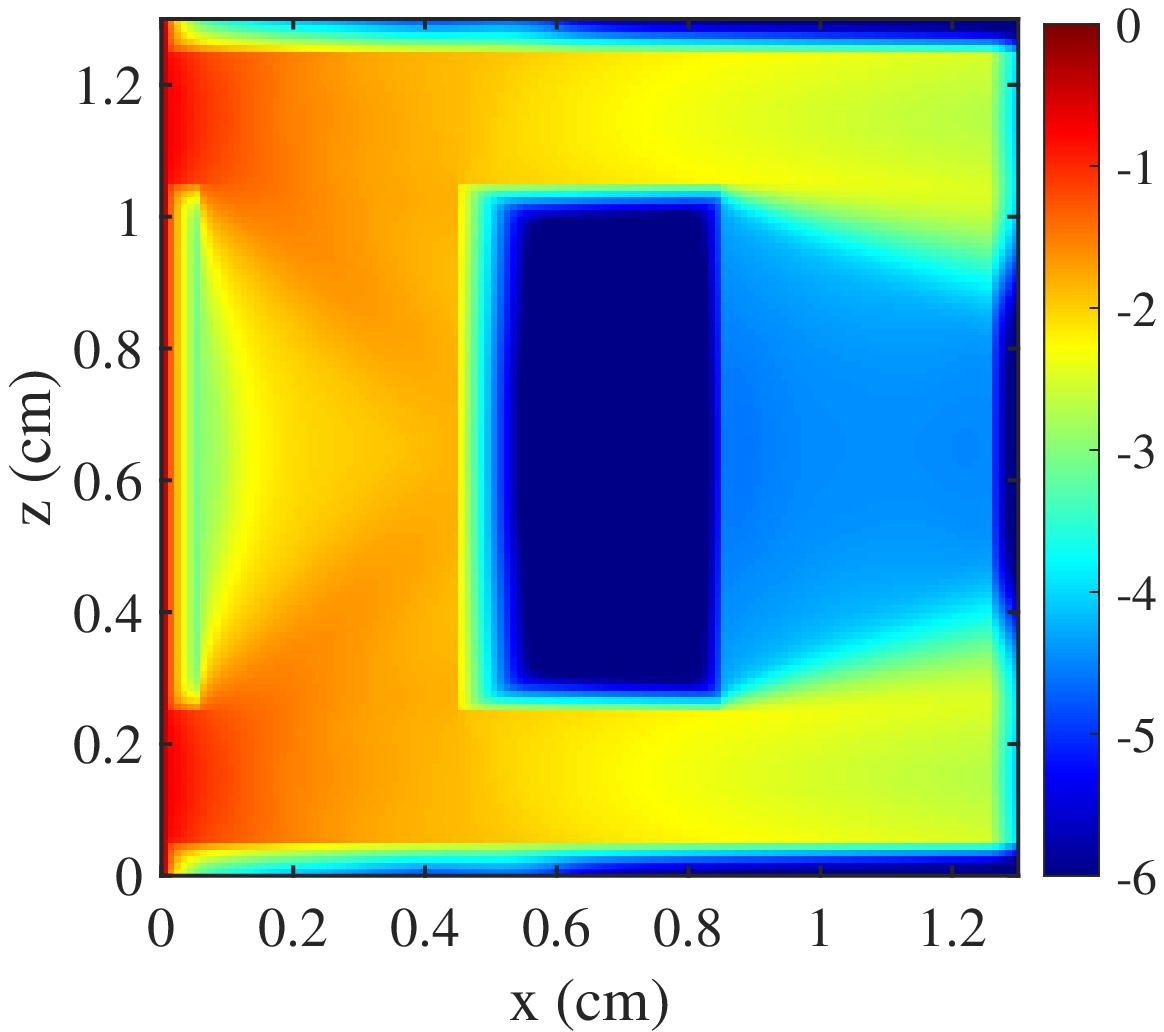}
	\subcaption{S$_{10}$, rank 25 (full rank)}
	\end{subfigure}
	\begin{subfigure}[b]{.327\linewidth}\centering
	\includegraphics[width=\textwidth]{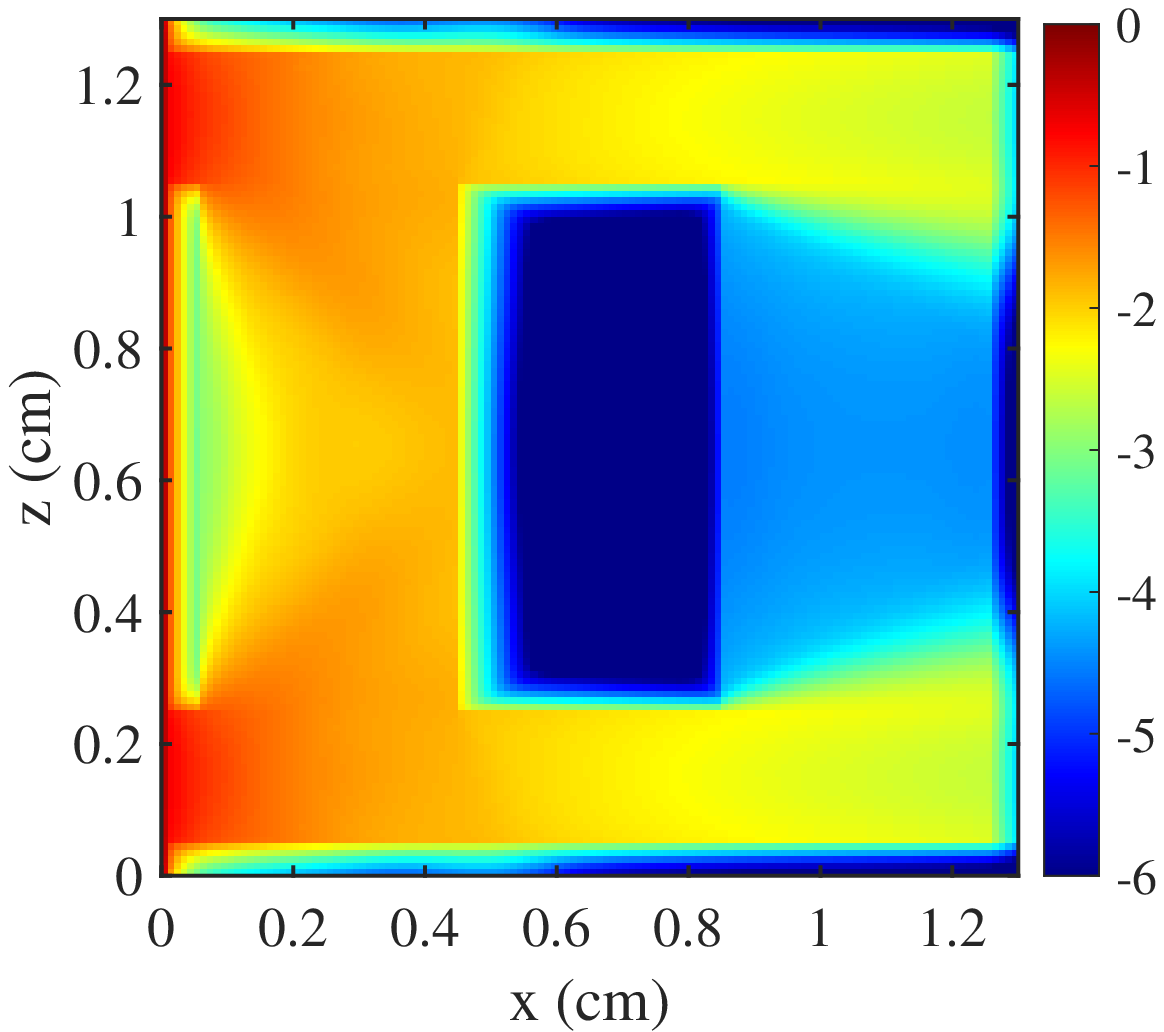}
	\subcaption{S$_{12}$, rank 36 (full rank)}
	\end{subfigure}
	\hfill
	\begin{subfigure}[b]{.327\linewidth}\centering
	\includegraphics[width=\textwidth]{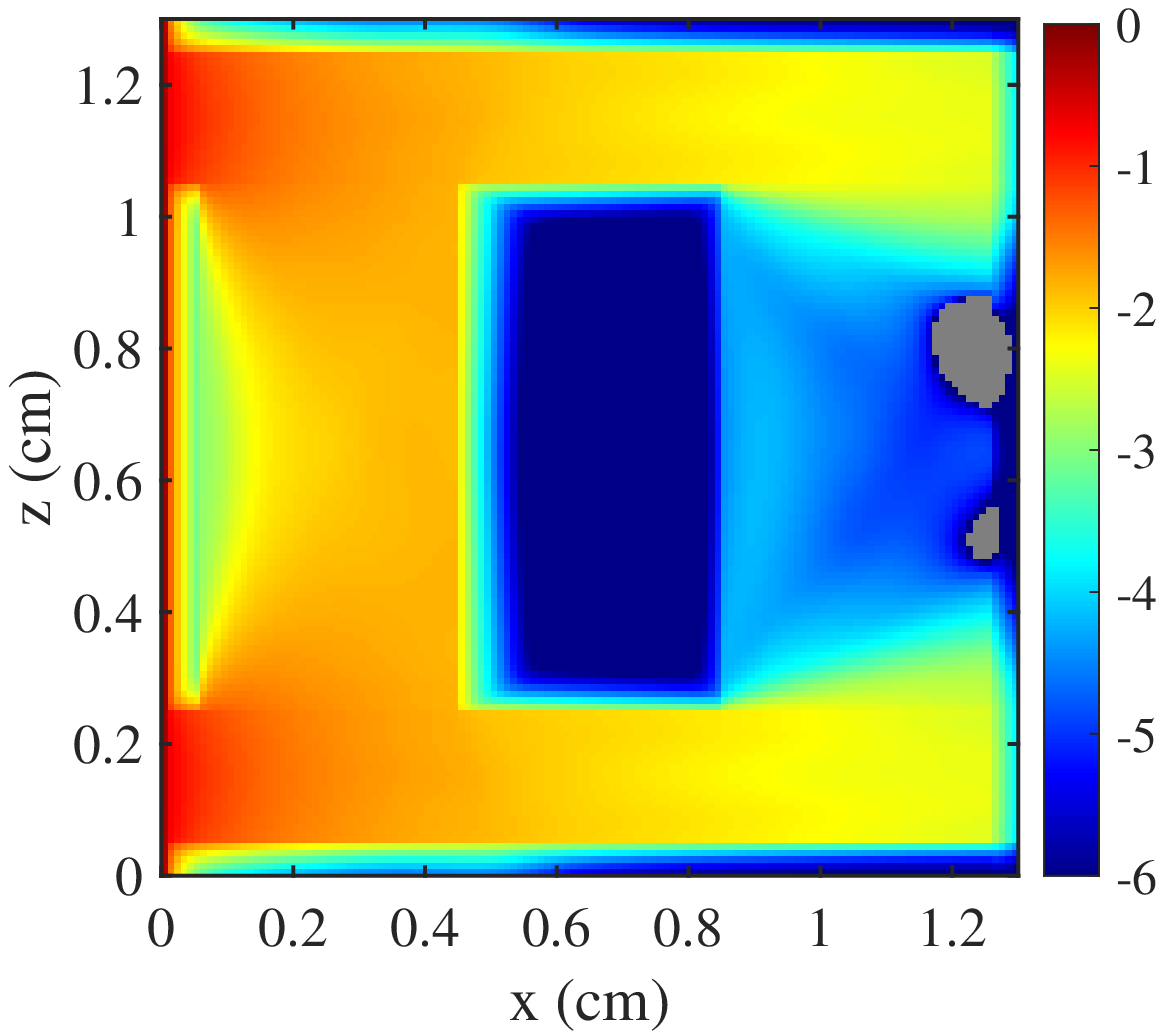}
    \subcaption{S$_{100}$, rank 16}
	\end{subfigure}
	\begin{subfigure}[b]{.327\linewidth}\centering
	\includegraphics[width=\textwidth]{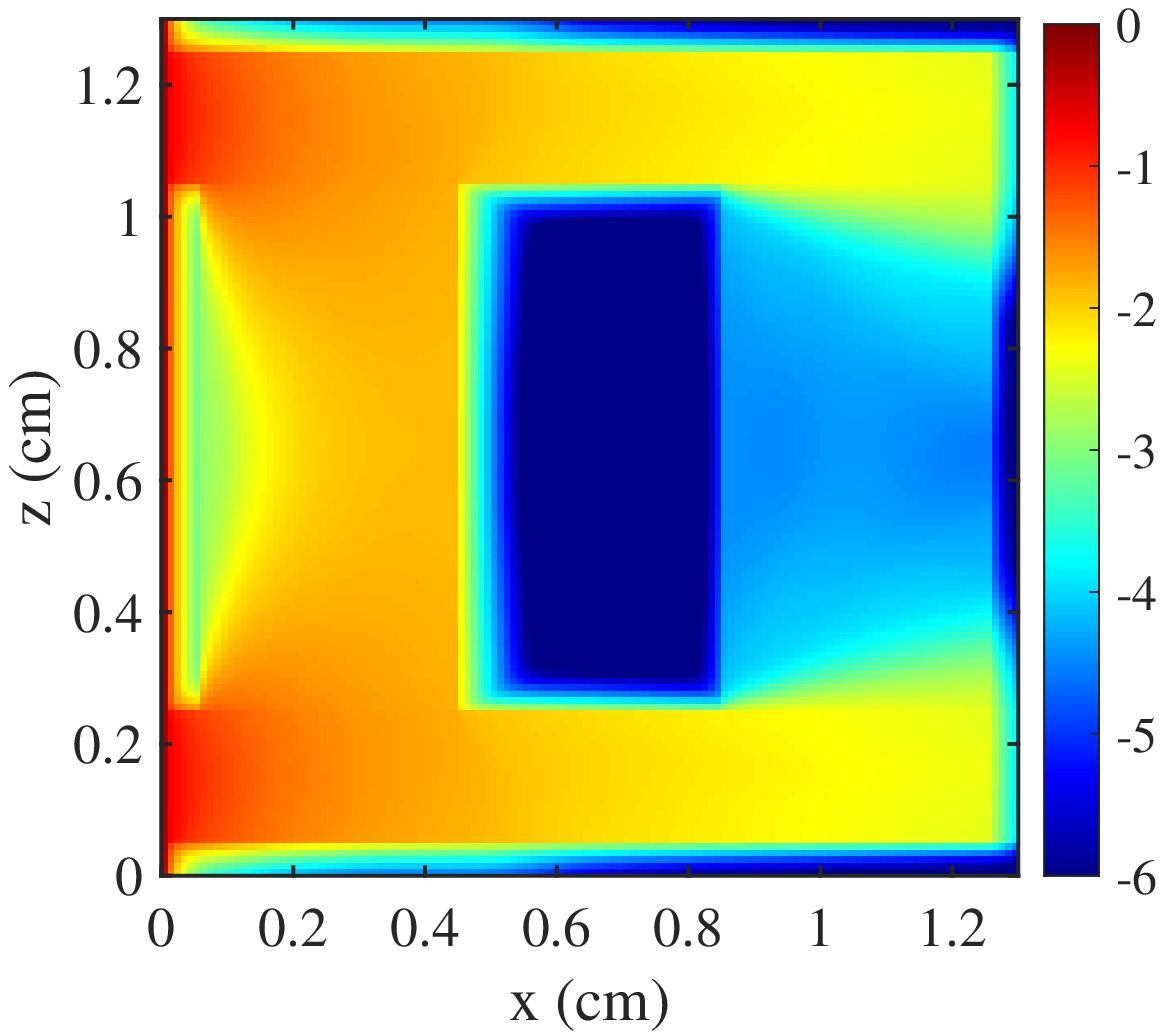}
	\subcaption{S$_{100}$, rank 25}
	\end{subfigure}
	\hfill
	\begin{subfigure}[b]{.327\linewidth}\centering
	\includegraphics[width=\textwidth]{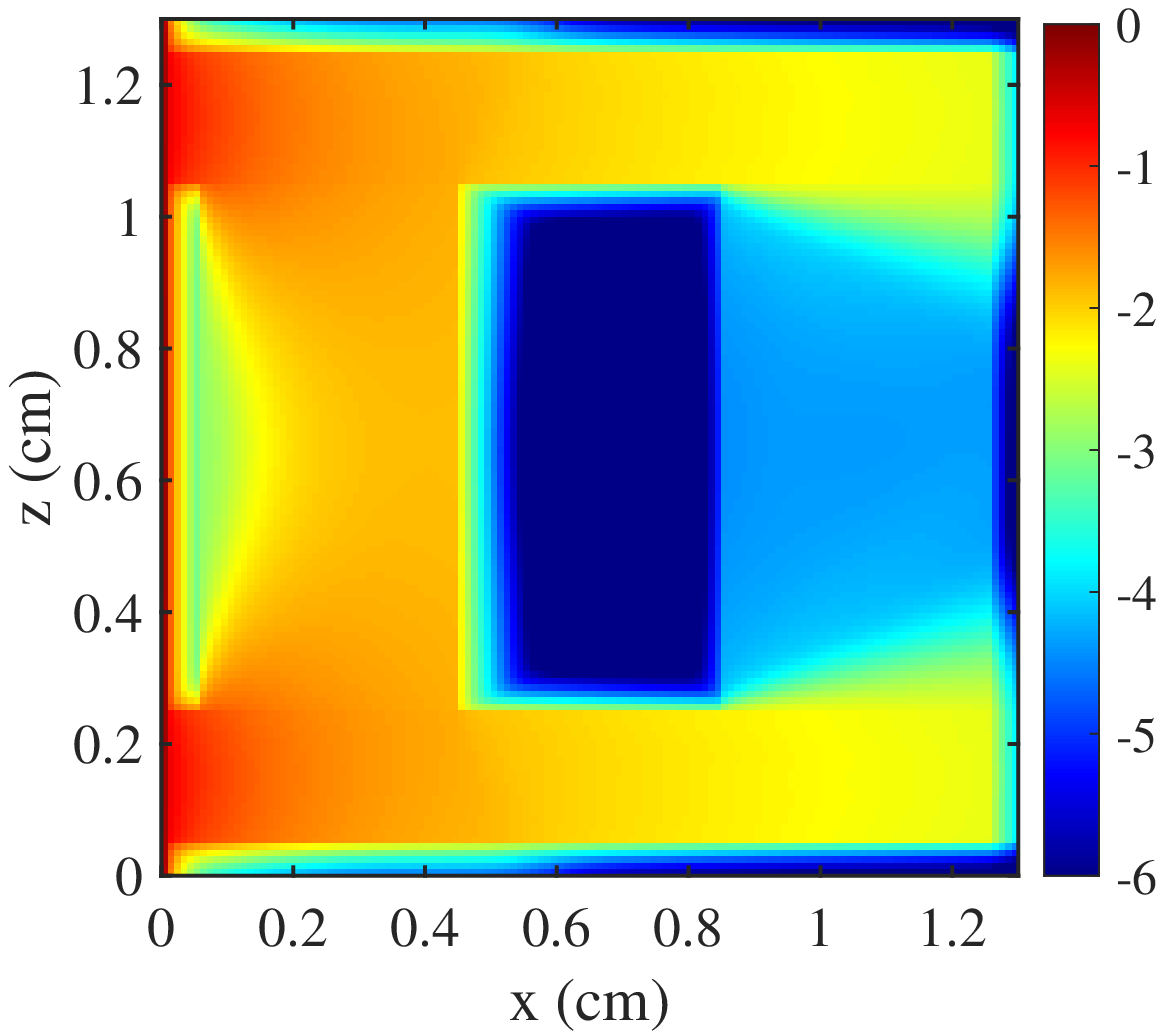}
	\subcaption{S$_{100}$, rank 36}
	\end{subfigure}
	\caption{{The scalar insentity to the Hohlraum problem calculated by the low-rank method with S$_{100}$ are compared to the same rank solutions without rank reduction. The color scale is logarithmic and negative regions are shaded gray.}}
	\label{fig: hr_compare}
\end{figure}

We make a similar comparison between the low-rank solutions to the full rank solutions with the same rank in Figure \ref{fig: hr_compare}. We observe beam-shaped shadows in the left part of the full rank S$_{8}$ solution, but not in the low-rank solution with S$_{100}$ and rank 16. There are negative regions in the low-rank solutions brought by the truncation from low-rank algorithms.

\begin{figure}[h!]
    \centering
	\begin{subfigure}[b]{.495\linewidth}\centering
	\includegraphics[width=\textwidth]{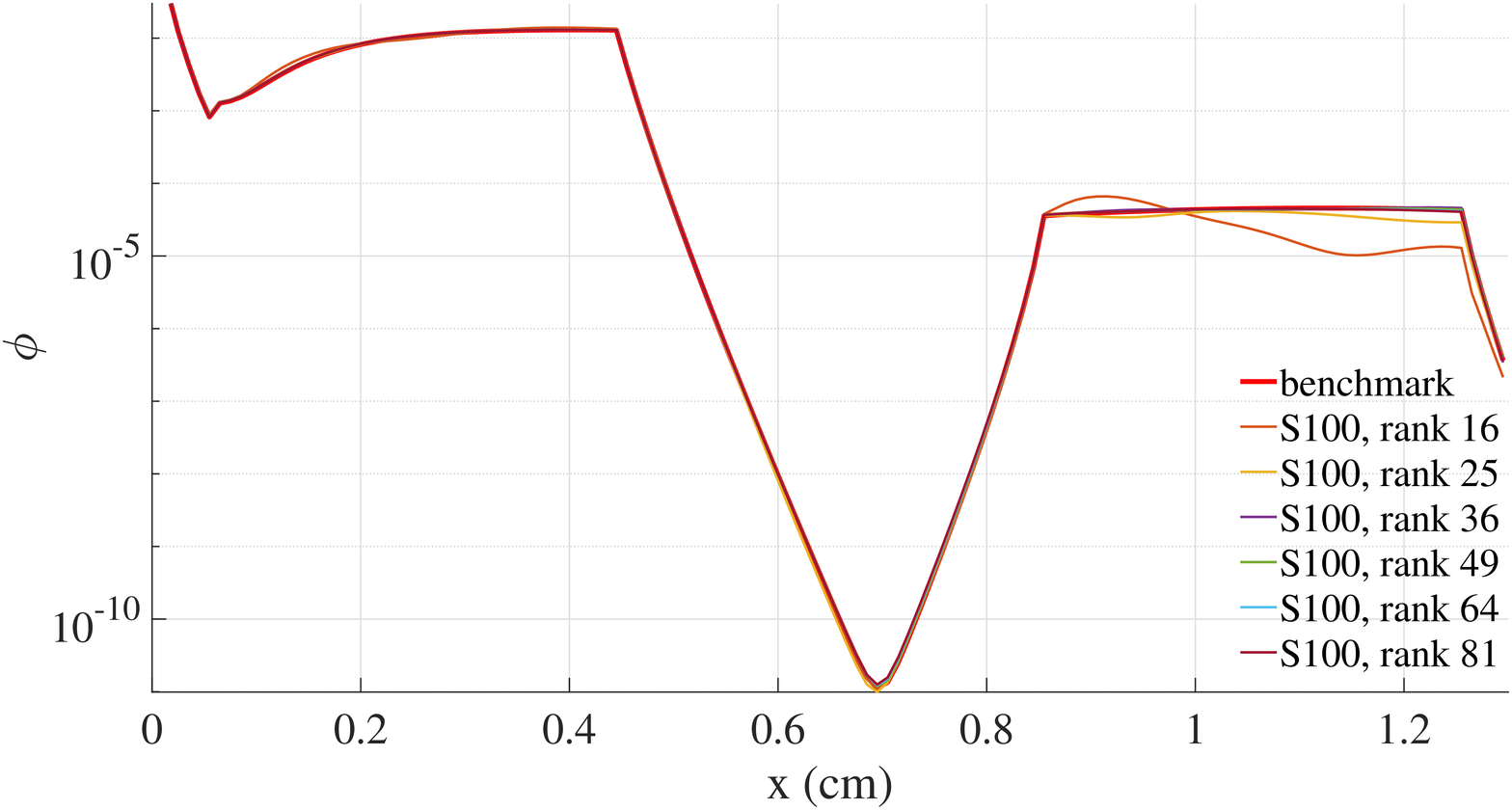}
	\subcaption{S$_{100}$ with varying rank}
	\end{subfigure}
	\hfill
	\begin{subfigure}[b]{.495\linewidth}\centering
	\includegraphics[width=\textwidth]{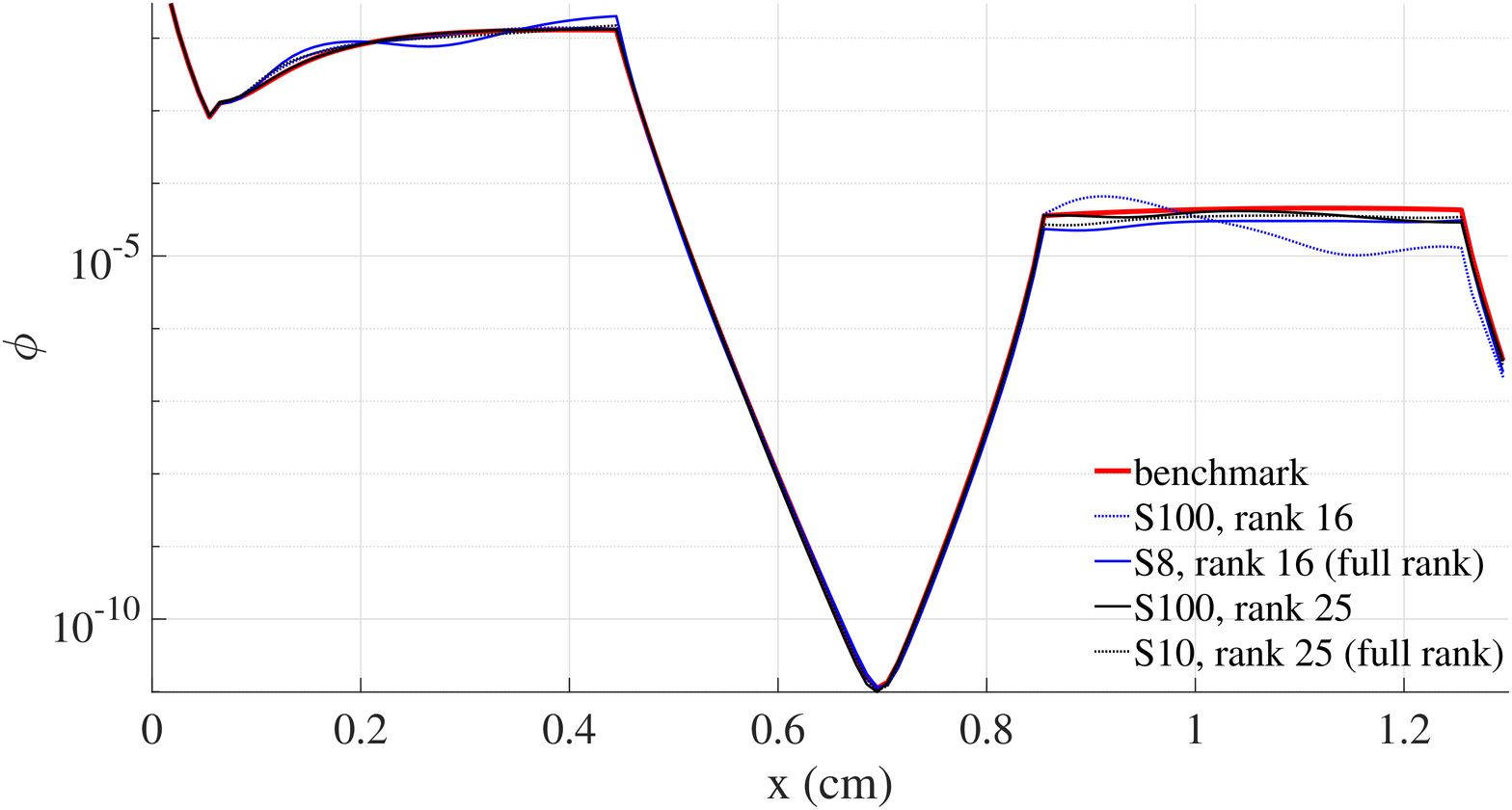}
	\subcaption{Same rank with varying S$_{N}$}
	\end{subfigure}
	\caption{{The logarithm of the scalar insentity along $z=0.65$ cm.}}
	\label{fig: hr_cut}
\end{figure}

Figure \ref{fig: hr_cut} presents the cut along $z = 0.65$ cm to compare the particle distribution behind the obstacle. The left plot shows that the solution with rank 36 is nearly identical to the full rank solution. The right plot shows that the low-rank solutions are closer to the benchmark than the full rank solution with the same rank. 

\subsection{Wires Problem}
\begin{figure}[h!]
    \centering
	\begin{subfigure}[b]{0.45\linewidth}\centering
	\includegraphics[width=\textwidth]{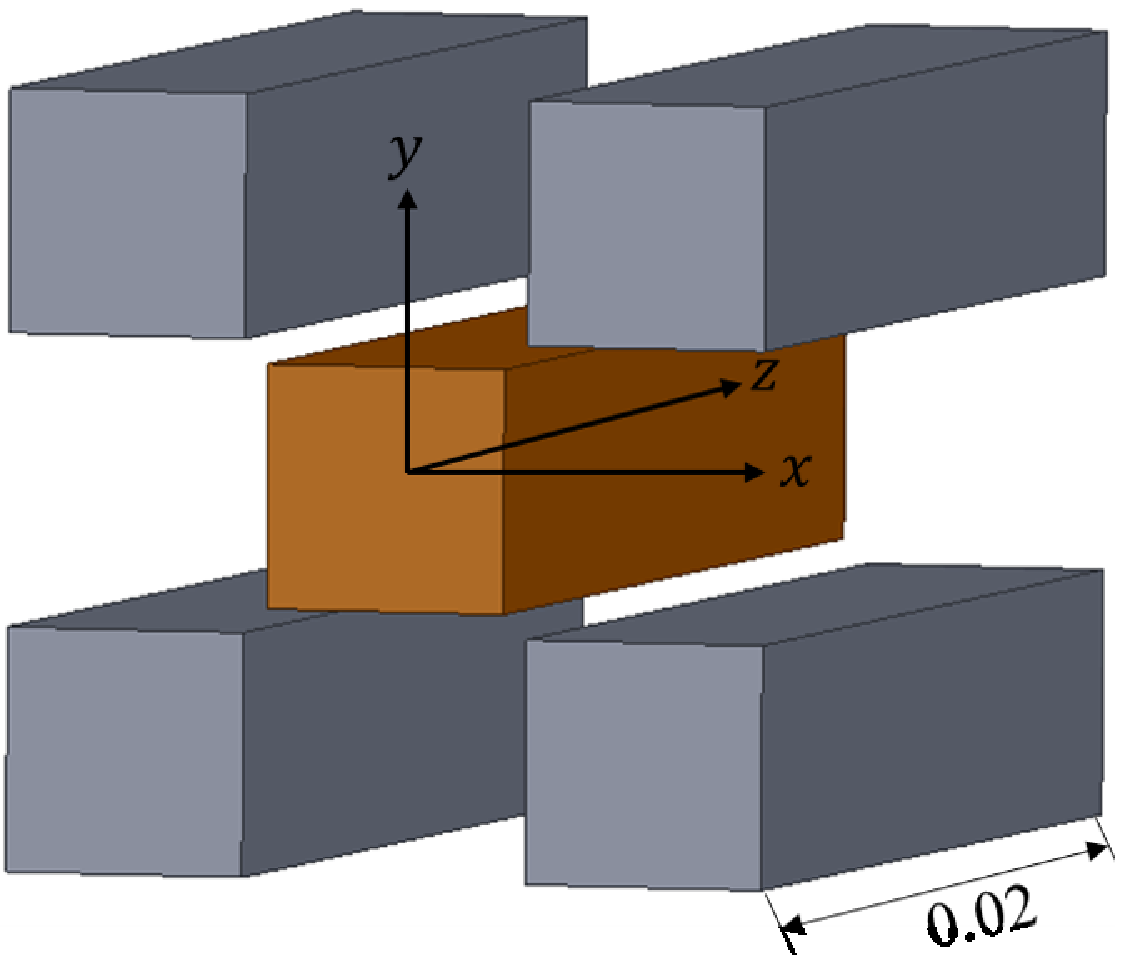}
	\subcaption{The wires}
	\end{subfigure} 
	\hfill
	\begin{subfigure}[b]{0.45\linewidth}\centering
	\includegraphics[width=\textwidth]{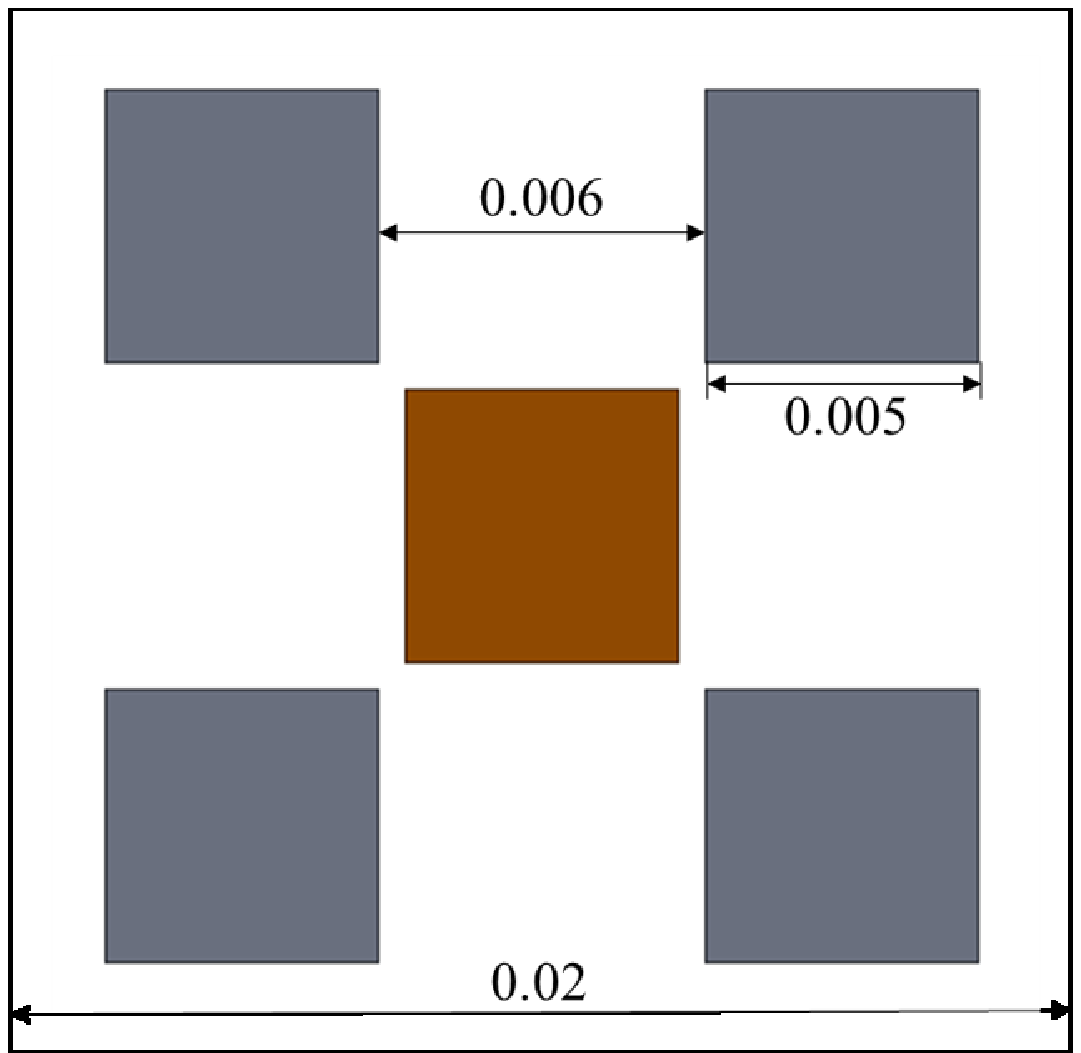}
	\subcaption{The geometry at the x-y plane}
	\end{subfigure} 
	\hfill
	\begin{subfigure}[b]{0.49\linewidth}\centering
	\includegraphics[width=\textwidth]{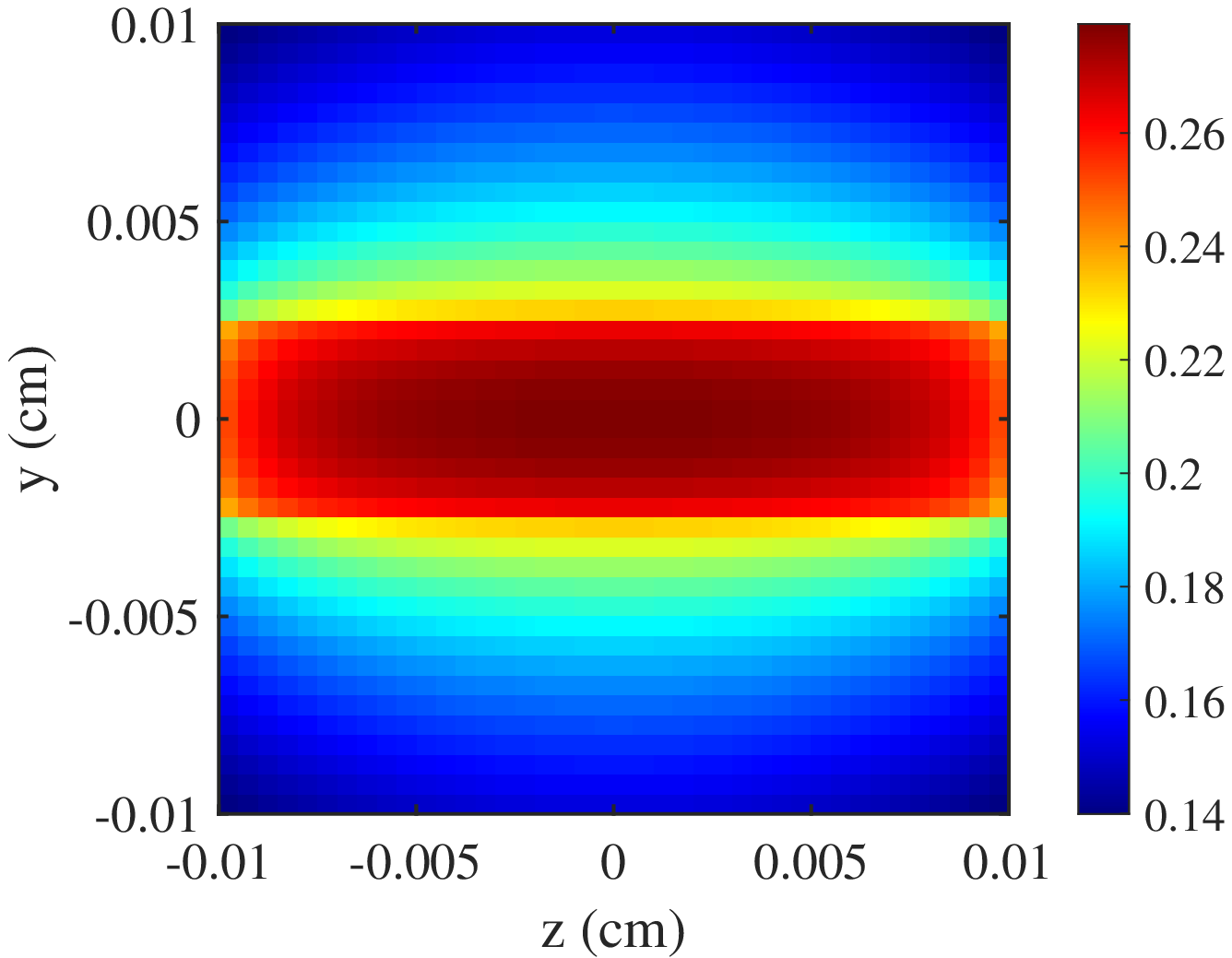}
	\subcaption{Benchmark in y-z along $z = 0.01$}
	\end{subfigure}
	\hfill
	\begin{subfigure}[b]{0.49\linewidth}\centering
	\includegraphics[width=\textwidth]{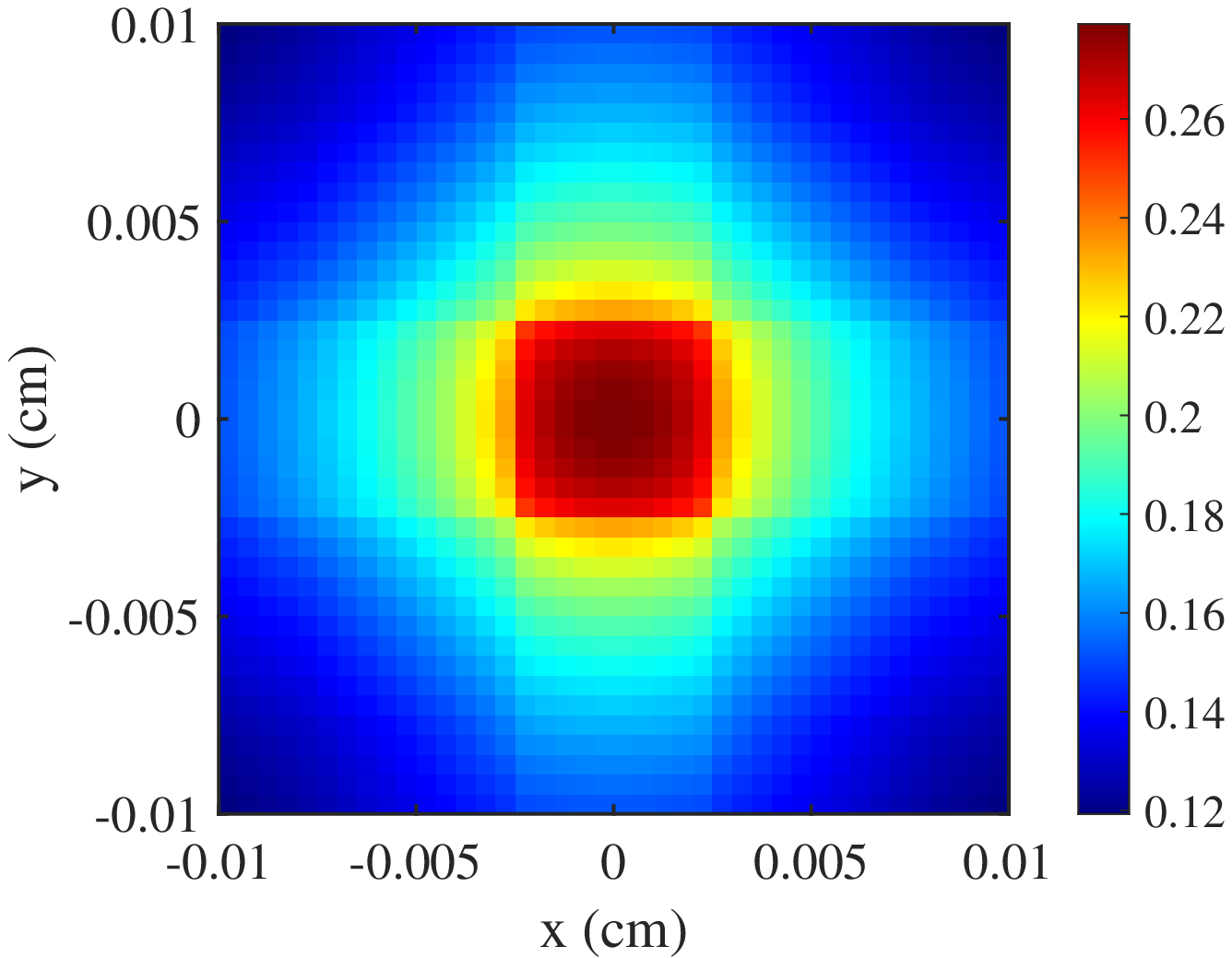}
	\subcaption{Benchmark in y-z along $x = 0$}
	\end{subfigure}
	\hfill
	\caption{{The layout of the wires problem and the high-order benchmark solution.}}
	\label{fig: wires_layout}
\end{figure}

As shown in Figure \ref{fig: wires_layout}, we introduce the 3-D wires problem, inspired by the 2-D problem in \cite{mcclarren2008solutions}, has an emitting dense material embedded vacuum. Particles are emitted from the central wire with $Q = 1$. All the five wires are made by the same material with $\sigmas =0.1$ and $\sigmat = 5$. The streaming regions has $\sigmas = \sigmat = 0$ and we use vacuum boundary conditions. The computational domain for this problem is set to $[-0.01, 0.01]\times[-0.01, 0.01]\times[0, 0.02]$ with a $40\times40\times40$ mesh grid. We simulate this problem to $t = 0.6$s using a time step $\Delta t = 0.2$ (CFL$ = 1200$). 

\begin{figure}[h!]
    \centering
	\begin{subfigure}[b]{.327\linewidth}\centering
	\includegraphics[width=\textwidth]{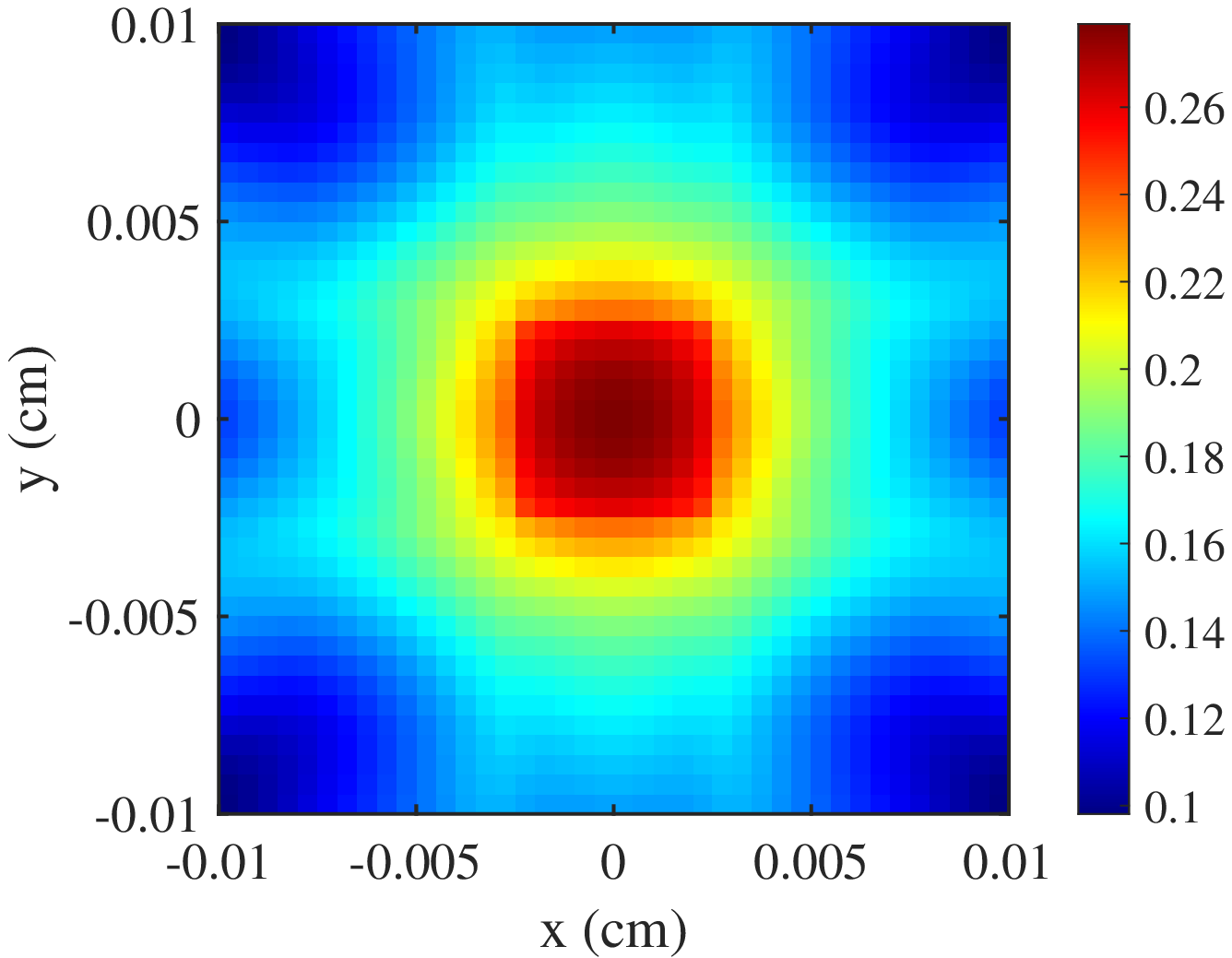}
	\subcaption{S$_{4}$, rank $4$ (full-rank)}
	\end{subfigure}
	\hfill
	\begin{subfigure}[b]{.327\linewidth}\centering
	\includegraphics[width=\textwidth]{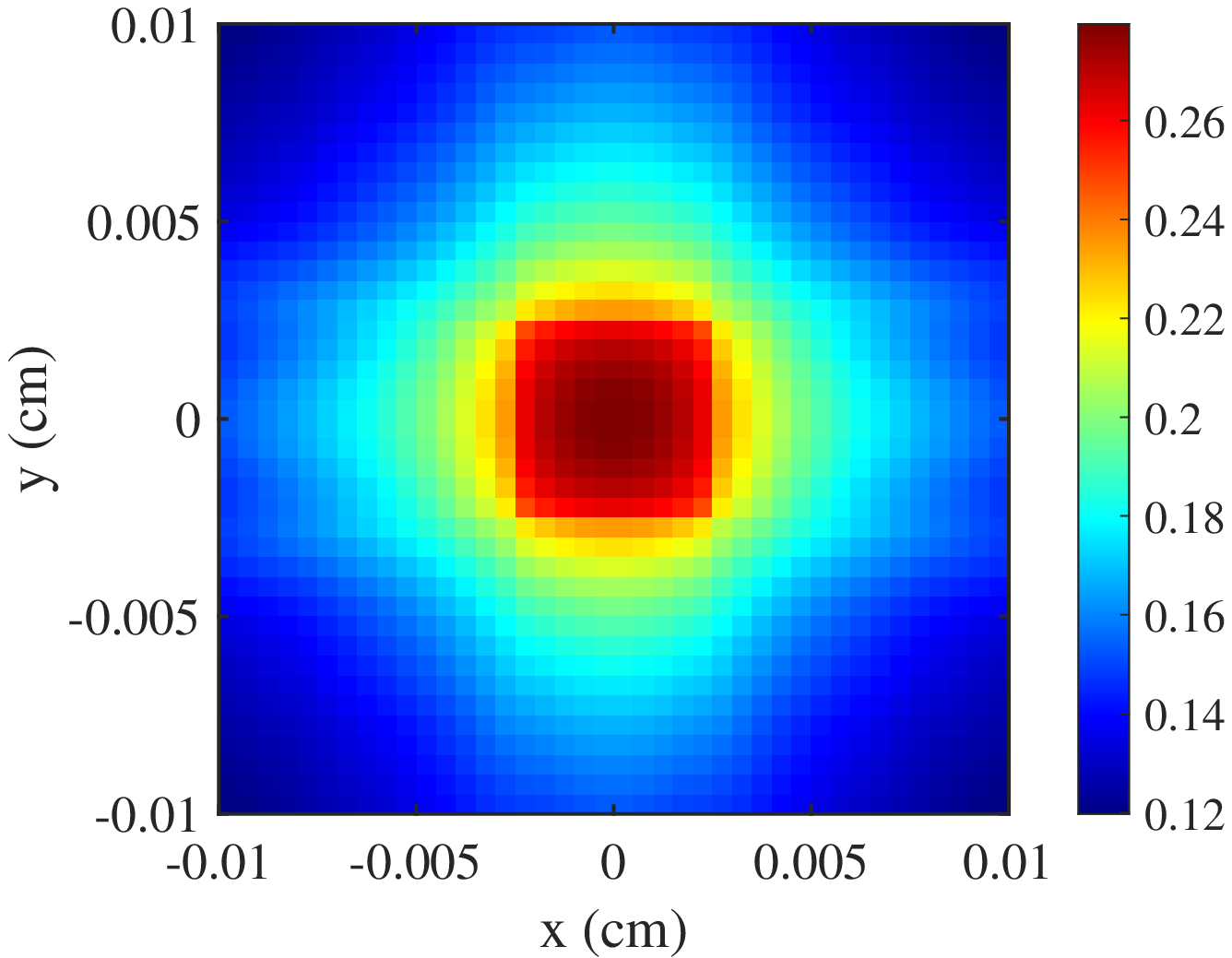}
	\subcaption{S$_{8}$, rank $16$ (full-rank)}
	\end{subfigure}
	\hfill
	\begin{subfigure}[b]{.327\linewidth}\centering
	\includegraphics[width=\textwidth]{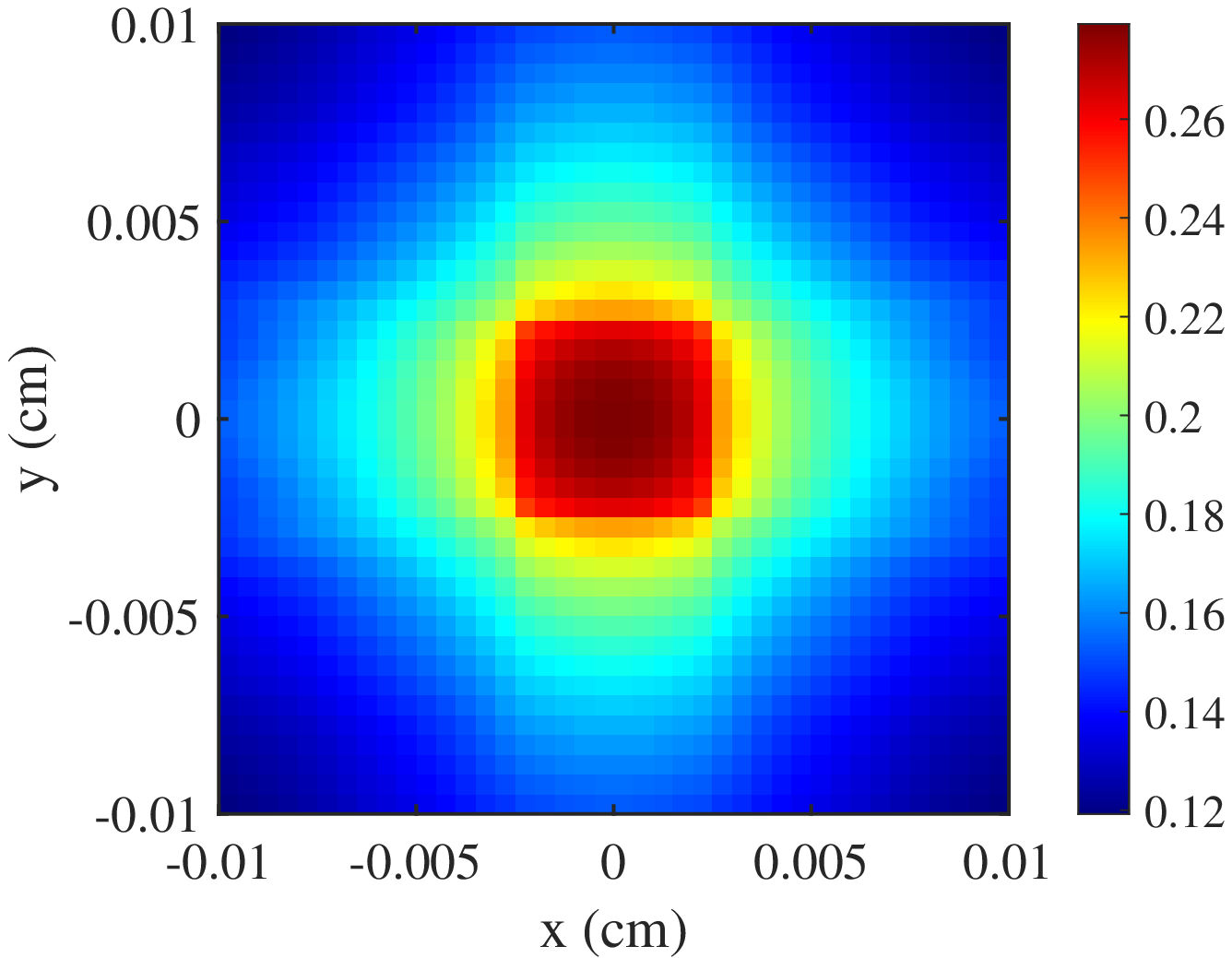}
	\subcaption{S$_{12}$, rank $36$ (full-rank)}
	\end{subfigure}
	\hfill
	\begin{subfigure}[b]{.327\linewidth}\centering
	\includegraphics[width=\textwidth]{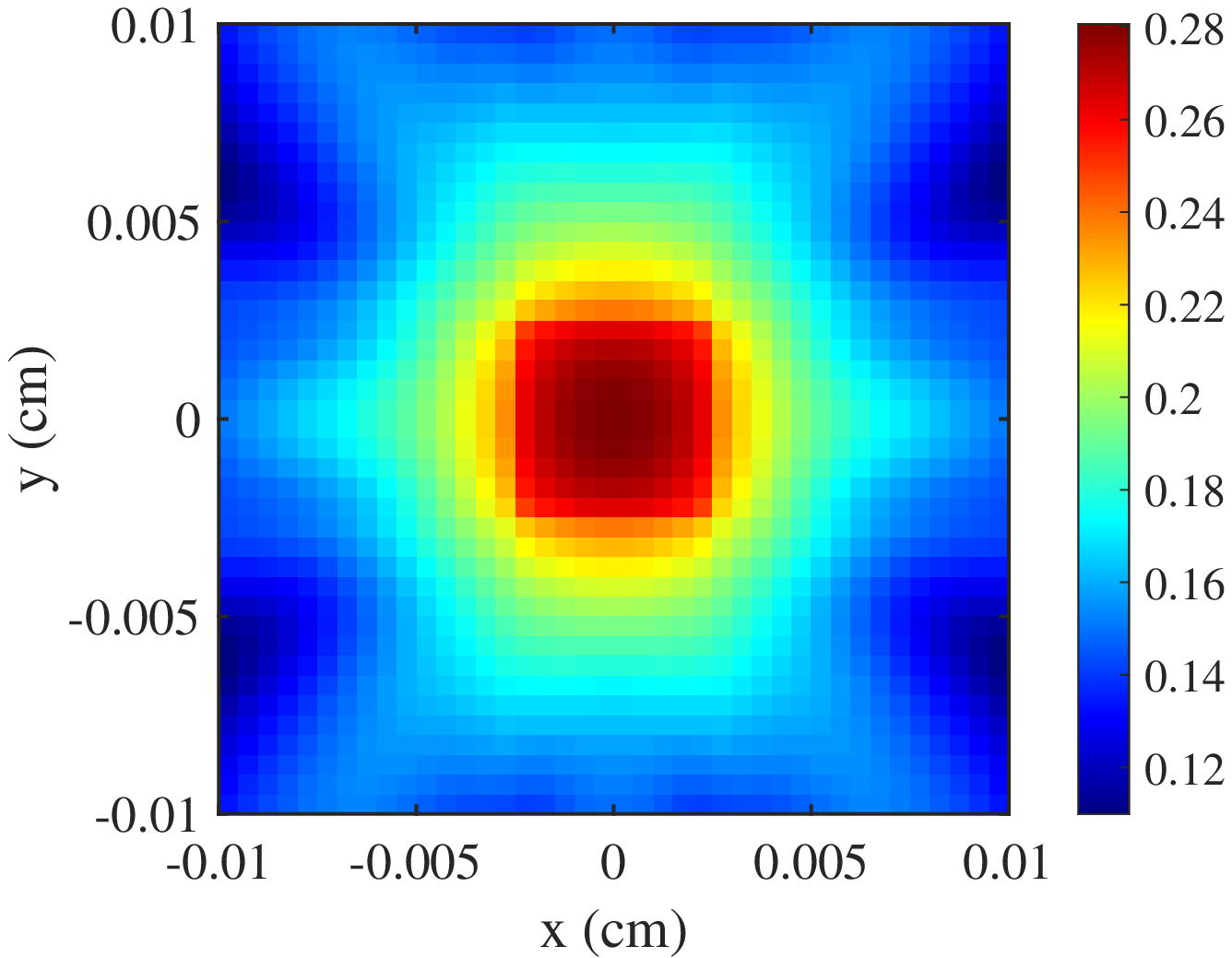}
	\subcaption{S$_{64}$, rank $4$}
	\end{subfigure}
	\hfill
	\begin{subfigure}[b]{.327\linewidth}\centering
	\includegraphics[width=\textwidth]{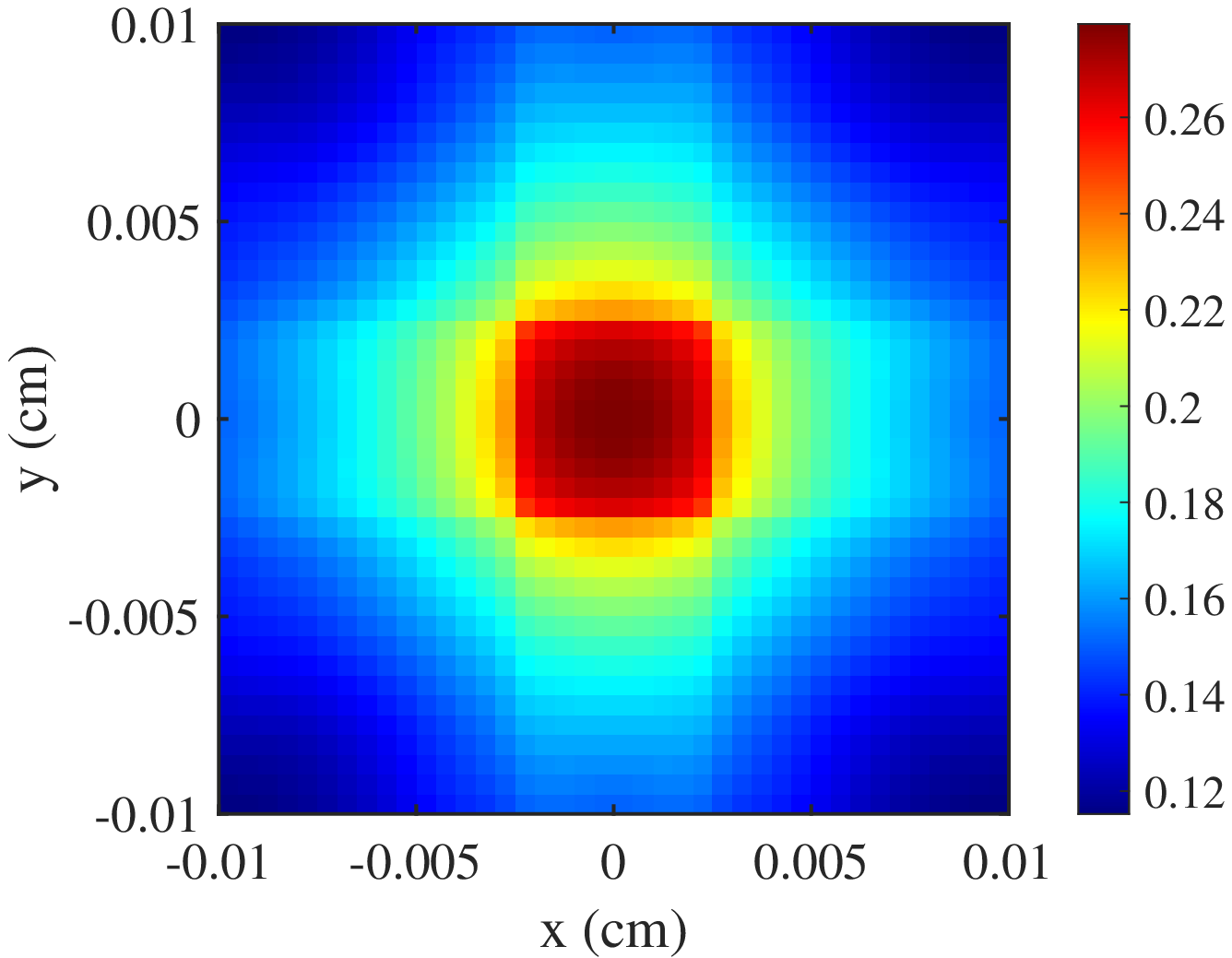}
	\subcaption{S$_{64}$, rank $16$}
	\end{subfigure}
	\hfill
	\begin{subfigure}[b]{.327\linewidth}\centering
	\includegraphics[width=\textwidth]{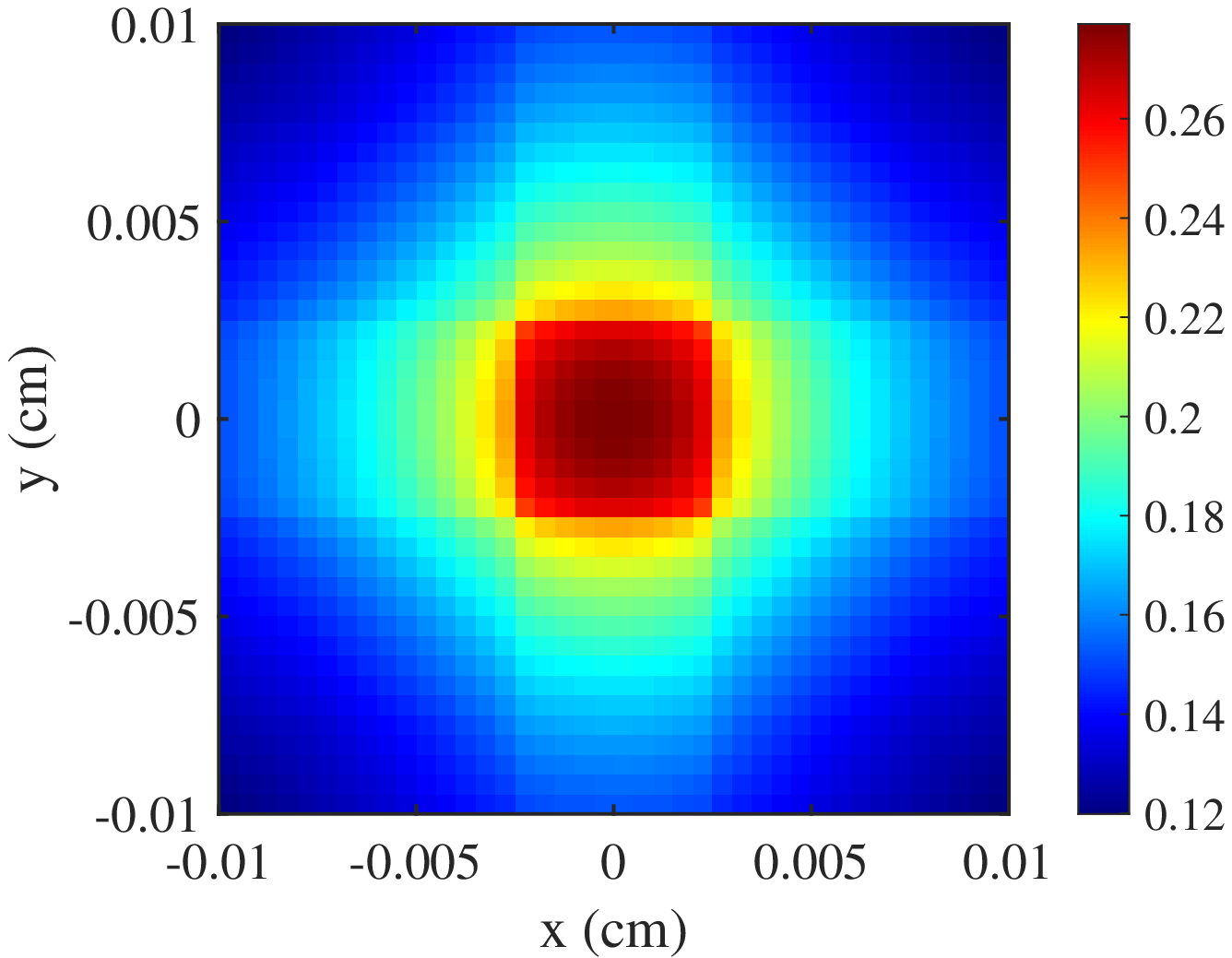}
	\subcaption{S$_{64}$, rank $36$}
	\end{subfigure}
	\caption{{The logarithmic scalar intensity solution  the wires problem calculated by the low-rank method with S$_{64}$ are compared to the same rank solutions without rank reduction.}}
	\label{fig: wires_compare_xy}
\end{figure}

The following results compare the low-rank solution and the full rank solution with the same rank. Figure \ref{fig: wires_compare_xy} shows the solution in the plane of $x-y$ along $z = 0.1$. First, we notice strong ray effects in the full rank S$_{4}$ solution, especially along $y = 0$. The low-rank solution with S$_{64}$ and rank 4 looks better in this aspect. The full rank solution with S$_{8}$ achieves better distributions than S$_{4}$ but is still unable to avoid the ray effects along $y = 0$. But we can see that the low-rank solution with rank 16 is close to the full rank solution. If we further increase the rank to 36, the low-rank solution is identical to the benchmark solution.

\begin{figure}[h!]
    \centering
	\begin{subfigure}[b]{.327\linewidth}\centering
	\includegraphics[width=\textwidth]{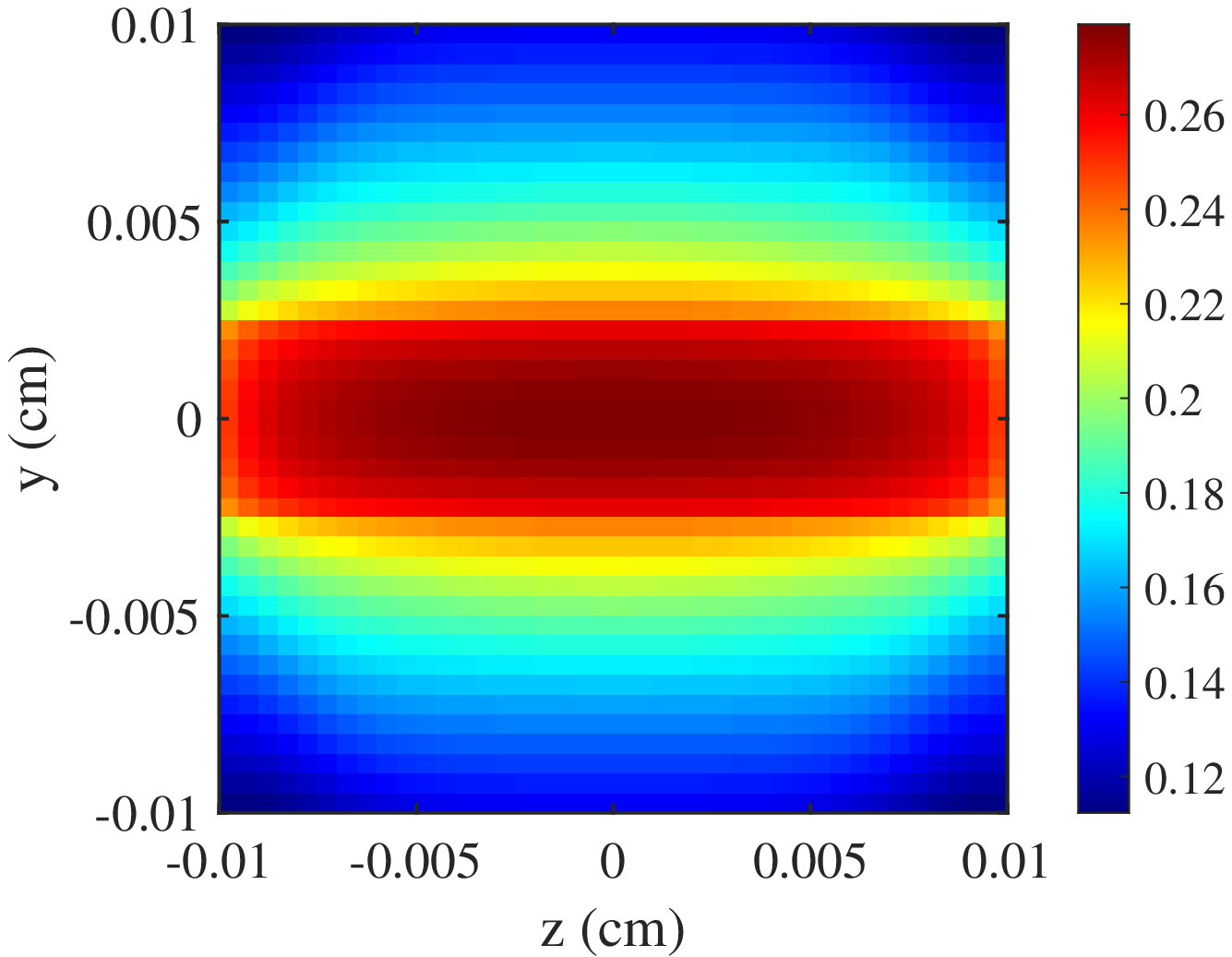}
	\subcaption{S$_{4}$, rank $4$ (full-rank)}
	\end{subfigure}
	\hfill
	\begin{subfigure}[b]{.327\linewidth}\centering
	\includegraphics[width=\textwidth]{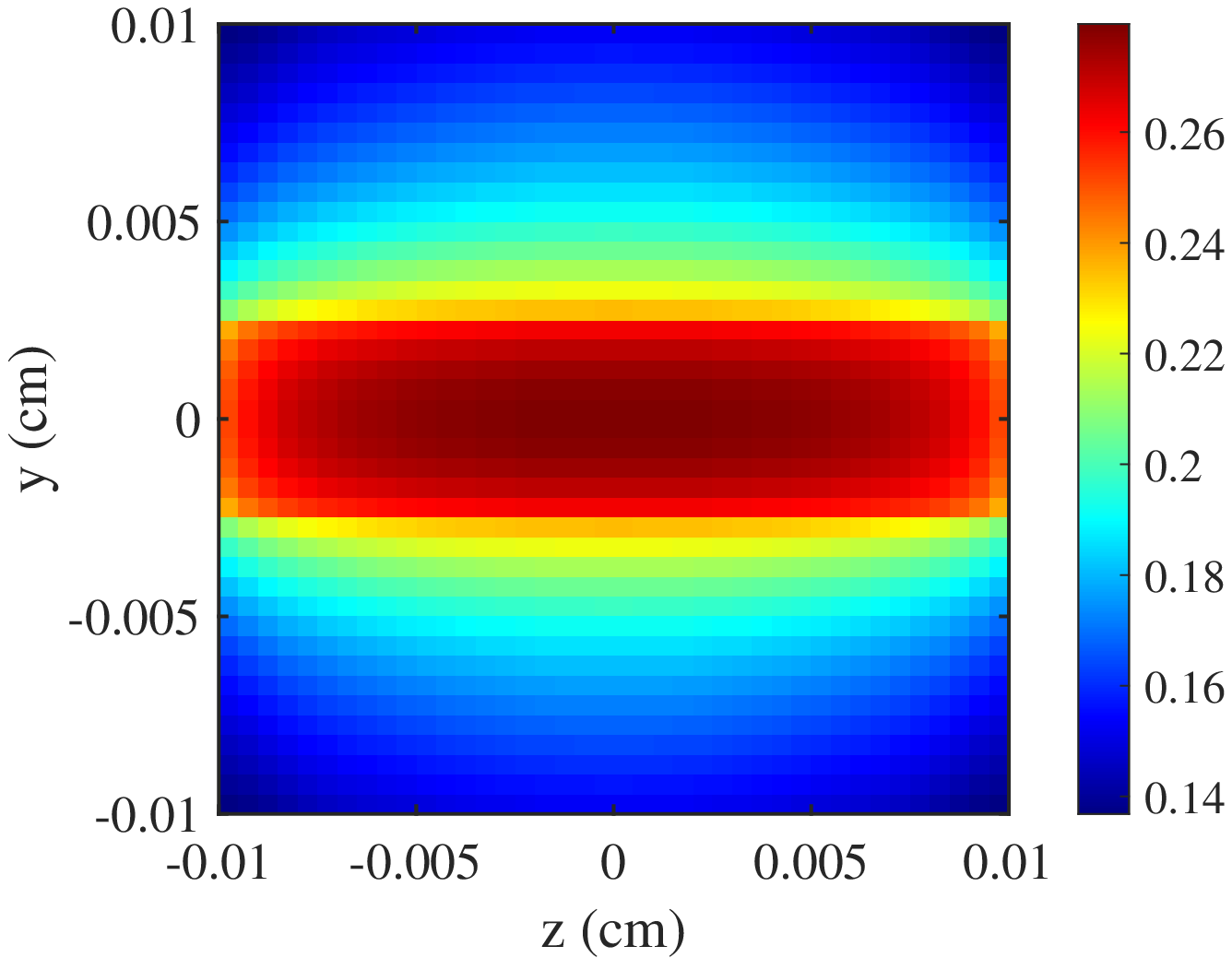}
	\subcaption{S$_{8}$, rank $16$ (full-rank)}
	\end{subfigure}
	\hfill
	\begin{subfigure}[b]{.327\linewidth}\centering
	\includegraphics[width=\textwidth]{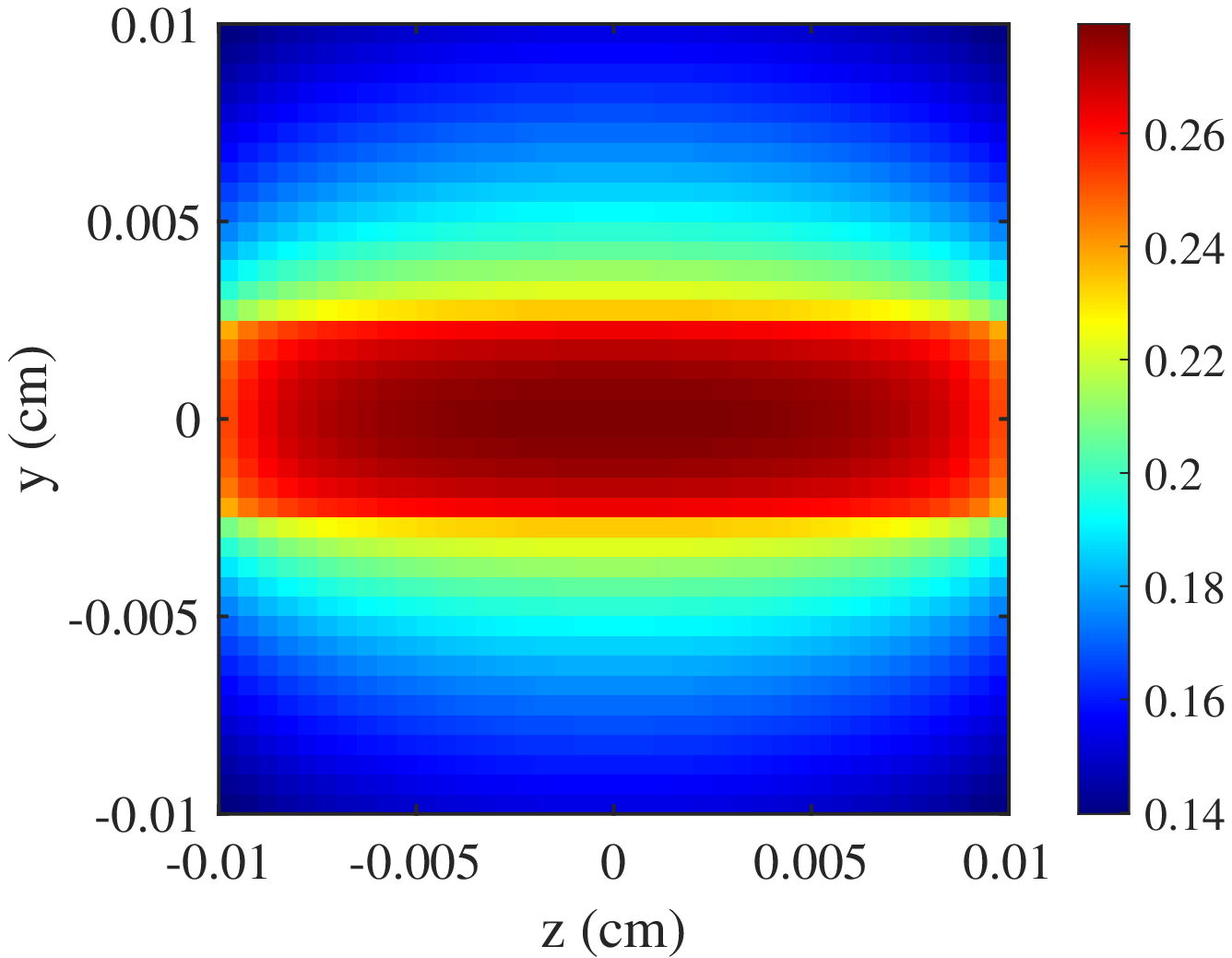}
	\subcaption{S$_{12}$, rank $36$ (full-rank)}
	\end{subfigure}
	\hfill
	\begin{subfigure}[b]{.327\linewidth}\centering
	\includegraphics[width=\textwidth]{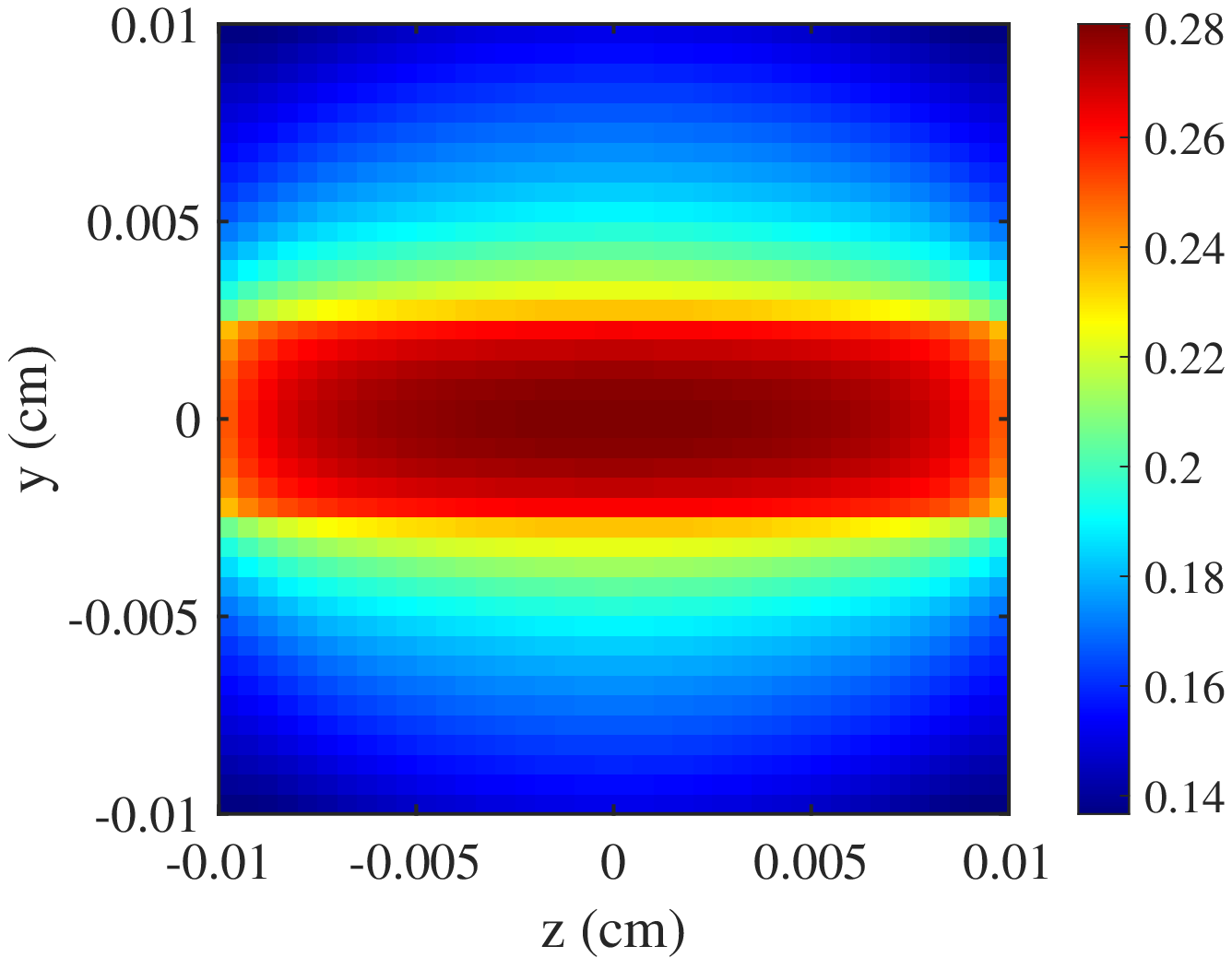}
	\subcaption{S$_{64}$, rank $4$}
	\end{subfigure}
	\hfill
	\begin{subfigure}[b]{.327\linewidth}\centering
	\includegraphics[width=\textwidth]{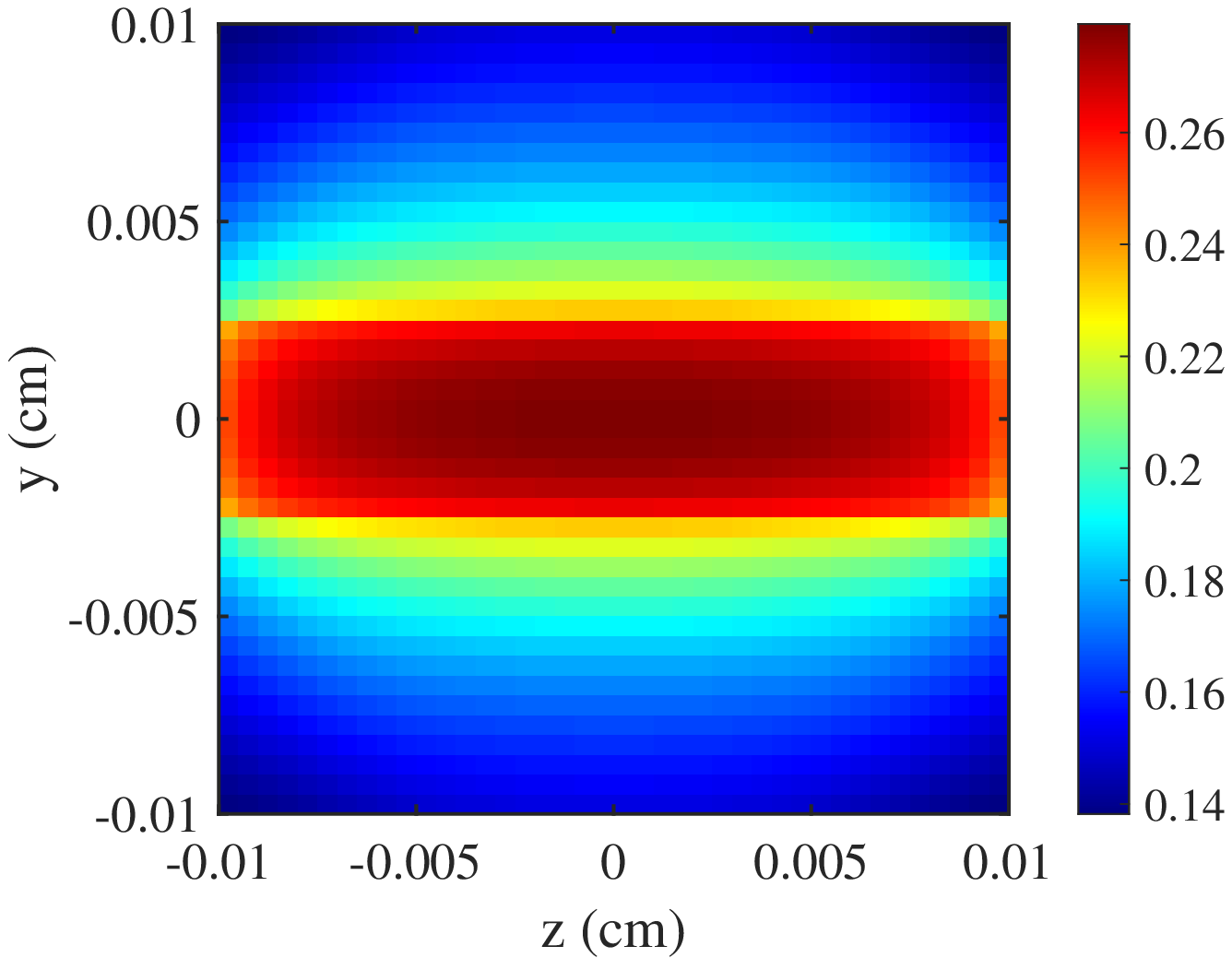}
	\subcaption{S$_{64}$, rank $16$}
	\end{subfigure}
	\hfill
	\begin{subfigure}[b]{.327\linewidth}\centering
	\includegraphics[width=\textwidth]{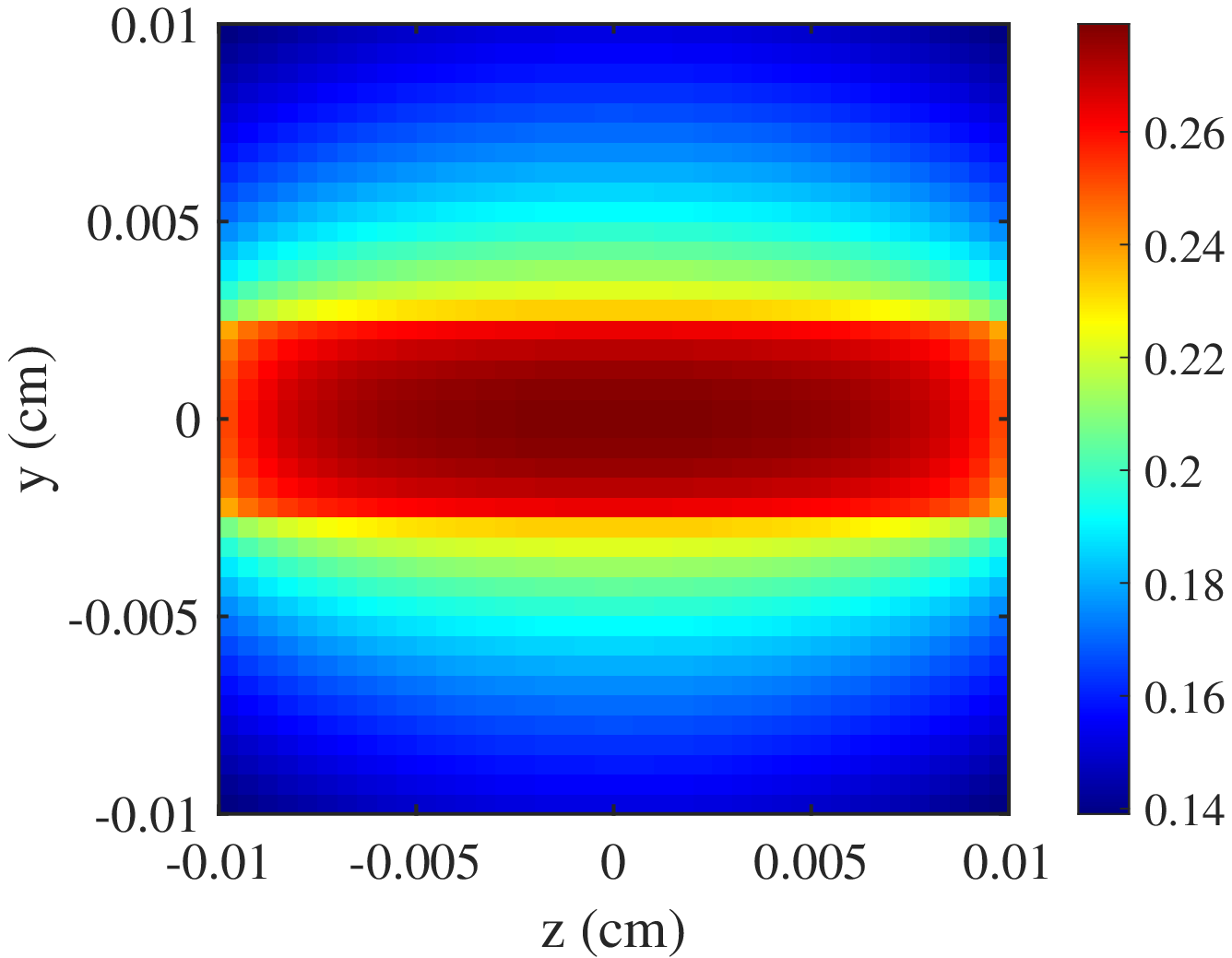}
	\subcaption{S$_{64}$, rank $36$}
	\end{subfigure}
	\caption{{The logarithmic scalar intensity solution to the wires problem calculated by the low-rank method with S$_{64}$ are compared to the same rank solutions without rank reduction.}}
	\label{fig: wires_compare_yz}
\end{figure}

We also plot the solution in the y-z plane along $x = 0$ shown in Figure \ref{fig: wires_compare_yz}. 
We can see that the solution with rank 4 is has large, qualitative errors, especially in the center source region. The low-rank solution significantly improves the situation with the errors greatly reduced.

\begin{figure} 
	\centering
	\includegraphics[width=\textwidth]{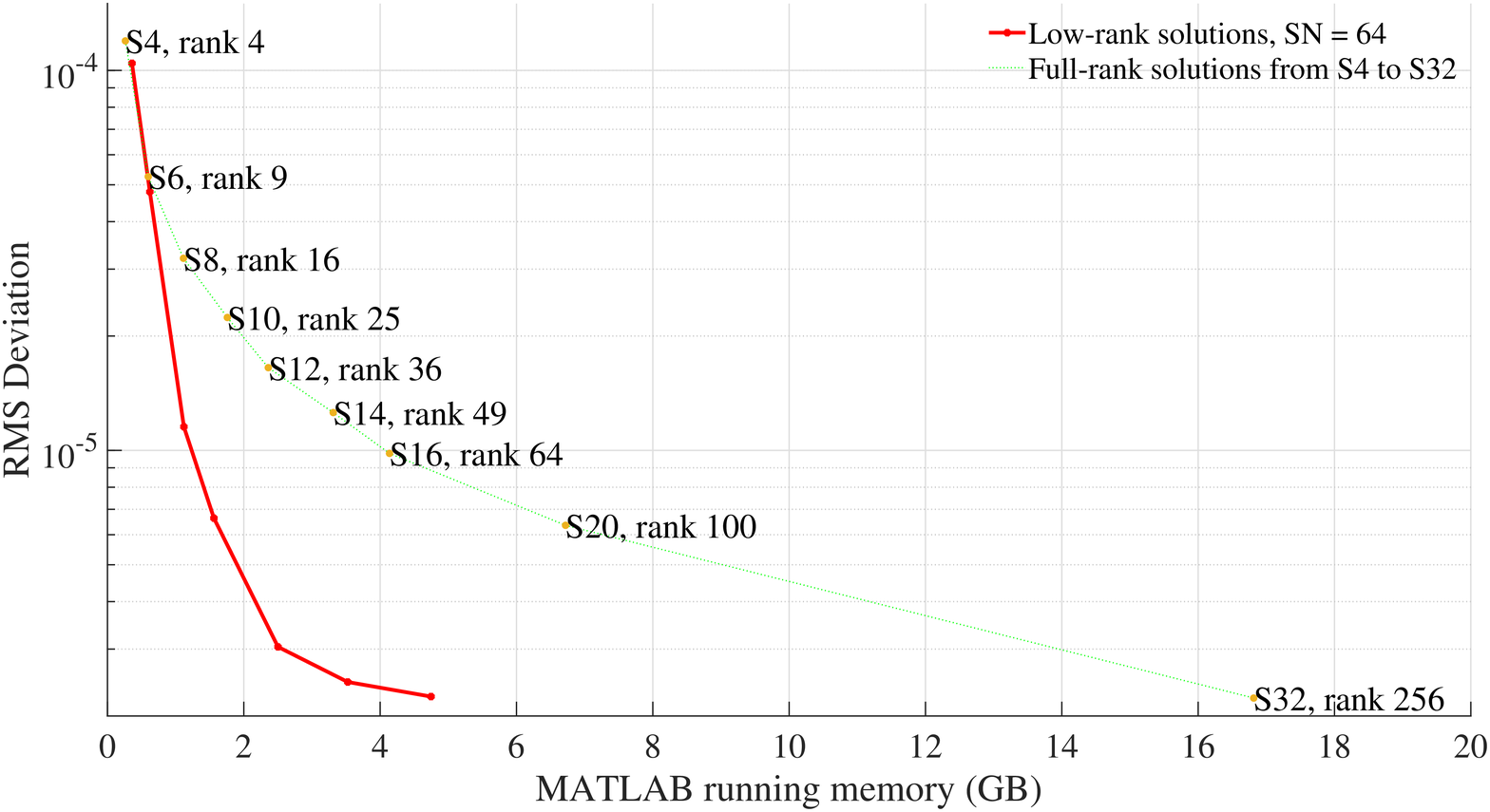}
	\caption{{The comparison of errors for the wires problem with different memory usage. The green dots represent the error of the full rank solutions that varies the number of discrete ordinates. The red solid line represents the error of the low-rank solutions with S$_{64}$ that varies the rank (rank = $[4, 9, 16, 25, 36, 49, 64]$).}}
	\label{fig: wires_memory}
\end{figure}

The running memory to carry out these simulations in MATLAB is shown in Figure \ref{fig: wires_memory}.  The rank 64 solution with S$_{64}$, represented by the rightmost dot in the solid red line, uses 25\% of the full rank $S_{32}$ solution but achieves the same accuracy. It also reaches five times lower error than the full rank $S_{16}$ solution with slightly more memory.

\section{Conclusion} 
In this work, we have presented a low-rank-S$_\mathrm{N}$ scheme to simulate discrete ordinate (\SN) radiative transfer equations using reduced computational costs. By applying the unconventional low-rank integrator, our scheme can be solved implicitly by transport sweeps, enabling more efficient radiation transport simulations on computing platforms with reduced memory per core. With the carefully chosen rank based on the intrinsic property of problems, this low-rank method can save up to $90\%$ of the computer memory and $95\%$ of the computational time. We also observe that if the rank is chosen to be too large, the solution does not suffer, but the calculation is not maximally efficient. 

We point out that the effect of the computational cost saving on a specific problem relies on the discovery of its ``true" rank. We are working on various radiative transfer problems simulations to find a practical strategy for selecting the rank. Another open question in low-rank methods is conserving the intrinsic properties of the underlying problem such as conservation of particles or the appropriate asymtotic limits. As in our previous work \cite{peng2021holo}, the high-order/low-order (HOLO) algorithm is one way to perform the fix. The implementation is straightforward since we can also calculate the Eddington tensor with the low-rank \SN~solution. Furthermore, we would be interested in applying the low-rank method to the multi-group transport problems for future study. This would give another dimension to potentially compress and potentially give even greater memory savings. Recent work on diffusion-based particle transport problems with multigroup indicates that this could be a fruitful investigation \cite{kusch2021}.

\nolinenumbers

\newpage


\clearpage
\bibliography{main}

\end{document}